\documentclass[aps,                
               prc,                
               showpacs,           
               notitlepage,
               onecolumn,          
               10pt,               
               twoside,            
               floatfix,           
               superscriptaddress] 
               {revtex4-1}

\usepackage{mathptmx}
\usepackage[T1]{fontenc}
\usepackage{geometry}         
\usepackage{fancyhdr}         
\usepackage{graphicx}         
\usepackage[svgnames]{xcolor} 
\usepackage{hyperref}         
\usepackage{amsmath}          
\usepackage{amssymb}          
\usepackage{amsfonts}         
\usepackage{amstext}          
\usepackage{textcomp}         
\usepackage{slashed}

\hypersetup{colorlinks=true,linkcolor=Blue,citecolor=Blue,urlcolor=Blue}

\DeclareMathAlphabet{\mathcal}{OMS}{cmsy}{m}{n}

\newcommand{\bs}[1]{p\!\!\!\!\! p \!\!\!\!\! p\!\!\!\!\! p}
\newcommand{\bsp}[1]{p\!\!\!\!\! p \!\!\!\!\! p\!\!\!\!\! p}
\newcommand{\bsq}[1]{q\!\!\!\!\! q}
\newcommand{\bsA}[1]{A\!\!\!\!\! A}
\newcommand{\bsP}[1]{P\!\!\!\!\! P}
\newcommand{\bsAA}[1]{A\!\!\!\!\!\!\! A}
\newcommand{\bsa}[1]{a\!\!\!\!\! a}
\newcommand{\bsu}[1]{u\!\!\!\!\! u}
\newcommand{\bsbt}[1]{\beta\!\!\!\!\!\! \beta \!\!\!\!\!\! \beta \!\!\!\!\!\! \beta}

\numberwithin{equation}{section}                                

\fancypagestyle{firststyle}
{%
  \fancyhf{}
  \fancyhead[C]{PHYSICAL REVIEW C {\bf 87}, 043912 (2013)} 
  \fancyfoot[L]{\small 0556-2813/2013/87(1)/043912(25)} 
  \fancyfoot[C]{\small 043912-\thepage} 
  \fancyfoot[R]{\small \textcopyright 2013 American Physical Society} 
}

\begin{document}

\title{\texorpdfstring{\begin{minipage}[c]{\textwidth}\centering Dynamical collisional energy loss and transport properties
of on- and off-shell heavy quarks in vacuum and in the Quark Gluon Plasma \end{minipage}}{Dynamical collisional energy loss and transport properties of on- and off-shell heavy quarks in vacuum and in the Quark Gluon Plasma}}

\author{H.~Berrehrah}
\email{berrehrah@fias.uni-frankfurt.de}
\affiliation{\begin{minipage}[c]{\textwidth}Frankfurt Institute for Advanced Studies and Institute for Theoretical Physics, Johann Wolfgang Goethe Universit\"at, Ruth-Moufang-Strasse 1,\end{minipage}\\
60438 Frankfurt am Main, Germany\\}

\author{P.B.~Gossiaux}
\email{gossiaux@subatech.in2p3.fr}
\affiliation{\begin{minipage}[c]{0.98\textwidth}Subatech, UMR 6457, IN2P3/CNRS, Universit\'e de Nantes, \'Ecole des Mines de Nantes, 4 rue Alfred Kastler, 44307 Nantes cedex 3, France\end{minipage}\\}

\author{J.~Aichelin}
\email{aichelin@subatech.in2p3.fr}
\affiliation{\begin{minipage}[c]{0.98\textwidth}Subatech, UMR 6457, IN2P3/CNRS, Universit\'e de Nantes, \'Ecole des Mines de Nantes, 4 rue Alfred Kastler, 44307 Nantes cedex 3, France\end{minipage}\\}

\author{W.~Cassing}
\email{wolfgang.cassing@theo.physik.uni-giessen.de}
\affiliation{\begin{minipage}[c]{\textwidth}Institut für Theoretische Physik, Universit\"at Giessen, 35392 Giessen, Germany\end{minipage}\\ \vspace{2.5mm}} 

\author{E.~Bratkovskaya}
\email{brat@th.physik.uni-frankfurt.de}
\affiliation{\begin{minipage}[c]{\textwidth}Frankfurt Institute for Advanced Studies and Institute for Theoretical Physics, Johann Wolfgang Goethe Universit\"at, Ruth-Moufang-Strasse 1,\end{minipage}\\
60438 Frankfurt am Main, Germany\\}


\begin{abstract}
   In this study we evaluate the dynamical collisional energy loss of heavy quarks, their interaction rate as well as the different transport coefficients (drag and diffusion coefficients, $\hat{q}$, etc). We calculate these different quantities for i) perturbative partons (on-shell particles in the vacuum with fixed and running coupling) and ii) for
dynamical quasi-particles (off-shell particles in the QGP medium at finite temperature $T$ with a running coupling in temperature as described by the dynamical quasi-particles model). We use the
    perturbative elastic $(q(g) Q \rightarrow q (g) Q)$ cross
    section for the first case, and the Infrared Enhanced Hard Thermal Loop cross sections for the second. The results obtained in this work
    demonstrate the effects of a finite parton mass and width on the heavy quark transport properties and provide the basic ingredients for
    an explicit study of the microscopic dynamics of heavy flavors in the QGP - as formed in relativistic heavy-ion collisions -
    within transport approaches developed previously by the authors.

\end{abstract}


\maketitle

\section{Introduction}
\label{Introduction}
\vskip -2mm
In ultrarelativstic heavy-ion collisions there is strong circumstantial evidence that gluons and light quarks ($u$,$d$,$s$) come to
an equilibrium and form a plasma of quarks and gluons (QGP) \cite{Adams:2005dq,Adcox:2004mh,Back:2004je,Arsene:2004fa,Zajc:2007ey,Cassing:2008sv}.
As a consequence, mesons and baryons  containing only light quarks are formed with a multiplicity which corresponds to that expected for a system
in statistical equilibrium. This observation limits strongly the use of such mesons and baryons to study the properties of the QGP during its expansion
because after thermal freeze out the information about the previous phase is lost.  Heavy quarks ($c$ and $b$) do not come to an equilibrium with the
plasma degrees of freedom and may therefore serve as a probe to study the properties of the deconfined system in the early phase of the reaction.

It has been shown in experiments at the Relativistic Heavy Ion Collider (RHIC) and at the CERN Large Hadron Collider (LHC) accelerators  that
high momentum charm and bottom quarks experience a remarkable quenching in the QGP despite of their large mass. This quenching is reflected
in the nuclear modification factor $R_{AA}$, i.e. the ratio of the transverse momentum spectra of heavy quarks in heavy-ion collisions to that
in proton-proton collisions, properly scaled by the number of binary proton-proton collisions in a heavy-ion reaction. Usually, such a quenching is
attributed to the energy loss of heavy quarks in the QGP, although alternative explanations are possible. Besides the $R_{AA}$, the elliptic flow $v_2$
of heavy quarks has turned out to be an important observable. Initially neither the heavy quarks, produced in hard collisions, nor the constituents
of the plasma have sizeable nonzero elliptic flow. In the hydrodynamically expanding plasma the spatial eccentricity of the overlap region of projectile
and target is converted into an elliptic flow in momentum space. In turn, heavy quarks may acquire an elliptic flow by interactions with the plasma
constituents. It is remarkable that experimentally the elliptic flow of heavy mesons (including c-quark degrees of freedom) equals almost that of light mesons.

These observations have triggered many studies
\cite{DuttMazumder:2004xk,Mustafa:2004dr,Svetitsky:1987gq,vanHees:2004gq,Moore:2004tg,vanHees:2005wb,Gossiaux:2004qw,Gossiaux:2006yu,Armesto:2005mz,vanHees:2007me,Molnar:2004ph,Zhang:2003wk,Gossiaux:2008jv,Gossiaux:2009hr,Linnyk:2008hp},
first by exploiting the single-electron data \cite{Wicks:2005gt,PhysRevC.78.014904,Bielcik:2005wu,Adler:2005xv,Rapp:2009my}, later by investigating
identified $D$-mesons. Despite of the many efforts a consensus on how to treat the interactions of heavy quarks with the plasma has not been reached yet.
Early predictions adopted the natural starting point to calculate the elastic interaction of the heavy quarks with the constituents of the plasma, the
light quarks and gluons, on the basis of perturbative QCD (pQCD) scattering cross sections, making a rather arbitrary choice of the coupling constant
and the infrared regulator of the gluon propagator. These calculations underestimate significantly the energy loss and the elliptic flow of heavy quarks, i.e.
the coupling of heavy quarks with the QGP. Later on radiative energy loss has been considered. It increases the energy loss but it is difficult to calculate
due to the interplay between Landau Pomeranschuck Migdal (LPM) effects and the rapid medium expansion.
In addition, it has also been suggested that heavy hadronic resonances could be formed in the plasma which might yield an additional energy loss \cite{Rapp:2009my}.

To be comparable with the experimental data these elementary interactions between heavy quarks and the plasma constituents have to be embedded
in a model which describes the expansion of the QGP itself. The description of this expansion is not unique and different expansion scenarios
may easily change the $R_{AA}$ and $v_2$ values by a factor of two \cite{Gossiaux:2011ea}. This observation, as well as the many parameters which
enter the calculations of the elementary interaction between heavy quarks and plasma constituents, which include quark and gluon masses, coupling constants,
spectral functions and infrared cutoffs, make it useful to study transport coefficients in which the complicated reaction dynamics is reduced to a couple
of coefficients which can be directly compared between different models (or with correlators from lattice QCD). It is the purpose of this article
to present these transport coefficients for
different models in continuation of Ref. \cite{Berrehrah:2013mua} in which the different elementary cross sections have been computed.

In line with Ref. \cite{Berrehrah:2013mua} we concentrate on the following models:
\begin{itemize}
\item HTL-GA (Hard Thermal Loop-Gossiaux Aichelin) approach. The details of this model can be found in Refs. \cite{Berrehrah:2013mua,Gossiaux:2008jv,Gossiaux:2009mk}.
As compared to former pQCD approaches this model differs in the description of the interaction between the heavy quarks and the plasma particles in two respects:

     \begin{itemize}
     \item an effective running coupling constant extended in the non-perturbative domain that remains finite at vanishing four-momentum transfer $t \rightarrow 0$
     in line with \cite{Dokshitzer:1995qm}.

     \item an infrared regulator $\mu$ in the $t$-channel which is determined from hard thermal loop calculations of Braaten and
     Thoma \cite{PhysRevD.44.R2625} including a running coupling $\alpha$. In practice, we use an effective gluon propagator with a 'global'
     $\mu^2 = \kappa \tilde{m}_D^2$, where $\tilde{m}_D^2 (T) = \frac{N_c}{3} \left(1 + \frac{N_f}{6} \right) 4 \pi \alpha (- \tilde{m}_D (T))
     \ T^2$, for $q  Q \rightarrow q Q$ and $g  Q \rightarrow g Q$ scattering. The parameter $\kappa$ is determined by requiring that the energy
     loss $d E/d x$, obtained with this effective propagator, reproduces the running coupling $\alpha$ result in the extended HTL calculation
     in the t-channel from Refs. \cite{PhysRevC.78.014904,Gossiaux:2009hr,Gossiaux:2009mk}. The resulting value is $\kappa \approx 0.2$.
     The details of this model are described in the appendix of Ref. \cite{Gossiaux:2008jv}.
     \end{itemize}

Using the latter  ingredients the cross section is increased as compared to the former pQCD studies, especially for small momentum
transfer \cite{Berrehrah:2013mua,Gossiaux:2008jv,Gossiaux:2009mk}.  In the HTL-GA approach we will also consider the case of a fixed coupling
constant ($\alpha = 0.3$) and of a Debye mass $m_D \approx \xi g_s T$ with $\xi = 1$ as infrared regulator, even if such naive pQCD calculations
are unable to reproduce the data, neither the energy loss nor the elliptic flow. The aim is to make contact with the previous calculations where
the gluon propagator in the $t$-channel Born matrix element has to be IR regulated by a fixed screening mass $\mu$ \cite{Combridge1979429,Svetitsky:1987gq}.
We will also report the results from standard pQCD calculations if available (cf. Moore and Teaney \citep{Moore:2004tg}).

In the HTL-GA approach, the heavy quarks are considered massive while the gluons and light quarks are massless. Nevertheless, we will study the effects
of finite masses of gluons and light quarks on the transport properties of the heavy quarks, too.

\item IEHTL (Infrared Enhanced Hard Thermal Loop) approach: This approach takes into account nonperturbative spectral functions and self-energies of the
quarks, antiquarks and gluons in the QGP at finite temperature. For this purpose, we use parametrizations of the quark and gluon masses and coupling constants
provided by the dynamical quasi-particle model (DQPM) which are  determined to reproduce lattice quantum chromodynamics (lQCD) results
\cite{Peigne:2008nd,PhysRevD.70.034016,0954-3899-31-4-046,PhysRevLett.110.182301,Wcassing2009EPJS,Cassing:2009vt,Cassing2007365,Cassing2009215} at finite temperature
and vanishing quark chemical potential $\mu$.
In this approach, the gluons and light and heavy quarks have finite masses and widths.

\item DpQCD (Dressed perturbative QCD) approach: In this approach the spectral functions of the IEHTL approach are replaced by the DQPM pole masses for
the incoming, outgoing and exchanged quarks and gluons. In this model the gluons and light and heavy quarks are massive but have zero widths.

\end{itemize}

The different models presented here aim to describe the heavy meson spectra and are related to the fundamental parameters of the microscopic
interactions of heavy quarks with the QGP partons.
The mesoscopic quantities, characterizing the transport properties of a heavy quark propagating in the QGP, are the transport coefficients formally expressed by
 the variable $\mathcal{X}$. Their time evolution is  given for on-shell partons by

{\setlength\arraycolsep{-10pt}
\begin{eqnarray}
\label{equ:Sec1.1}
& & \displaystyle \frac{d <\!\!\mathcal{X}^{\textrm{on}}\!\!>}{d \tau} = \sum_{q,g} \frac{1}{(2 \pi)^5 2 M_Q} \int \frac{d^3 q}{2 E_q} f (\bsq{q})
\int \frac{d^3 q'}{2 E_{q'}} \int \frac{d^3 p'}{2 E_{p'}} \delta^{(4)} (P_{in} - P_{fin}) \
\mathcal{X} \ \frac{1}{g_Q g_p} \sum_{k,l} |\mathcal{M}_{2,2}^{\textrm{on}} (p, q; i, j |p', q'; k,
l)|^2 .
\end{eqnarray}}
Here the indices $(i,j)$ stand for the discrete quantum numbers (spin $s$, flavor $f$ and color $c$) of the initial heavy quark (with the momentum $p$,
mass $M_Q$, energy $ E_p=\sqrt{\mathbf{p}^2 + m_Q^2}$ and velocity $v_Q$) and the initial light quark or gluon (with the momentum $q$, mass $m_q$,
energy $E_q=\sqrt{\bsq{q}^2 + m_q^2}$ and velocity $v_q$) while $(k,l)$ denote those of the heavy quark (with the momentum $p'$) and of the light
quark/gluon (with the momentum $q'$) in the final state. $g_Q$ is the degeneracy factor of the heavy quark ($g_Q = 6$) and $g_p$ is the degeneracy
factor of the parton ($g_p = 6$ for light quarks, $g_p = 16$ for massless gluons and $g_p = 24$ for massive gluons). Furthermore,
$\displaystyle \sum_{q,g}$ denotes the sum over the light quarks and gluons of the medium while  $|\mathcal{M}_{2,2} (p, q; i, j |p', q'; k, l)|^2$ is
the transition matrix-element squared, related to the cross section $\sigma_{2 \rightarrow 2}$ by:
{\setlength\arraycolsep{-10pt}
\begin{eqnarray}
\label{equ:Sec1.2}
& & \frac{1}{(2 \pi)^6} \! \int \! \frac{d^3 q'}{2 E_{q'}} \frac{d^3 p'}{2 E_{p'}} \ (2 \pi)^4 \delta^{(4)} (P_{in} - P_{fin})
\ \frac{1}{g_Q g_p} \sum_{k,l} |\mathcal{M}_{2,2} (p, q; i, j |p', q'; k, l)|^2 = 2\lambda(s,m_q^2,M_Q^2) \ \sigma_{2 \rightarrow 2} (\sqrt{s}),
\end{eqnarray}}
with
\begin{eqnarray}
\label{equ:Sec1.2a}
2\lambda(s,m_q^2,M_Q^2) = 4 E_p E_q|v_Q-v_q|
\end{eqnarray}
denoting the flux of the incoming particles and $\lambda(a,b,c) = a^2+b^2+c^2-2ab-2ac-2bc$.

Depending on the choice of $\mathcal{X}$ one can address different transport coefficients: For $\mathcal{X}$ = 1, $\mathcal{X}$ = $(E - E')$,
$\mathcal{X}$ = $(\bsp{p} - \bsp{p}')$, $\mathcal{X}$ = $(p - p')^{\mu}(p - p')^{\nu}$, the formula (\ref{equ:Sec1.1}) corresponds, respectively,
to the interaction rate $\mathcal{R} (p, T) = \frac{d N_{coll}}{d \tau}$, the energy loss $\frac{d E}{d \tau} (p, T)$, the drag coefficient
$\mathcal{A} (p, T)$ and finally the diffusion coefficient $\mathcal{B}^{\mu \nu} (p, T)$, all evaluated per unit proper time $\tau$ in the
rest system of the heavy quark.

In our study we use (\ref{equ:Sec1.1}) for the case of on-shell heavy quarks and partons with the transition amplitudes $\sum |\mathcal{M}_{2,2}|^2$
$\equiv \sum |\mathcal{M}_{2,2}^{HTL-GA}|^2, \ \ \sum |\mathcal{M}_{2,2}^{DpQCD}|^2$. Replacing the on-shell phase-space measure in (\ref{equ:Sec1.1})
by the off-shell equivalent
{\setlength\arraycolsep{-10pt}
\begin{eqnarray}
\label{equ:Sec1.3}
& & Tr_{(i)}^{\textrm{on}} = \int \frac{d^3 p_i}{(2 \pi)^3 2 E_i} \ \ \rightarrow \ \ Tr_{(i)}^{\textrm{off}} = \int \frac{d^4 p_i}{(2 \pi)^4} \rho_i (p_i) \Theta (\omega_i),
\nonumber\\
& & {} \displaystyle \frac{1}{2 E_p} \ \ \rightarrow \ \ \int \frac{d \omega}{2 \pi} \rho_p (\omega) \theta (\omega),
\end{eqnarray}}
 where $\rho_i$ is the spectral function of the off-shell particle $i$ and $\Theta$ denotes the Heaviside function, one can extend (\ref{equ:Sec1.1})
 to the case of off-shell partons to
{\setlength\arraycolsep{-6pt}
\begin{eqnarray}
\label{equ:Sec1.4}
\displaystyle \frac{d <\!\!\mathcal{X}^{\textrm{off}}\!\!>}{d \tau} = \sum_{q,g} \frac{1}{(2\pi)^{9}} Tr_{q}^{\textrm{off}} \ Tr_{p}^{\textrm{off}} \ Tr_{q'}^{\textrm{off}} \ Tr_{p'}^{\textrm{off}} \ f(\bsq{q}) \delta^{(4)} (P_{in} - P_{fin}) \ \mathcal{X} \ \frac{1}{g_Q g_p} \sum_{k,l} |\mathcal{M}_{2,2}^{\textrm{off}} (p, q; i, j |p', q'; k, l)|^2,
\end{eqnarray}}
  with $\sum |\mathcal{M}_{2,2}^{\textrm{off}}|^2 \equiv \sum |\mathcal{M}_{2,2}^{IEHTL}|^2$ as explained above.

In this study we will address the different transport coefficients of a heavy quark in a static medium at finite temperature. They give the  response of
the medium to the propagation of a heavy quark under fixed thermodynamical conditions. For out-of-equilibrium microscopic dynamics of the heavy quarks
in the QGP, a description of the medium evolution is required. This study is beyond the scope of this paper and can be realized within the
Parton-Hadron-String-Dynamics (PHSD) or Monte-Carlo $\alpha_s$ Heavy Quarks (MC@sHQ) transport approaches.

The paper is organized as follows: In Sec. \ref{DQPM_Reminder} we first briefly recall the Dynamical Quasi-Particle Model in order to fix the
main ingredients used in the DpQCD/IEHTL approaches. In Section \ref{ElasticInteractionRate} we evaluate the elastic interaction rate and the relaxation time
for the approaches introduced above.  The calculation of the transport coefficients is performed on the basis of these cross sections.
In Sections \ref{DragCoefficient} and \ref{EnergyLoss} we calculate the drag coefficient as well as the energy loss per unit length and the spatial
diffusion coefficient. Both the drag and diffusion coefficients are the transport coefficients which are necessary for Fokker-Planck transport approaches describing the time evolution of the heavy-quark  momentum distribution. In Section \ref{DiffusionCoefficient} we discuss the transport coefficient $\hat{q}$, the average change
of the square of the transverse momentum per unit length which is strongly related to the transverse diffusion. Finally, in Section \ref{TransportCrossSection} we
evaluate the transport cross section and perform a critical analysis of the applicability of the independent collision approach. We conclude in Section \ref{Summary} with a summary of our findings.
\section{Reminder of the dynamical quasi-particle model}
\label{DQPM_Reminder}

The dynamical quasi-particle model \cite{PhysRevD.70.034016,0954-3899-31-4-046,Wcassing2009EPJS} describes QCD properties in terms of ``resummed'' single-particle Green's functions
(in the sense of a two-particle irreducible (2PI) approach) and leads to a quasi-particle equation of state, which reproduces the QCD equation of state
extracted from lattice QCD calculations in Ref. \cite{Borsanyi:2010bp,Borsanyi:2010cj}. According to the DQPM, the constituents of the strongly coupled Quark
Gluon Plasma (sQGP) are strongly interacting massive partonic quasi-particles. Details of the approach can be found in \cite{Berrehrah:2013mua}. Due to the
finite imaginary parts of the selfenergies the partons of type $``i"$ are described by spectral functions $\rho_i (p)$ for which we can assume one of the
following forms:

\begin{itemize}

\item[$i)$] Lorentzian-$\omega$ form: the spectral function $\rho_i (p)$, with $p = (\omega, \bsp{p})$, is given by
\cite{PhysRevLett.110.182301,Wcassing2009EPJS,Cassing:2009vt,Cassing2007365,Cassing2009215}

{\setlength\arraycolsep{-10pt}
\begin{eqnarray}
\label{equ:Sec2.1.5}
& & \rho_i^L (\omega) = \frac{\gamma_i}{\tilde{E}_i} \biggl(\frac{1}{(\omega - \tilde{E}_i)^2 + \gamma_i^2} - \frac{1}{(\omega + \tilde{E}_i)^2 + \gamma_i^2} \biggr) \equiv \frac{4 \omega \gamma_i}{(\omega^2 - \bs{p}^2 - M_i^2)^2 + 4 \gamma_i^2 \omega^2}.
\nonumber\\
& & {} \textrm{with:} \ \tilde{E}_i^2 (\bs{p}) = \bs{p}^2 + M_i^2 - \gamma_i^2, \ \ \textrm{and} \ \ i \in [g, q, \bar{q}, Q, \bar{Q}].
\end{eqnarray}}

This spectral function (\ref{equ:Sec2.1.5}) is antisymmetric in $\omega$ and normalized as

{\setlength\arraycolsep{-10pt}
\begin{eqnarray}
\label{equ:Sec2.1.6}
& & \int_{- \infty}^{+ \infty} \frac{d \omega}{2 \pi} \ \omega \ \rho_i^L (\omega, \bs{p}) = \int_0^{+ \infty} \frac{d \omega}{2 \pi} \ 2 \omega
\ \rho_i^L (\omega, \bs{p}) = 1.
\end{eqnarray}}
$M_i$, $\gamma_i$ are the particle pole mass and width, respectively.

\item[$ii)$] Breit-Wigner-$m$ form: the spectral function $\rho_i (p)$ is given by the Breit-Wigner distribution:

{\setlength\arraycolsep{0pt}
\begin{eqnarray}
& & \rho_i^{BW} (m_i) = \frac{2}{\pi} \frac{m_i^2 \gamma_i^{\star}}{\left(m_i^2 - M_i^2 \right)^2 + (m_i \gamma_i^{\star})^2}, \ \ \textrm{with:} \ \ \int_0^{\infty} d m_i \ \rho_i^{BW} (m_i, T) = 1,
\label{equ:Sec2.1.7}
\end{eqnarray}}
where $M_i$ is the dynamical quasi-particle (DQPM) mass (i.e pole mass). Here $m \equiv m_i$ is the independent variable; it's related to
$\omega$ by $\omega^2 = m^2 + \bs{p}^2$ while  $\gamma^{\star}$ is related to $\gamma$ by the relation $2 \omega \gamma = m \gamma^{\star}$.

\end{itemize}

\begin{figure}[h!]
\begin{center}
 \hskip -1em
  \includegraphics[height=6.15cm, width=6.05cm]{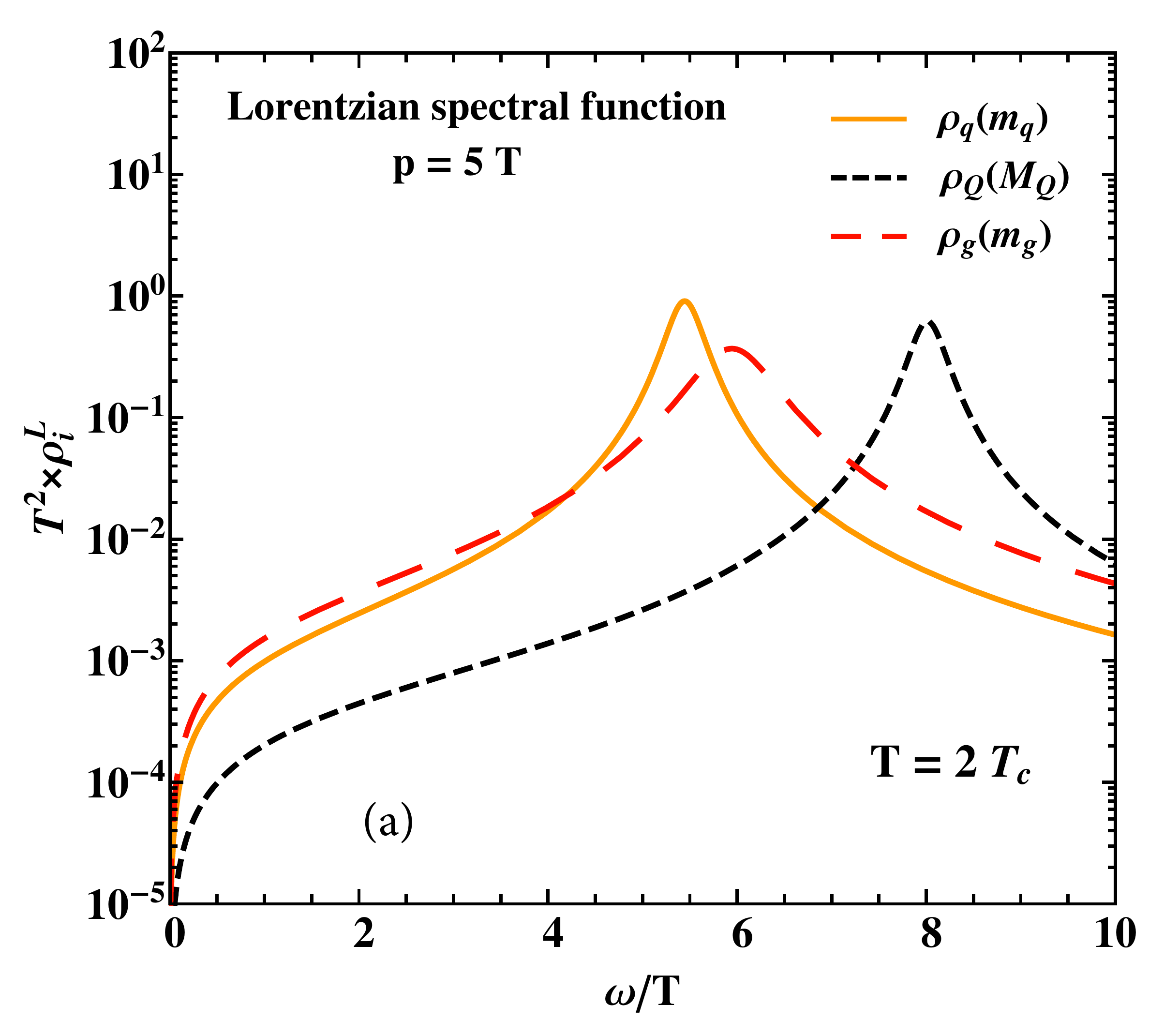}
  \includegraphics[height=6.15cm, width=6.05cm]{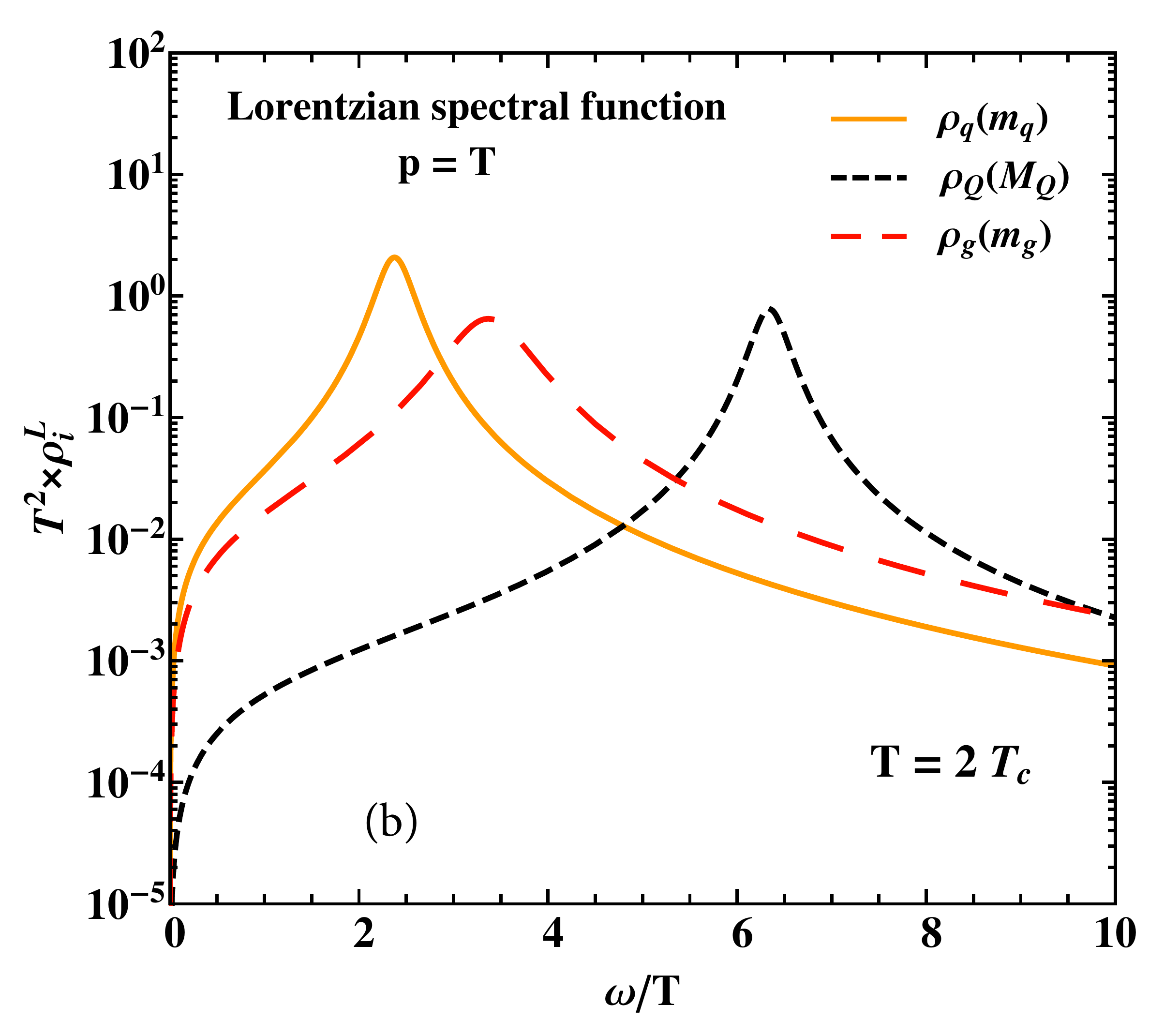}
  \includegraphics[height=6.15cm, width=6.05cm]{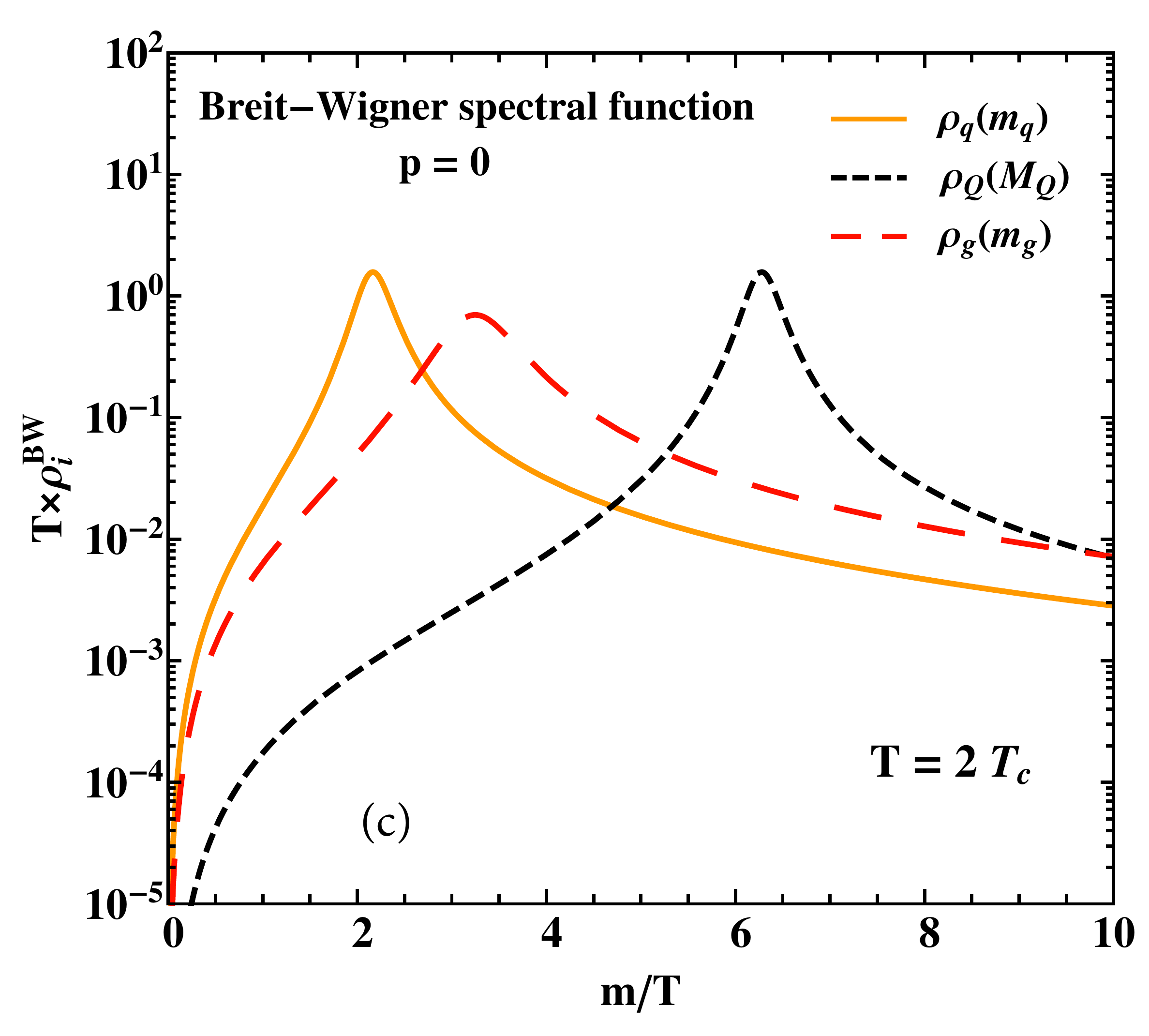}
\vskip -0.3cm
 \caption{\emph{(Color online) The DQPM spectral functions of light quarks, heavy quarks and gluons of (\ref{equ:Sec2.1.5})-(\ref{equ:Sec2.1.7})
 for $T/T_c = 2$ ($T_c =0.158$ GeV). $ \gamma_i$ and $M_i$ can be found in \cite{Berrehrah:2013mua}.  Left (a) (Center (b)):
 The Lorentzian spectral function as a function of $\omega/T$ for $p=5$ T ($p=$ T). Right (c): The Breit-Wigner spectral function as a function
 of $m/T$ for $p = 0$.}}
\label{fig:2DSpectralFunctionDQPM}
\end{center}
\end{figure}

The Breit-Wigner-$m$ form of the spectral function is a non-relativistic approximation of the Lorentzian-$\omega$ form
(neglecting $\parallel \!\! \bs{p} \!\! \parallel$ compared to $m$, $m \gg \parallel \!\! \bs{p} \!\! \parallel$).
Figure \ref{fig:2DSpectralFunctionDQPM} (l.h.s and center) shows the Lorentzian-$\omega$ form as a function of $\omega/T$ for different
values of the momentum $p$ (the Breit-Wigner-$m$ form as a function of $m/T$) of the spectral function of gluons, light and heavy quarks for $T = 2 T_c$ (r.h.s).
One sees that the peaks and the widths of these distributions are different. The Lorentzian-$\omega$ shape depends on the momentum
$\parallel \!\! \bs{p} \!\! \parallel$, whereas the Breit-Wigner-$m$ form is independent of $\parallel \!\! \bs{p} \!\! \parallel$. Comparing the
figures \ref{fig:2DSpectralFunctionDQPM}-left and center, we notice that an increasing momentum leads to a shift of poles masses to larger values of $\omega/T$.

\begin{figure}[h!] 
\begin{center}
   \begin{minipage}[c]{.495\linewidth}
     \includegraphics[height=7.8cm, width=8cm]{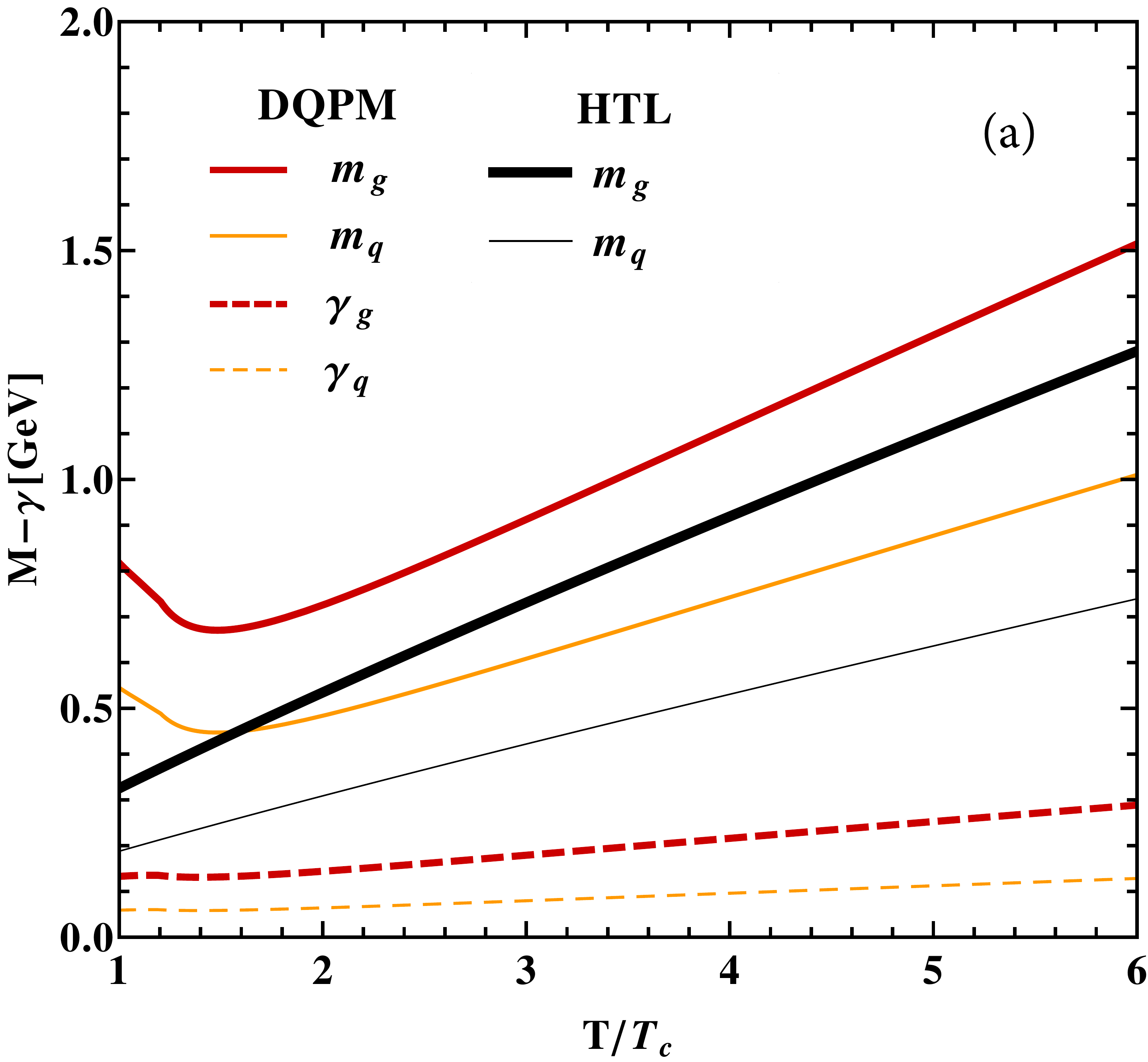}
   \end{minipage} 
   \begin{minipage}[c]{.495\linewidth}
      \includegraphics[height=7.8cm, width=8cm]{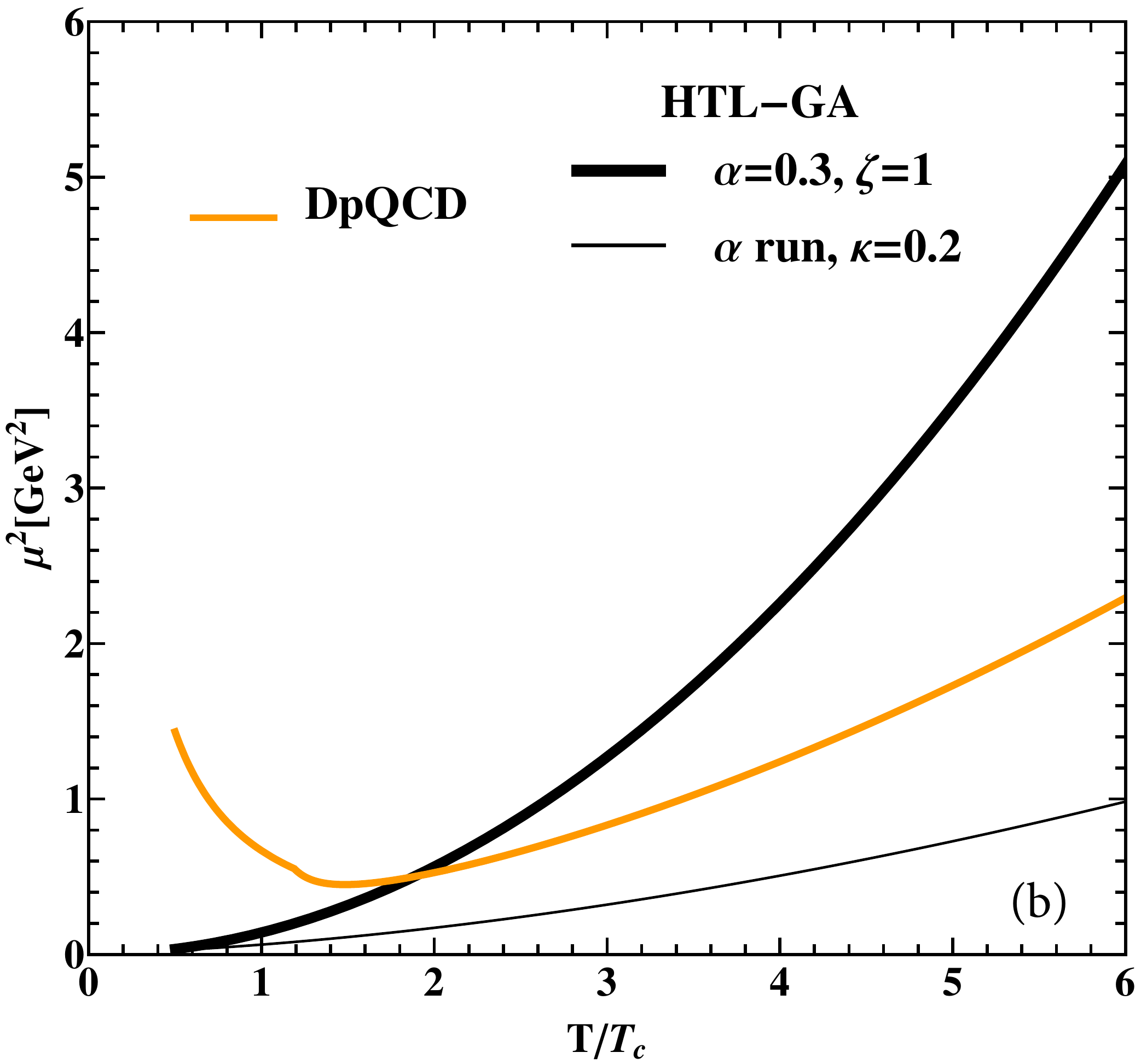}
\end{minipage}
\vspace{-0.1cm}
\caption{\emph{(Color online) Masses and widths of light quarks and gluons in the DQPM \cite{PhysRevC.87.024901} and masses in the HTL as a function of $T/T_c$ ($T_c = 0.158$ GeV) (left). The infrared regulator $\mu^2$ as a function of the temperature $T/T_c$ given in HTL-GA as well as in the DpQCD model (right). The parameters $\xi$, $\kappa$ and $\alpha$ in the HTL-GA model are described in the introduction.}}
\label{fig:MassDQPM-Mu2}
\end{center}
\end{figure}

The pole masses and widths for light quarks and gluons used in the DQPM are displayed in Fig. \ref{fig:MassDQPM-Mu2}-(a) as a function of $T/T_c$.
The functional forms for the dynamical pole masses of quasiparticles (gluons and quarks) are chosen in a way that they become identical to the perturbative thermal
masses in the asymptotic high-temperature regime \cite{Wcassing2009EPJS,Cassing:2009vt,Cassing2007365,Cassing2009215}. Fig. \ref{fig:MassDQPM-Mu2}-(a) gives also
the light quark and gluon masses in HTL, where $m_q = m_g/\sqrt{3}$ and $m_g = \tilde{m}_D (T)/\sqrt{3}$, with $\tilde{m}_D^2 (T) = \frac{N_c}{3}
\left(1 + \frac{N_f}{6} \right) 4 \pi \alpha (- \tilde{m}_D (T)) \ T^2$.

Figure \ref{fig:MassDQPM-Mu2}-(b) shows the infrared regulator $\mu^2$ used in the t-channel as a function of the medium temperature following the
different models presented in the introduction. We recall that $\mu^2$ is identical to the DQPM gluon mass squared in the DpQCD model and is given by
$\mu^2 = \frac{N_c}{3} \left(1 + \frac{N_f}{6} \right) 4 \pi \alpha T^2$ for the model HTL-GA with constant $\alpha$ and by $\mu^2 = \kappa \tilde{m}_D^2$,
where $\kappa \approx 0.2$ for the HTL-GA model with running $\alpha$. The small value of $\mu^2$ observed in HTL-GA with $\alpha$ running as
compared to the DpQCD value will have a strong effect on the transport coefficients (see below).

\section{Elastic Interaction rate and the associated relaxation time}
\label{ElasticInteractionRate}

We start out with the elastic interaction rate, i.e. the rate of the interaction of a heavy quark with momentum $\bsp{p}$ propagating through a QGP in thermal
equilibrium at a given temperature $T$. The quarks of the plasma are described by a Fermi-Dirac distribution $f_q (\bsq{q}) = \frac{1}{e^{E_q/ T} + 1}$
whereas the gluons follow a Bose-Einstein distribution $f_{g} (\bsq{q}) = \frac{1}{e^{E_g/ T} - 1}$. Therefore, the interaction rates are temperature dependent.
In our calculation we use the elastic scattering cross sections of Ref. \cite{Berrehrah:2013mua} for $q (\bar{q})Q$ and $g Q$ collisions, for on-shell as well
as for off-shell partons.

\subsection{Elastic interaction rates for on-shell partons}

For on-shell particles (and in the reference system in which the heavy quark has the velocity $\bsbt{\beta} = \bs{p}/E_p$) the interaction rate for
$2 \rightarrow 2$ collisions is given by $R^{\textrm{on}} (\bs{p}) = \frac{d N_{coll}^{2 \rightarrow 2}}{dt} =
\frac{M_Q}{E_p} \frac{d N_{coll}^{2 \rightarrow 2}}{d \tau}$ (see eq.(\ref{equ:Sec1.1})). We introduce the invariant quantity


{\setlength\arraycolsep{0pt}
\begin{eqnarray}
\label{equ:Sec3.3}
r^{\textrm{on}} (s) &=& \frac{1}{(2\pi)^2}\int \frac{d^3 q'}{2E_{q'}} \int \frac{d^3 p'}{2E_{p'}} \delta^{(4)} (P_{in} - P_{fin}) \times \frac{1}{g_Q g_p} \sum_{k,l} |\mathcal{M}_{2,2} (p, q; i, j |p', q'; k, l)|^2
\nonumber\\
&=& \frac{p_{cm}}{16 \pi^2 \sqrt{s}} \!\! \int \!\! d \Omega_{q'} \ \frac{1}{g_Q g_p} \sum_{k,l} |\mathcal{M}_{2,2} (p, q; i, j |p', q'; k, l)|^2 (s, \Omega_{q'})
\nonumber\\
& =& \frac{ \pi }{16 \pi^2 \sqrt{s}}\frac{1}{ p_{cm} (s)} \int_{- 4 p_{cm}^2}^0 d t \ \frac{1}{g_Q g_p} \sum_{k,l} |\mathcal{M}_{2,2} (p, q; i, j |p', q'; k, l)|^2 (s,t),
\end{eqnarray}}
  where
{\setlength\arraycolsep{0pt}
\begin{eqnarray}
\label{equ:Sec3.3a}
p_{cm}=\lambda^{1/2}(s,m_q^2,M_Q^2)/(2\sqrt{s}),
\end{eqnarray}}
 is the momentum of the scattering partners in the c.m. frame. The interaction rate in the plasma rest system for a heavy quark with momentum $\bsp{p}$ is given,
 following (\ref{equ:Sec1.1}), by

\begin{eqnarray}
\label{equ:Sec3.4}
R^{\textrm{on}} (\bs{p}) = \sum_{q,g} \frac{1}{2 E_p} \int \frac{d^3 q}{(2 \pi)^3 \ 2 E_q} f (\bsq{q}) \ r^{\textrm{on}} (s)= \sum_{q,g} \int \frac{d^3 q}{(2 \pi)^3} \ f (\bsq{q}) |v_Q-v_q| \sigma (s),
\end{eqnarray}
 where $\displaystyle \sum_{q,g}$ denotes the sum over the light quarks and gluons of the medium. The integral in the middle of (\ref{equ:Sec3.4}) does not
 depend on the reference frame and therefore it  is convenient to perform  the calculation in the rest frame of the heavy quark,

\begin{eqnarray}
\label{equ:Sec3.5}
R^{\textrm{on}} (\bs{p}=0) = \sum_{q,g} \frac{1}{2 M_Q} \int \frac{d^3 q}{(2 \pi)^3 \ 2 E_q} f_r (\bsq{q}) \ r^{\textrm{on}} (s),
\end{eqnarray}
 where $f_r (\bsq{q})$ is the invariant distribution of the plasma constituents in the rest frame of the heavy
 quark:

{\setlength\arraycolsep{0pt}
\begin{eqnarray}
\label{equ:Sec3.8}
\int d\Omega\ f_r (\bsq{q})= 2\pi \int d cos\theta \frac{1}{e^{(u^0 E_q - u \ q \cos \theta)/T}\pm 1},
\end{eqnarray}}
 with $u\equiv (u^0,\bsu{u}) = \frac{1}{M_Q}(E_p,-\bs{p})$ being the fluid 4-velocity measured in the heavy quark rest frame,
 while $\theta$ is the angle between $\bsq{q}$ and $\bsu{u}$.

\subsection{Elastic interaction rates for off-shell partons}

For off-shell partons the elastic interaction rate (\ref{equ:Sec1.4}) is obtained by replacing  $\int \frac{d^3 p}{(2 \pi)^3} \frac{1}{2 E_p}$ by
$ \int \frac{d^4 p}{(2 \pi)^4} \rho (p) \Theta (p_0)$ with $\rho (p)$ being the spectral function which can be specific for each particle species
(\ref{equ:Sec1.3}). For the two forms of spectral functions,  the Lorentzian-$\omega$ form and the  Breit-Wigner-$m$ form, the explicit results are given in
appendix \ref{AppendixA} for completeness.



\begin{figure}[h!] 
\begin{center}
   \begin{minipage}[c]{.495\linewidth}
     \includegraphics[height=8.8cm, width=9cm]{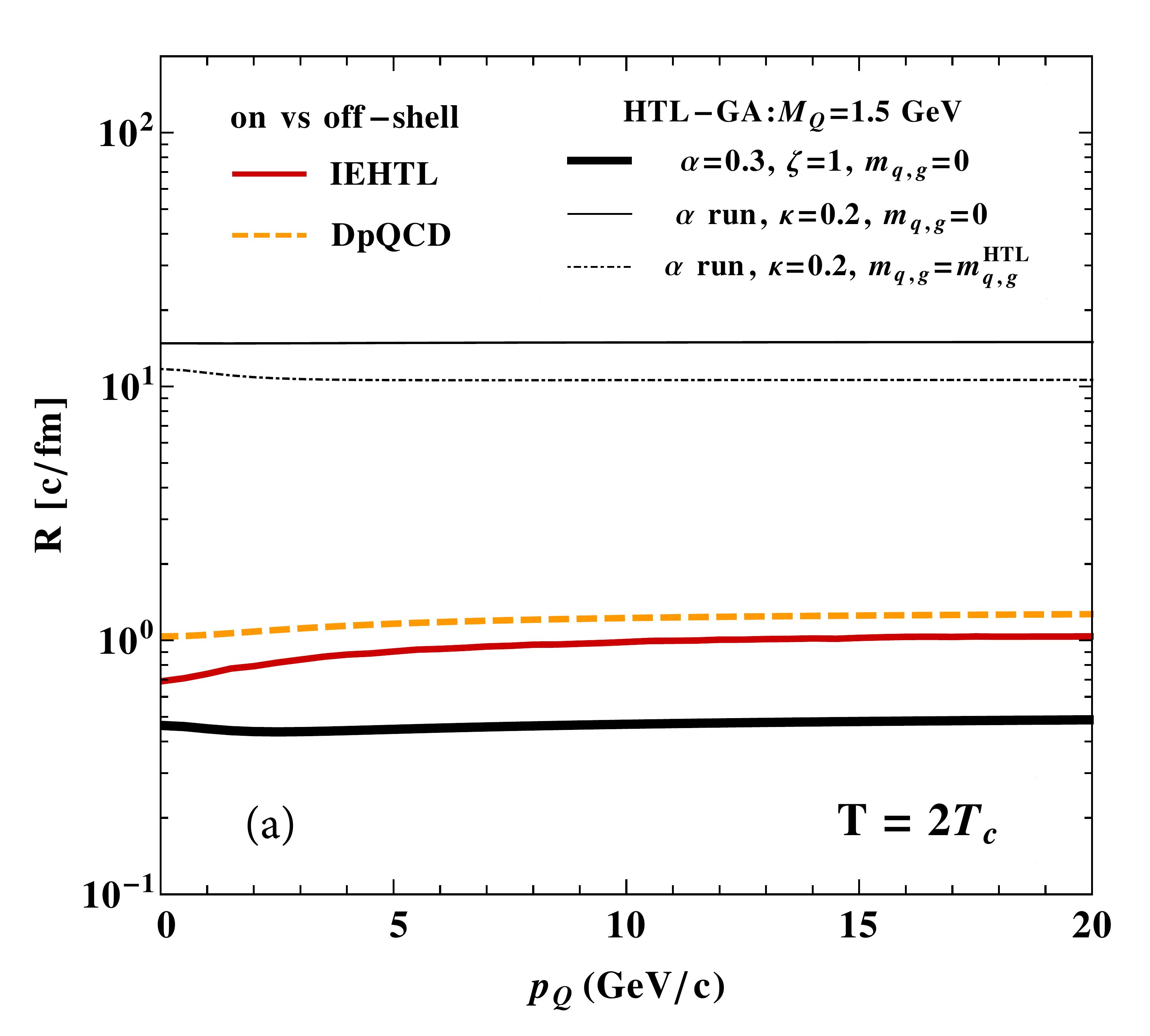}
   \end{minipage} 
   \begin{minipage}[c]{.495\linewidth}
      \includegraphics[height=8.8cm, width=9cm]{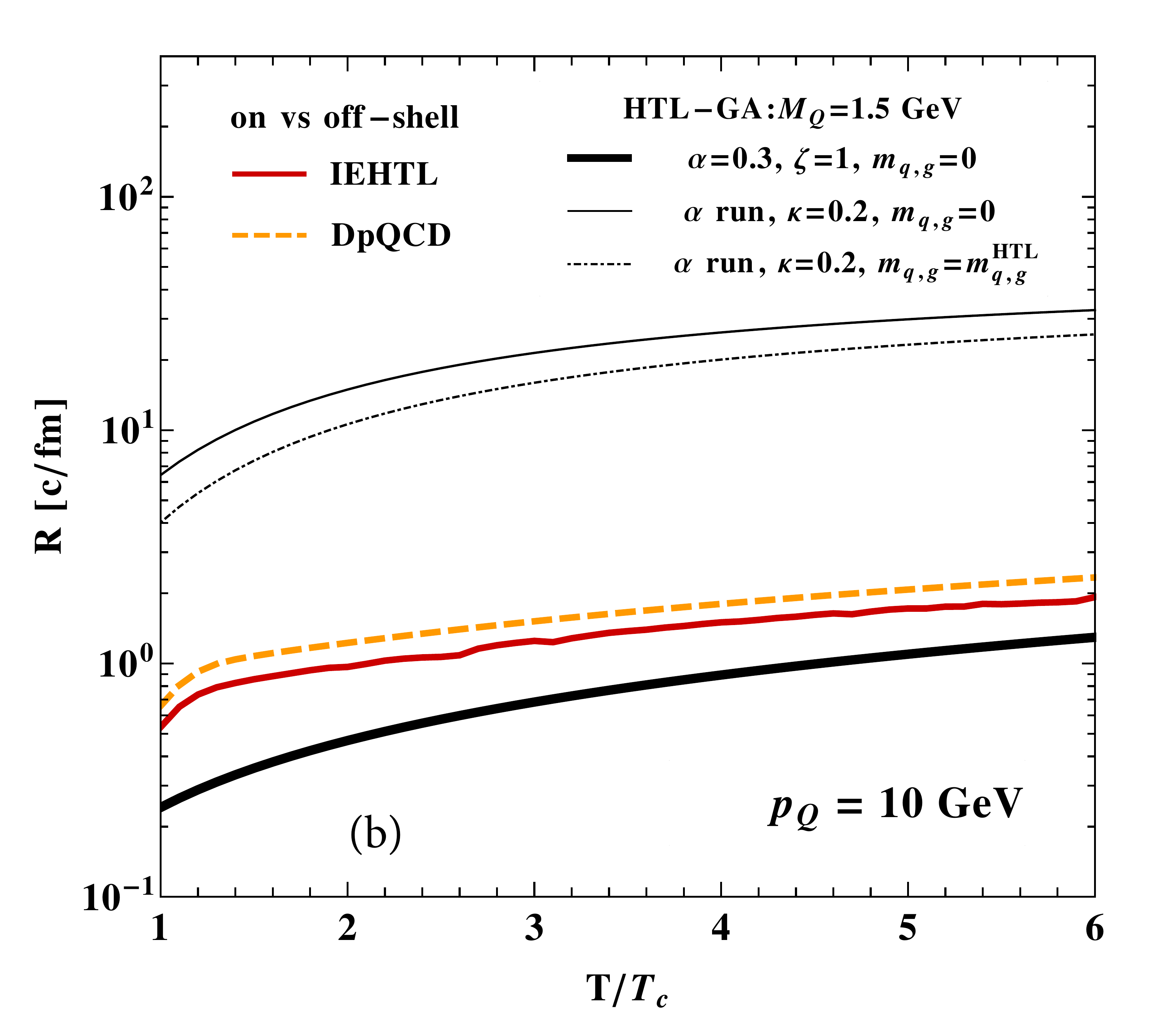}
\end{minipage}
\vspace{-0.1cm}
\caption{\emph{(Color online) $R$, the total elastic interaction rate of c-quarks in the plasma rest frame for the three different approaches as a function
of the heavy quark  momentum for $T = 2 T_c$ ($T_c = 0.158$ GeV) (left) and  as a function of the temperature $T$ for a heavy quark momentum $p_Q = 10$ GeV (right).
The parameters $\xi$, $\kappa$ and $\alpha$ in the HTL-GA model are described in the introduction.}}
\label{fig:cRatevsDifferentModel}
\end{center}
\end{figure}

The total interaction rate of the off-shell approach (IEHTL) in the plasma rest system is compared to that of the on-shell calculations (HTL-GA and DpQCD)
(\ref{equ:Sec3.4}) in figure \ref{fig:cRatevsDifferentModel}-(a) and (b) as a function of the momentum of the heavy quark, $p_Q$ and the temperature, respectively.
We assume here a Breit-Wigner spectral function and a Boltzmann-J\"utner distribution for both, the light quarks and the gluons. Our results are rather
independent of the choice of the spectral function and of whether the  Boltzmann-J\"utner distribution is replaced by a Fermi/Bose distribution.
In figures \ref{fig:cRatevsDifferentModel}-(a) and (b), the thick black line refers to the case where a constant $\alpha$ and a Debye mass $m_D = \xi g_s T$
with $\xi = 1$ as infrared regulator $\mu$ are used, whereas the thin black line gives the result for a  constant  running coupling and
$\mu^2 = \kappa \tilde{m}_D^2$, with $\tilde{m}_D^2 (T) = \frac{N_c}{3} \left(1 + \frac{N_f}{6} \right) 4 \pi \alpha (- \tilde{m}_D (T^2)) \ T^2$
and $\kappa =0.2$ in HTL-GA model. If we give a finite mass to the light quark and gluon ($m_{q,g} = m_{q,g}^{HTL}$) as shown in figure \ref{fig:MassDQPM-Mu2}-a)
in the HTL-GA model the total rate changes only moderately as can be seen from the black dotted dashed line. The yellow dashed and red solid lines in
figure \ref{fig:cRatevsDifferentModel}-(a) and (b) display the results for the DpQCD and IEHTL models, respectively.
As expected, the rates decrease when $m_{q,g}$ increases because the number of plasma collision particles becomes smaller. The reduction factor due to a finite mass
is almost independent of $p_Q \geq M_Q$ and therefore the ratio Rate(massive $q,g$)/Rate(massless $q,g$) in the HTL-GA model shows only a week temperature dependence.

A running coupling  with an effective Debye mass of $\kappa \tilde{m}_D$ yields a larger cross section than a fixed coupling constant $\alpha = 0.3$.
Naturally this increases the interaction rate if the mass of the particles does not change. Even in regions where the masses of gluons and quarks in the
HTL-GA model are not very different from those in the DpQCD approach (cf. Fig.\ref{fig:MassDQPM-Mu2}-a) the rates differ significantly. Therefore,
the finite mass of light quark and gluons explains only  part of the much lower interaction rate observed for the former model. The remaining difference
is due to the cross sections which are lower in the DpQCD approach as compared to HTL-GA, Ref.\cite{Berrehrah:2013mua}.

The description of the interaction between the heavy quark and the partons of the medium differs between the models. Whereas the coupling constant is strong
in the HTL-GA (with $\alpha$ running) and DpQCD/IEHTL models, the coupling is weak in HTL-GA for constant coupling $\alpha$. On the other hand the
interaction has a large range in HTL-GA (with $\alpha$ running) due to a small infrared regulator and it has an intermediate range in the DpQCD/IEHTL and
HTL-GA for constant coupling $\alpha$.

We see from Figs. \ref{fig:cRatevsDifferentModel}-(a) and (b) that the spectral function decreases the interaction rate of heavy quarks with the medium on
the order of 20\%. This modification is rather independent of the heavy quark momentum and of the temperature of the plasma.

The difference between the DpQCD and IEHTL rates is related on one side to the propagator, which contains an additional imaginary part proportional to the
gluon width in the IEHTL model, and on the other side to the asymmetry of the Breit-Wigner spectral function. For finite $\gamma_{g,q,Q}$ width,
the contribution of larger masses in the Breit-Wigner spectral function (right side of the pole mass) is larger compared to smaller masses
(left side of the pole mass), i.e. with broader gluon and light/heavy quark in the mass distribution, the average mass is shifted to higher values
of $m_{g,q,Q}$. This implies that the heavy quark is getting more massive on average and consequently is slowing down in momentum/velocity. Therefore,
the number of interactions/per unit length becomes smaller, as shown in figure \ref{fig:cRatevsDifferentModel}-(a) and (b).

The HTL-GA approach (with running coupling) and massless light quarks and gluons gives by far the largest rate.  Besides having the largest cross
section \citep{Berrehrah:2013mua}, the massless nature of the QGP constituent increase the number of possible elastic interactions. Even a finite mass
lowers the rate not substantially.

The dependence of the rates on the medium temperature $T$ for an intermediate heavy quark momentum ($p_Q = 10$ GeV) is illustrated in figure
\ref{fig:cRatevsDifferentModel}-(b). All models show an increase of the interaction rate with temperature. The linear dependence seen for the HTL-GA model
with $\alpha = 0.3$ and $\xi = 1$ can be easily explained since $R \propto n \times \sigma \propto T$, where $n\propto T^3$ is the parton density
and $\sigma \propto T^{-2}$ is the total cross section in this model. As studied in Ref. \citep{Berrehrah:2013mua} the temperature dependence of the total
cross section differs from one model to another due to the different coupling constants and infrared regulators employed. The extra decrease of the rate
in the DPQCD/IEHTL models for small temperature is related to the temperature dependence of the DQPM parton masses (cf. Fig.\ref{fig:MassDQPM-Mu2}-a).

The number of heavy quark elastic collisions per Fermi at $T = 2 T_c$ and for an intermediate heavy quark momentum ($p_Q = 10$) is 18.4 for HTL-GA
(running coupling constant and $m_q=m_g=0$) model, 1.37 for DpQCD and 1.06 for IEHTL. These numbers are larger at $T = 5 T_c$ with 29.84 elastic
collisions per Fermi in the HTL-GA model, 2.07 for DpQCD and 1.72 for IEHTL.

\subsection{Relaxation time for on- and off-shell partons}

For the transport coefficients one needs to specify the relaxation time $\tau$ which can be deduced from the elastic interaction rate $R (T)$ by averaging
over the initial heavy-quark momentum. The relaxation time $\tau$ is given for the on- and off-shell partons by

{\setlength\arraycolsep{-6pt}
\begin{eqnarray}
\label{equ:Sec3.6}
& & \displaystyle(\tau_c^{-1})^{\textrm{on}} = \frac{1}{n_Q^{\textrm{on}}} \!\!  \sum\limits_{j \in{q,\bar{q},g}} g_Q \int \frac{d^3 p_Q}{(2 \pi)^3} f_Q (p_Q, M_Q, T) \ R_{\textrm{j c}}^{\textrm{on}} (p_Q, M_Q^i, m_j^i, M_Q^f, m_j^f).
\nonumber\\
& & {} \displaystyle(\tau_c^{-1})^{\textrm{off}} = \frac{1}{n_Q^{\textrm{off}}} \!\! \sum\limits_{j \in{q,\bar{q},g}} g_Q \int \!\!\!\! \int \!\! \frac{d^3 p_Q}{(2 \pi)^3} \ \rho_j (m_j^{i,f}) \rho_j (M_Q^{i,f}) \ d m_j^{i,f} d M_Q^{i,f} \ f_Q (p_Q, M_Q, T) \ R_{\textrm{j c}}^{\textrm{on}} (p_Q, M_Q^i, m_j^i, M_Q^f, m_j^f),
\end{eqnarray}}
   with $\rho_{q} (m_q^{i,f})$ and $\rho_{Q} (M_Q^{i,f})$ being the Breit-Wigner-$m$ spectral functions \citep{Berrehrah:2013mua} for the incoming and
   outgoing light and heavy quark. Furthermore, $g_Q$ is the heavy quark degeneracy factor while $f_Q (p_Q, M_Q, T)$ is the Fermi-Dirac distribution and $n_Q$ the heavy quark density given for the on-shell case $n_Q^{\textrm{on}}$ and the off-shell case $n_Q^{\textrm{off}}$ by

{\setlength\arraycolsep{-6pt}
\begin{eqnarray}
\label{equ:Sec3.7}
& & n_Q^{\textrm{on}} (M_Q, T) = g_Q \int \frac{d^3 p_Q}{(2 \pi)^3} f_Q (p_Q; M_Q, T)
\nonumber\\
& & {} \displaystyle n_Q^{\textrm{off}} (T) = g_Q \int \!\!\!\! \int \!\! \frac{d^3 p}{(2 \pi)^3} \ \rho_Q (M_Q) \ d M_Q\  f_Q (p_Q; T, m_Q).
\end{eqnarray}}
 The in-medium elastic cross section for $qQ$, $\bar{q}Q$ and $gQ$ scattering processes have been studied in detail in Ref. \cite{Berrehrah:2013mua}.
The processes involving chemical equilibration  ($Q \bar{Q} \rightarrow g$, $Q \bar{Q} \rightarrow g g$ and $Q \bar{Q} \rightarrow q \bar{q}$) will not be
included in the computation of the relaxation time for heavy quarks since their contribution is negligible due to the small probability that 2 of the few heavy
quarks interact.

\begin{figure}[h!] 
\begin{center}
   \begin{minipage}[c]{.497\linewidth}
     \includegraphics[height=8.8cm, width=8.9cm]{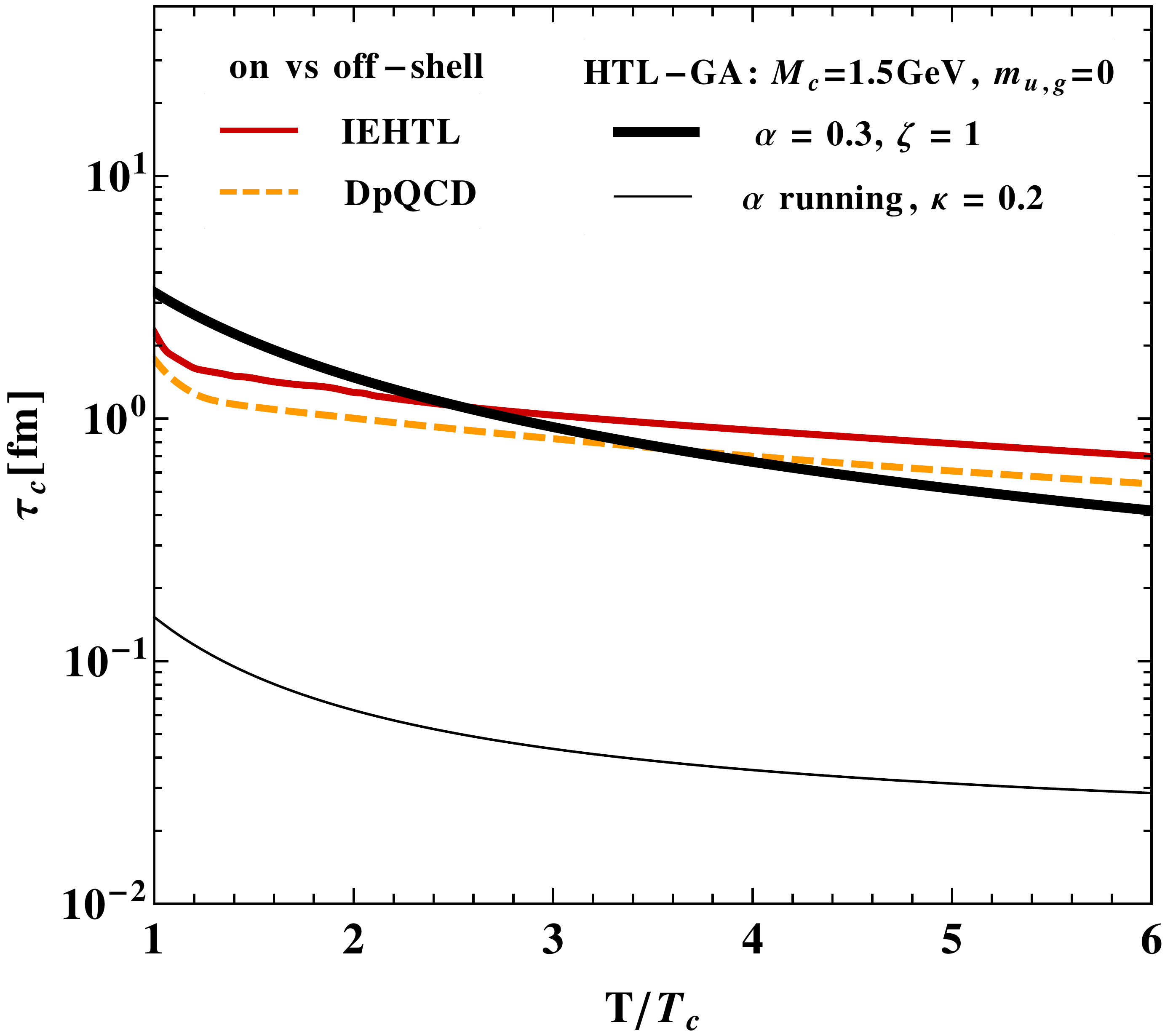}
   \end{minipage} 
   \begin{minipage}[c]{.497\linewidth}
      \includegraphics[height=8.8cm, width=8.9cm]{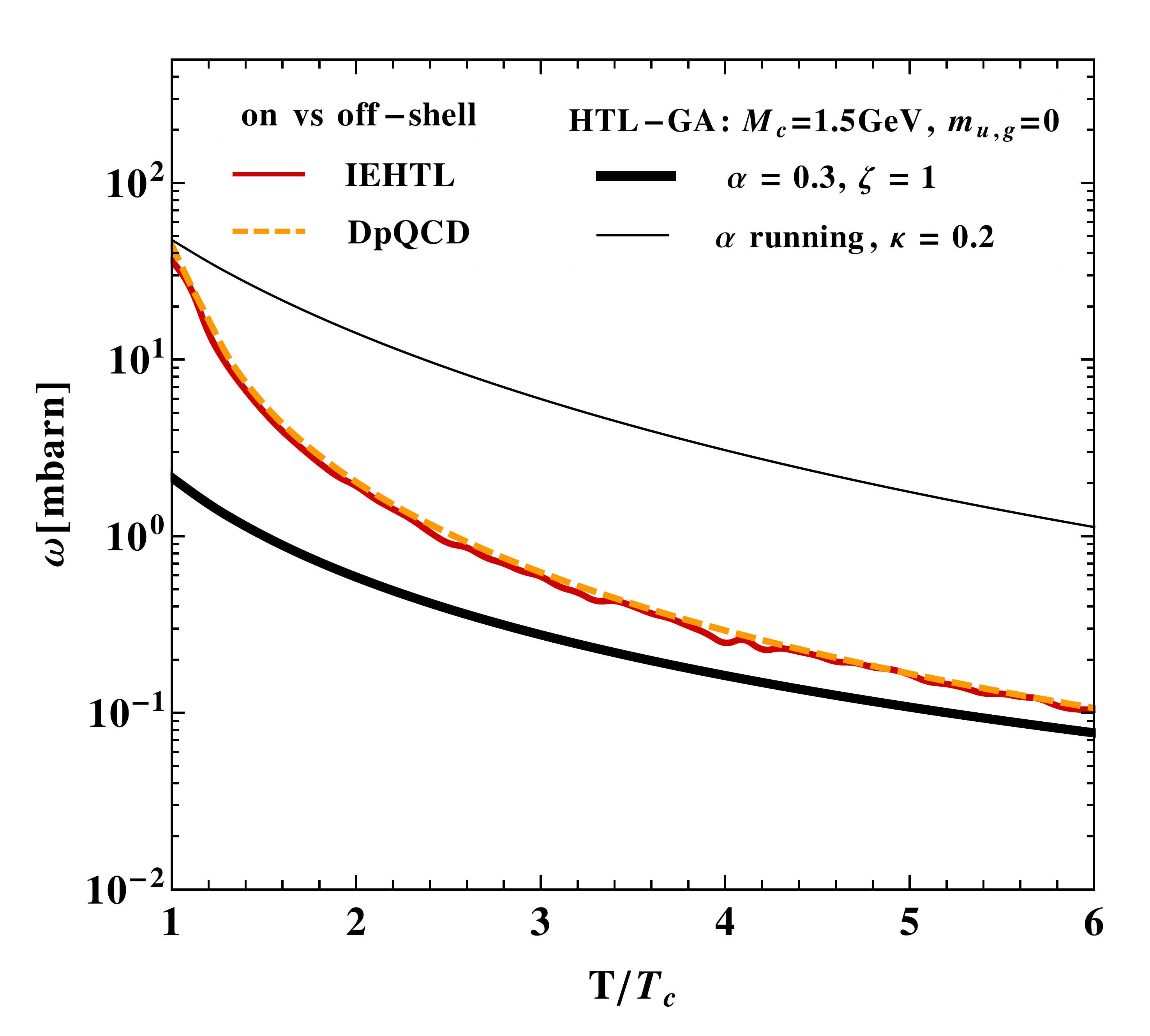}
\end{minipage}
\vspace{-0.1cm}
\caption{\emph{(Color online) The relaxation time $\tau_c$ (left) and the total transition rate $\omega (T)$ (right) for different models as a function of $T/T_c$ with $T_c = 0.158$ GeV at $\mu = 0$ for off-shell (IEHTL, red solid line) and on-shell (DpQCD, yellow dashed line) partons. We consider the DQPM pole masses for the on-shell partons in the DpQCD model and the DQPM spectral functions for the off-shell IEHTL approach. Also shown is a comparison with results in the HTL-GA approaches with constant and running coupling.}\label{fig:RelxationTime}}
\end{center}
\end{figure}

The results for the  relaxation time $\displaystyle \tau_c$ of charm quarks are shown in figure \ref{fig:RelxationTime}-(a) for the on-shell and
off-shell cases at finite temperature $T$ and at $\mu = 0$ in a medium composed by light quarks/antiquarks and gluons. We display the results
for the DpQCD/IEHTL and HTL-GA models. From  Fig. \ref{fig:RelxationTime}-(a) one deduces that the heavy quark relaxation time is about the same for
 DpQCD/IEHTL and HTL-GA with fixed coupling. A much shorter relaxtion
time is seen for HTL-GA with a running coupling.

Figure \ref{fig:RelxationTime}-(a) shows also that $\tau_c$  decreases with temperature since the quark and gluon densities are increasing functions
of the temperature. We can evaluate -in terms of powers of $T$- the behavior of this relaxation time for the different approaches. The density is
proportional to $T^3$ for large temperatures in case of $m \ll T$. The transition rate then is proportional to a power law in $T$, i.e.
$\sim T^{- \beta}$ with $\beta^{T < 1.2T_c} \sim 2$, $\beta^{ T > 1.2T_c} \sim 1.7$ for the HTL-GA model and $\beta^{T < 1.2T_c} \sim 4$, $\beta^{ T > 1.2T_c}
\sim 2$ for the DpQCD/IEHTL models. Hence, the relaxation time $\displaystyle \tau_c^{T<1.2 T_c} \propto T$ and $\displaystyle \tau_c^{T>1.2 T_c}
\propto T^{-1}$ for the DpQCD/IEHTL models. The large exponent in the relaxation times for $T<1.2 \ T_c$ in the DpQCD/IEHTL models can be traced back to the
infrared enhancement of the effective coupling.

With the etablished results for the relaxation time we can now calculate  transport coefficients in the quark gluon medium, like the viscosities and conductivities, and study their sensitivity to the off-shellness of the medium partons. Furthermore, from the interaction rate one can also deduce the transition rate $\omega (T)$ for particles in a heat bath of a given temperature $T$. In contradistinction to the elastic interaction rate, here both scattering partners are assumed to have a thermal distribution. The transition rate $\omega (T)$ is derived for the case of on- and off-shell partons by

{\setlength\arraycolsep{-6pt}
\begin{eqnarray}
\label{equ:Sec3.8}
& & \omega^{\textrm{on}} = \frac{1}{n_Q^{\textrm{on}}} \!\! \sum\limits_{j \in{q,\bar{q},g}} \int \frac{d^3 p_Q}{(2 \pi)^3} f_Q (p_Q, M_Q, T) \ \times \frac{1}{n_j^{\textrm{on}}}\ R_{\textrm{j c}}^{\textrm{on}} (p_Q, M_Q^i, m_j^i, M_Q^f, m_j^f),
\nonumber\\
& & {} \omega^{\textrm{off}} = \frac{1}{n_Q^{\textrm{off}}} \!\! \sum\limits_{j \in{q,\bar{q},g}} \int \!\!\!\! \int \!\! \frac{d^3 p_Q}{(2 \pi)^3} \ \rho_j (m_j^{i,f}) \rho_j (M_Q^{i,f}) \ d m_j^{i,f} d M_Q^{i,f} \ f_Q (p_Q, M_Q, T) \ \times \frac{1}{n_j^{\textrm{on}}} R_{\textrm{j c}}^{\textrm{on}} (p_Q, M_Q^i, m_j^i, M_Q^f, m_j^f),
\end{eqnarray}}
where $n_j^{\textrm{on}}$ is the density of the on-shell parton $j$. The results for the total on- and off-shell transition rates  are displayed in
Fig. \ref{fig:RelxationTime}-(b). Due to the small DQPM parton widths, the influence of the spectral function is negligible and we obtain for DpQCD and IEHTL
about the same transition rate $\omega_{qQ}^{\textrm{off}} (T)$. Parametrizing $\omega_{q/gQ}$ by $T^{-\beta}$ we find differences between HTL-GA and DpQCD/IEHTL as discussed above.
This different behaviour leads to different transport coefficients as we will show below.

\section{Drag Force and Coefficient}
\label{DragCoefficient}

We continue our study with the drag and diffusion coefficient. For heavy masses of $c$ and $\bar{c}$ quarks the relaxation times are large as
compared to the typical time of an individual $q c \rightarrow q c$ and $g c \rightarrow g c$ collision. Therefore we might describe the time
evolution of heavy quarks in momentum space by a Fokker-Planck (FP) equation \cite{Svetitsky:1987gq,GolamMustafa:1997id}:

{\setlength\arraycolsep{-6pt}
\begin{eqnarray}
\label{equ:Sec8.1}
\frac{\partial f(\bs{p}, t)}{\partial t} = \frac{\partial}{\partial p_i} \bigl[A_i (\bs{p}) f(\bs{p}, t) + \frac{\partial}{\partial p_j} (B_{i j} (\bs{p}) f (\bs{p}, t)) \bigr].
\end{eqnarray}}
  The drag ($A$) and diffusion ($B$) coefficients are evaluated according to \cite{Svetitsky:1987gq,GolamMustafa:1997id,Gossiaux:2004qw} by a
Kramers-Moyal power expansion of the collision integral kernel of the Boltzmann equation. Note that the diffusion tensor $B$ admits a
transverse-longitudinal decomposition (perpendicular and along the direction of the heavy quark in the frame where the fluid is at rest) and
contains two independent coefficient $B_{\parallel}$ and $B_{\perp}$. As realized in Refs. \cite{Gossiaux:2004qw,Walton:1999dy}, the asymptotic distribution
coming out of the FP evolution deviates from a (relativistic) Maxwell-Boltzmann distribution, which is a consequence of  the truncation of the Kramers-Moyal series.

\subsection{Drag coefficient for on-shell partons}

The drag coefficient $A$  describes the time evolution of the mean momentum of the heavy quark. For on-shell partons it has been defined in the plasma
rest frame by Svetitsky \cite{Svetitsky:1987gq,GolamMustafa:1997id}. It is obtained by multiplying (\ref{equ:Sec1.1}) by $M_Q/E_p$, using for
 $\mathcal{X}$ the longitudinal component of the momentum transfer, $\mathcal{X}=(p - p')^l$.


In the absence of diffusion, ($B_{ij} = 0$), the Fokker Planck equation (\ref{equ:Sec8.1}) allows to write

{\setlength\arraycolsep{-6pt}
\begin{eqnarray}
\label{equ:Sec8.3}
\frac{\partial f}{\partial t} = \frac{\partial }{\partial p_i}(A_i f) \to
\frac{d <\!\! p_i \!\!>_f}{d t} = - \int d^3 p \ A_i (\bs{p}) f(\bs{p}, t).
\end{eqnarray}}
 In particular, assuming $f = \delta (\bsp{p} - \bsp{p}_0)$, we obtain

{\setlength\arraycolsep{-6pt}
\begin{eqnarray}
\label{equ:Sec8.4}
\frac{d p_{0,i}}{d t} = - A_i (\bsp{p}_0)  \rightarrow  \frac{d p_{0,i}}{d \tau} = - \frac{E_p}{M_Q} A_i (\bsp{p}_0) \equiv \mathcal{A}_{i} (\bsp{p}_0),
\end{eqnarray}}
 where $\tau$ is the time measured in the heavy-quark rest system. $\mathcal{A}^i = \frac{E_p}{M_Q} A^i (\bsp{p})$ is the spatial part of the
 covariant $\mathcal{A}^{\mu}$

{\setlength\arraycolsep{-6pt}
\begin{eqnarray}
\label{equ:Sec8.5}
\mathcal{A}^\mu = - \sum_{q,g} \frac{1}{2 M_Q} \int \frac{d^3 q}{(2\pi)^3 2E_q} f(\bsq{q}) a^\mu (\bsp{p},\bsq{q})
\end{eqnarray}}
with

{\setlength\arraycolsep{-6pt}
\begin{eqnarray}
\label{equ:Sec8.6}
a^{\mu} (\bsp{p},\bsq{q}) := \frac{1}{(2\pi)^2}\int \frac{d^3 q'}{2E_{q'}} \int \frac{d^3 p'}{2E_{p'}} (q - q')^{\mu} \delta^{(4)} (P_{in} - P_{fin})
\frac{1}{g_p g_Q} \sum_{k,l} |\mathcal{M}_{2,2} (p, q; i, j |p', q'; k, l)|^2
\end{eqnarray}}
using $(p - p')^{\mu} = - (q - q')^{\mu} $. We evaluate $a^{\mu}$ in the c.m. frame as

{\setlength\arraycolsep{-6pt}
\begin{eqnarray}
\label{equ:Sec8.7}
a^{\mu}_{cm} (\bsp{p}_{cm}) \ :&=& \frac{1}{(2\pi)^2}\int \frac{d^3 q'}{2E_{q'}} \int \frac{d^3 p'}{2E_{p'}} (q - q')^{\mu} \delta^{(3)} (\bsq{q}'-\bsp{p}')\delta(\sqrt{s}-\sqrt{m_q^2+q'^2}-\sqrt{m_Q^2+q'^2}) \frac{1}{g_p g_Q} \sum_{k,l} |\mathcal{M}_{2,2} (p, q; i, j |p', q'; k, l)|^2.
\nonumber\\
&=&  \frac{1}{(2\pi)^2}\int \frac{d^3 q'}{2E_{q'}2E_{p'}} (q - q')^{\mu}\frac{E_{q'}E_{p'}}{\sqrt{s}p_{cm}}  \delta(|\mathbf{q'}|-|\mathbf{p}_{cm}|) \ \frac{1}{g_p g_Q} \sum_{k,l} |\mathcal{M}_{2,2} (p, q; i, j |p', q'; k, l)|^2.\nonumber \\
&=&  \frac{1}{(2\pi)^2}\frac{p_{cm}^2}{4 \sqrt{s}}\int \int d\Omega_{q'} (\hat q_{cm} - \hat q'_{cm})^{\mu}| \ \frac{1}{g_p g_Q} \sum_{k,l} \mathcal{M}_{2,2} (p, q; i, j |p', q'; k, l)|^2,
\end{eqnarray}}
 where $E_{p'} = \sqrt{M_Q^2 + p_{cm}^2}$, $E_{q'} = \sqrt{m_q^2 + p_{cm}^2}$. $\hat{\bsq{q}}_{cm}$ is the unit vector in the direction of $\bsq{q}$
 which we choose to have only a $z$ component and $\hat{\bsq{q}}_{cm}'$ is the unit vector in the direction of $\bsq{q}'$  with a $z$ and a $x$ component;
 $q^0_{cm} = q'^0_{cm}$. Therefore we find

{\setlength\arraycolsep{-6pt}
\begin{eqnarray}
\label{equ:Sec8.8}
(\hat q_{cm}-\hat q_{cm}')^\mu =
 \begin{pmatrix}
  0 \\
  -\sin{\theta} \\
  0  \\
  1-\cos{\theta}
 \end{pmatrix}.
\end{eqnarray}}
 The integration over the transverse component of $\hat q'$ gives zero and we are left with a four-vector in which only the $z$ component differs
 from zero. Using $ 1-\cos{\theta} = -t/(2p_{cm}^2)$  (with $\theta$ denoting the angle between $\bsq{q}$ and $\bsq{q}'$ in the c.m frame)
 we can replace the $\cos{\theta}$ integration by an integration over $t$ and obtain

{\setlength\arraycolsep{-6pt}
\begin{eqnarray}
\label{equ:Sec8.9}
a_{cm}^0 = 0, \hspace{0.3cm} \bsa{a}_{cm} = \frac{ p_{cm}^2 (s) m_1 (s)}{4 \pi \sqrt{s}} \times \hat{\bsq{q}}_{cm},  \hspace{1cm} \textrm{with:} \hspace{0.3cm}  m_1 (s) := \frac{1}{8 p_{cm}^4} \int_{-4 p_{cm}^2}^0 \ \sum |\mathcal{M}|^2 (s, t) (- t) d t.
\end{eqnarray}}
We have to calculate now $\mathcal{A}^\mu$ in the heavy quark rest frame. For this we have to boost $a_{cm}^\mu $ to this system.
Using $\gamma = \frac{M_Q+E_q}{\sqrt{s}}$ and $\gamma \ \bsbt{\beta}= \frac{\bsq{q}_{rest}}{\sqrt{s}}$, with $E_q (\bsq{q}_{rest})$ being the
energy (momentum) of the light quark in the heavy quark rest system we find

{\setlength\arraycolsep{-6pt}
\begin{eqnarray}
\label{equ:Sec8.10}
a_{rest}^\mu = \frac{ p_{cm}^2(s) m_1 (s)}{4 \pi s} \, \begin{pmatrix}
{q}_{rest} \\
(E_{q_r} + M_Q) \ \hat{\bsq{q}}_{cm}
\end{pmatrix}.
\end{eqnarray}}
By construction $\bs{q}_{rest}$ has the same direction as $\hat{\bsq{q}}_{cm}$. To calculate

{\setlength\arraycolsep{-6pt}
\begin{eqnarray}
\label{equ:Sec8.11}
\mathcal{A}^{\mu}_{rest} = - \frac{1}{8 (2\pi)^4 M_Q} \int \frac{p_{cm}^2 (s) m_1 (s)}{s} \Biggl[ \int d \Omega_{\bsq{q}} \left(q \delta^{\mu 0} + (E_q + M_Q) \hat{q}_{cm\ i} \delta^{\mu i}  \right) f_{r} (\bsq{q}) \Biggr] \frac{q^2 d q}{E_q},
\end{eqnarray}}
  we need $f_{r} (\bsq{q})$ , the invariant distribution of the light quark in the rest frame of the heavy quark. The heat bath velocity in the heavy
  quark rest system is given by $u \equiv (u^0, \bsu{u}) = (E_p/M_Q, - \bsp{p}/M_Q)$.  If we define $f (E_q/T)$ as the invariant distribution in the rest
system of the heat bath, the light quark distribution in the heavy quark rest system is given by
$f_r \left(\frac{u^0 E_q - uq \cos \theta_{r}}{T} \right)$  with $\theta_{r}$ denoting the angle between $\bsq{q}$ and $\bsu{u}$
(see eq. (\ref{equ:Sec3.5})). The expression (\ref{equ:Sec8.11}) then becomes

{\setlength\arraycolsep{-6pt}
\begin{eqnarray}
\label{equ:Sec8.12}
\mathcal{A}^{\mu}_{rest} & & = - \frac{1}{4 (2\pi)^3 M_Q} \int \frac{p_{cm}^2 (s) m_1 (s)}{s} \Biggl[q f_0(q) \delta^{\mu 0} + (E_q + M_Q) f_1 (q) \delta^{\mu i}  (\hat{u})_i  \Biggr] \frac{q^2 d q}{E_q} =
 \begin{pmatrix}
  A_{rest}^0  \\
 (A_{rest}^{\nu} \hat{u})^{i}\\
 \end{pmatrix},
\end{eqnarray}}
where

{\setlength\arraycolsep{0pt}
\begin{eqnarray}
\label{equ:Sec8.13}
\hspace{-0.5cm} f_0 (q) = \frac{1}{2} \int d \cos \theta_{r} \ f \left(\frac{u^0 E_q - uq \cos \theta_{r}}{T} \right), \hspace{1cm}  f_1 (q) = \frac{1}{2} \int d \cos \theta_{r} \ f \left(\frac{u^0 E_q - uq \cos \theta_{r}}{T} \right) \cos \theta_{r}.
\end{eqnarray}}
 Using $p_{cm}^2 = \frac{q^2 M_Q^2}{s}$, $A_{rest}^0$ and $A_{rest}^{\nu}$ reduce to  a one dimensional integral:

{\setlength\arraycolsep{-6pt}
\begin{eqnarray}
\label{equ:Sec8.14}
A_{rest}^0 = \frac{M_Q}{4 (2 \pi)^3} \int_0^{\infty} \frac{q^5 m_1 (s) f_0 (q)}{s^2 E_q} \ d q,  \hspace{1.5cm} A_{rest}^{\nu} = \frac{M_Q}{4 (2 \pi)^3} \int_0^{\infty} \frac{q^4 (M_Q + E_q) m_1 (s) f_1 (q)}{s^2 E_q} \ d q.
\end{eqnarray}}
 We note in passing that $A_{rest}^0 \rightarrow 0$ as $u \rightarrow 0$ so that $A_{rest}^0$ dominates $A_{rest}^{\nu}$ for small heavy quark momenta.
For $p \rightarrow + \infty$, one observes the opposite trend.

 $\mathcal{A}^{\mu}_{\textrm{heat\ bath}}$ in the heat bath rest frame is obtained by a Lorentz transformation ($\gamma_u \ \bsbt{\beta}_u= \bsp{p}/M_Q, \gamma_u= E_p/M_Q$)

{\setlength\arraycolsep{-6pt}
\begin{eqnarray}
\label{equ:Sec8.15}
\mathcal{A}^{\mu}_{\textrm{heat \ bath}} = \begin{pmatrix}
\frac{E_p}{M_Q} A_{rest}^{0} - A_{rest}^{\nu} \frac{\bsp{p}}{M_Q} \hat{\bsp{p}} \\
\frac{E_p}{M_Q} A_{rest}^{\nu} - A_{rest}^0 \frac{\bsp{p}}{M_Q} \hat{\bsp{p}}
\end{pmatrix}.
\end{eqnarray}}
which yields for the drag force in the heat bath rest frame

{\setlength\arraycolsep{-6pt}
\begin{eqnarray}
\label{equ:Sec8.16}
\frac{d<\!\!\bsp{p}\!\!>}{dt} = \bsA{A}_{\textrm{heat \ bath}} = \frac{M_Q}{E_p} \mathcal{\bsAA{A}}_{\textrm{heat\ bath}} = \Biggl(A_{rest}^{\nu} - \frac{\bsp{p} \hat{\bsp{p}}}{E_p}  A_{rest}^0 \Biggr) \hat{\bsp{p}},
\end{eqnarray}}
and for the energy component
{\setlength\arraycolsep{-6pt}
\begin{eqnarray}
\label{equ:Sec8.17}
\frac{d<\!\!E\!\!>}{dt}=A^0_{\textrm{heat \ bath}} = \frac{M_Q}{E_p} \mathcal{A}^0_{\textrm{heat\ bath}} = A_{rest}^{0} - \frac{\bsp{p} \hat{\bsp{p}}}{E_p}  A_{rest}^\nu .
\end{eqnarray}}

\subsection{Drag coefficient for off-shell partons} 

For the off-shell drag force, $A_i^{\textrm{off}}$, we apply eq.(\ref{equ:Sec1.4}) and find:


{\setlength\arraycolsep{-6pt}
\begin{eqnarray}
\label{equ:Sec9.22}
& & \bsA{A}_{\textrm{heat\ bath}}^{\textrm{off}} = \Biggl(A_{rest}^{\nu,\textrm{off}} - \tilde{A}_{rest}^{0,\textrm{off}} \Biggr) \hat{\bsp{p}},
\end{eqnarray}}
where $A_{rest}^{\nu,\textrm{off}}$ and $\tilde{A}_{rest}^{0,\textrm{off}}$ are defined by

{\setlength\arraycolsep{0pt}
\begin{eqnarray}
\label{equ:Sec9.23}
& & \tilde{A}_{rest}^{0,\textrm{off}} = \frac{4}{(2 \pi)^7} \ \Pi_{i \in{p,q,p',q'}} \int \frac{p \ m_p}{E_p} m_i d m_i \ \rho_i^{WB} (m_i) \ \int_0^{\infty} \frac{q^5 m_1 (s) f_0 (q)}{s^2  E_q} d q,
\\
& & {} A_{rest}^{\nu,\textrm{off}} = \frac{4}{(2 \pi)^7} \ \Pi_{i \in{p,q,p',q'}} \int m_p  m_i d m_i \ \rho_i^{WB} (m_i) \ \int_0^{\infty} \frac{q^4 (m_Q + E_q) m_1 (s) f_1 (q)}{s^2 E_q} d q.
\nonumber
\end{eqnarray}}
The details are relegated to appendix \ref{AppendixB}.

\subsection{Results}

\begin{figure}[h!] 
\begin{center}
   \begin{minipage}[c]{.495\linewidth}
     \includegraphics[height=8.9cm, width=9cm]{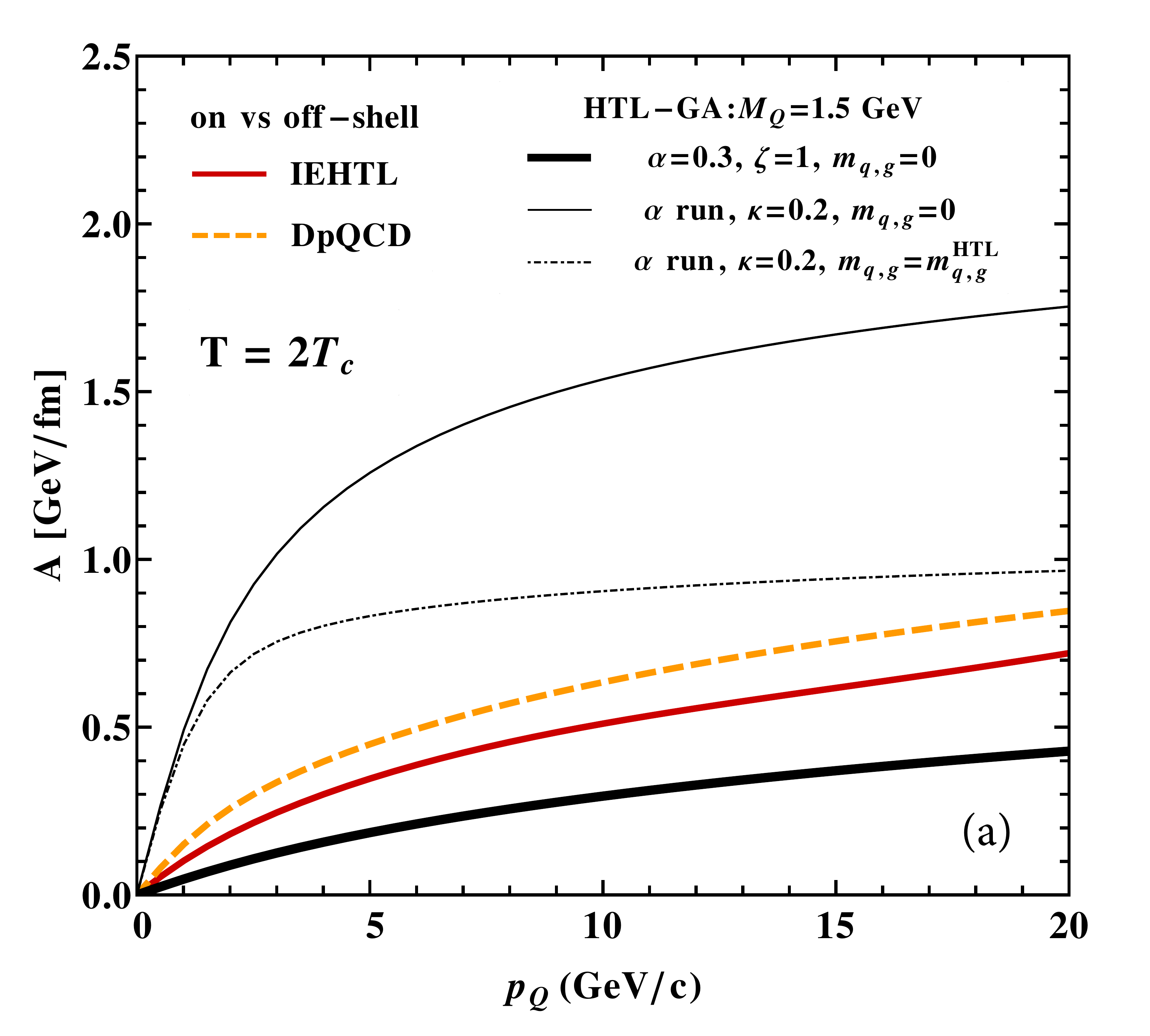}
   \end{minipage} 
   \begin{minipage}[c]{.495\linewidth}
      \includegraphics[height=8.9cm, width=9cm]{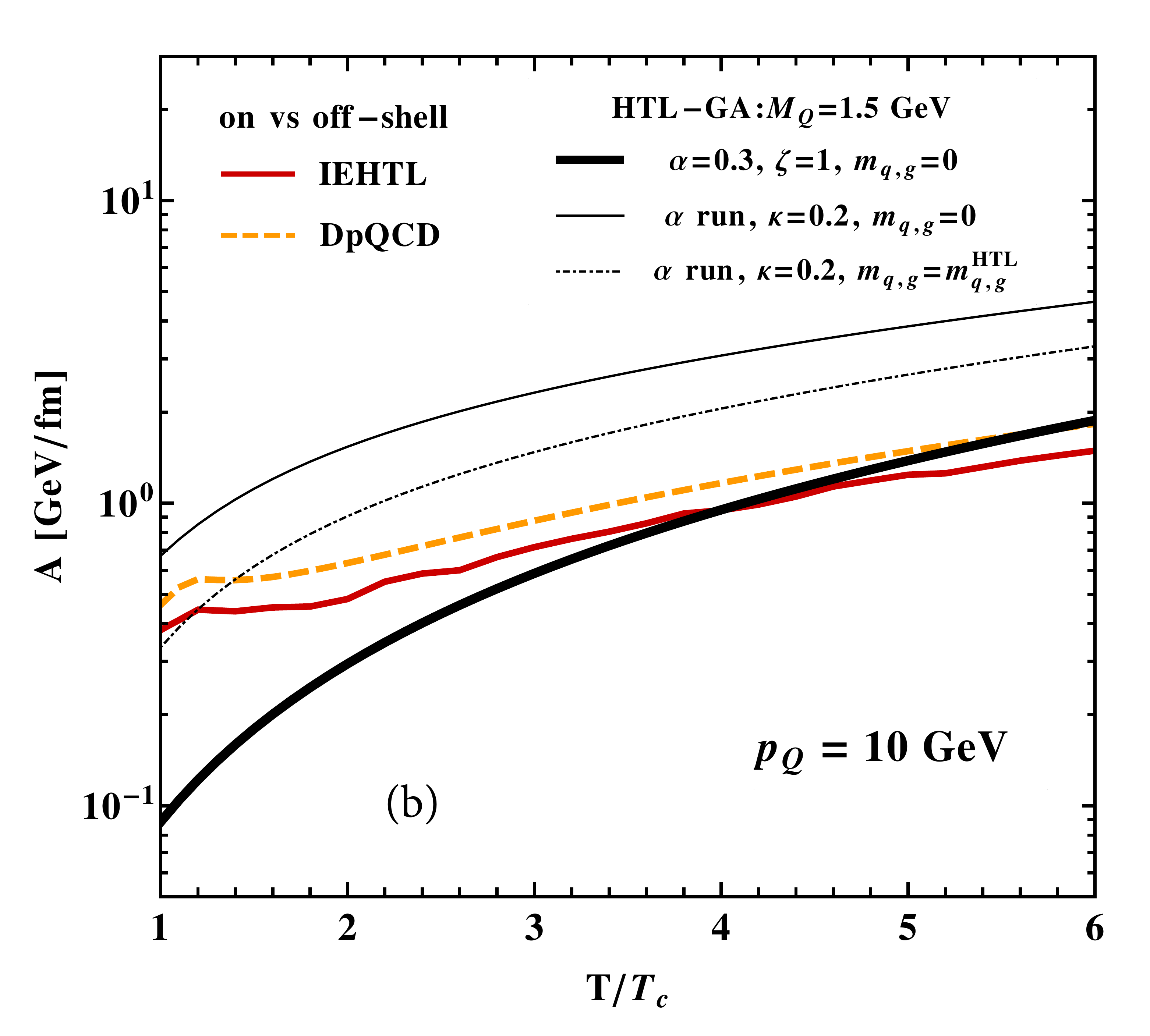}
\end{minipage}
\vspace{-0.2cm}
\caption{\emph{(Color online) The drag coefficient $A$ of c-quarks in the plasma rest frame (\ref{equ:Sec8.16}) and (\ref{equ:Sec9.22})
for the three different approaches as a function of the heavy quark momentum $p_Q$ for $T = 2 T_c$  (a) and  as a function of the
temperature $T/T_c$ ($T_c = 0.158$ GeV) for an intermediate heavy quark momentum, $p_Q = 10$ GeV (b).}}
\label{fig:cDragDifferentModel}
\end{center}
\end{figure}

We discuss now the drag coefficients for the models HTL-GA, DpQCD and IEHTL from Ref. \cite{Berrehrah:2013mua}. Here we extend the study
from Ref. \cite{Berrehrah:2013mua} and consider finite light quark and gluon masses in HTL-GA model for a running coupling $\alpha$.
In the DpQCD approach, the light and heavy quark and gluon masses are given by the DQPM pole masses. Fig. \ref{fig:cDragDifferentModel}-(a) shows the
drag coefficient (\ref{equ:Sec8.16}) and (\ref{equ:Sec9.22}) of heavy quarks as a function of the heavy quark momentum. We display the results
for the HTL-GA model for a constant and a running $\alpha$ and for massless and massive light quarks and gluons, where $m_{q,g}^{HTL}$ denote the light
quark and gluon mass given by HTL and plotted in figure \ref{fig:MassDQPM-Mu2}-(a). The temperature of the heath bath is chosen as $T=2T_c$.
In the HTL-GA approach a finite mass of the light quark leads to a reduction of the drag coefficient for a given temperature. This effect is similar for both,
the constant and running coupling. However, the reduction of the drag coefficient is not uniform as a function of $p_Q$ and dominantly affects the
largest momentum. The drag coefficient in the DpQCD model is in between the extremes of the  HTL-GA approach.

Figure \ref{fig:cDragDifferentModel}-(a) illustrates also the influence of a finite parton width on the heavy quark drag coefficient, where the drag
coefficient is displayed for on-shell partons (DpQCD) and off-shell partons (IEHTL) as a function of the heavy quark momentum and for a temperature
of 2 $T_c$. For off-shell partons the drag coefficient is slightly lower than for the corresponding on-shell case. We see in
Fig. \ref{fig:cDragDifferentModel}-(a) that for thermal heavy quark $p_Q\sim2\ $ GeV the IEHTL and the on-shell DpQCD give a drag coefficient which is
slightly above the one from the HTL-GA drag coefficient for a fixed $\alpha$. For a running coupling  in HTL-GA the drag coefficient (\ref{equ:Sec8.16})
is about a factor of 2.5 larger.

In order to understand the reduction of the drag coefficient in the HTL-GA and DpQCD/IEHTL approaches, we plot in Fig. \ref{fig:cDragDifferentModel}-(b)
the drag coefficient as a function of the temperature for an intermediate (10 GeV/c) heavy quark momentum. From the temperature dependence of the drag
coefficient in the HTL-GA model with masseless/massive light quarks and gluons, one sees that the ratio $A$(massive $q$,$g$)/$A$(massless $q$,$g$) shows
almost no dependence on the temperature. Comparing the different results, one notices that in the HTL-GA model the drag is always increasing as a function
of the temperature whereas for the DpQCD/IEHTL models we observe a minimum at 1-2 $T_c$. The bump observed around 1-2 $T_c$ is directly related to the
temperature dependence of the DQPM parton masses. The large drag at low temperatures in DpQCD/IEHTL is due to the strong increase of the running coupling
(infrared enhancement) at these temperatures.

In the HTL-GA model (with massless partons) an increasing temperature increases the density of scattering partners in the QGP
$f \propto \exp (-\frac{\sqrt{q^2}}{T})$. For quarks with zero mass this increase is $\propto T^3$ what is partially counterbalanced by the
decrease of the cross section $\propto T^{-2}$. In the  DpQCD/IEHTL approaches we have in addition to take into account that the light quark masses
and the coupling constants increase with temperature. This lowers on the one side the increase of the parton density
($f \propto \exp (-\frac{\sqrt{m_{q,g}^2+q^2}}{T})$) with temperature, allows on the other side for a larger momentum transfer in a single collision.
The drag coefficient increases with the heavy quark momentum for both, the HTL-GA and DpQCD/IEHTL model. The running coupling  in HTL-GA leads to larger
values of the heavy quark drag coefficient. The difference of $A$ between the HTL-GA and DpQCD/IEHTL models is on one side due to the different masses
and on the other side
due to the fundamental ingredients used in the corresponding cross sections, i.e. the coupling and the infrared regulator \cite{Berrehrah:2013mua}
as  explained above. The low temperature enhancement of DpQCD/IEHTL as compared to HTL-GA  is essentially due to the strong increase of the coupling
$\alpha(T/T_c)$ for temperatures close to $T_c$.

The drag coefficient $A$ for the DpQCD/IEHTL models can be parametrized as a function of $p_Q$ by the functional form
$A (p_Q) = a p_Q - b p_Q^2 + c p_Q^3 - d p_Q^4$, where $a$ is the dominant coefficient. For
$A^{T=T_c} (p_Q) = 0.03609 p_Q - 0.00146 p_Q^2 + 0.000037 p_Q^3 - 5.469 \times 10^{-7} p_Q^4$.

\begin{figure}[h!] 
\begin{center}
   \begin{minipage}[c]{.495\linewidth}
     \includegraphics[height=8.8cm, width=9cm]{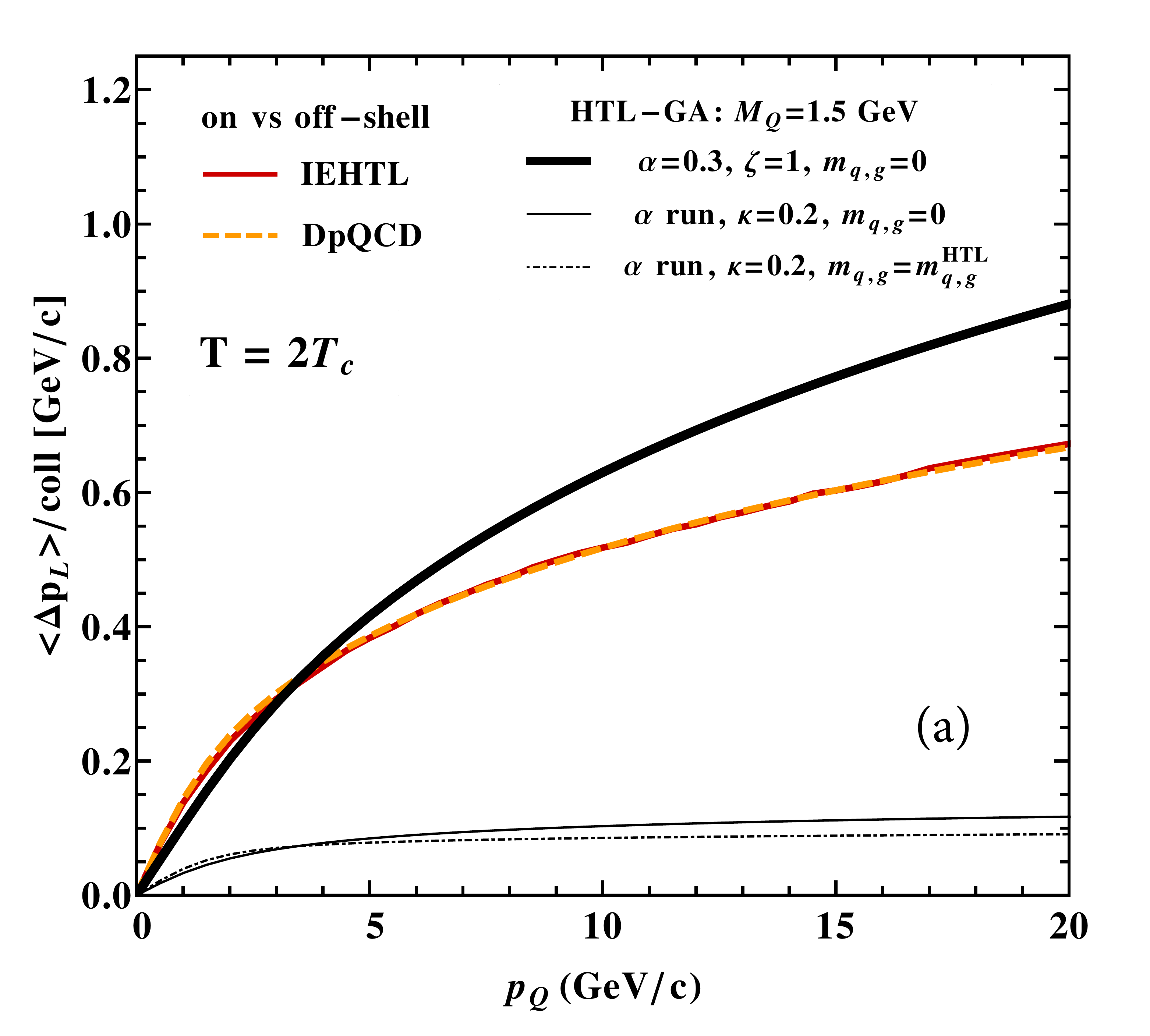}
   \end{minipage} 
   \begin{minipage}[c]{.495\linewidth}
     \includegraphics[height=8.8cm, width=9cm]{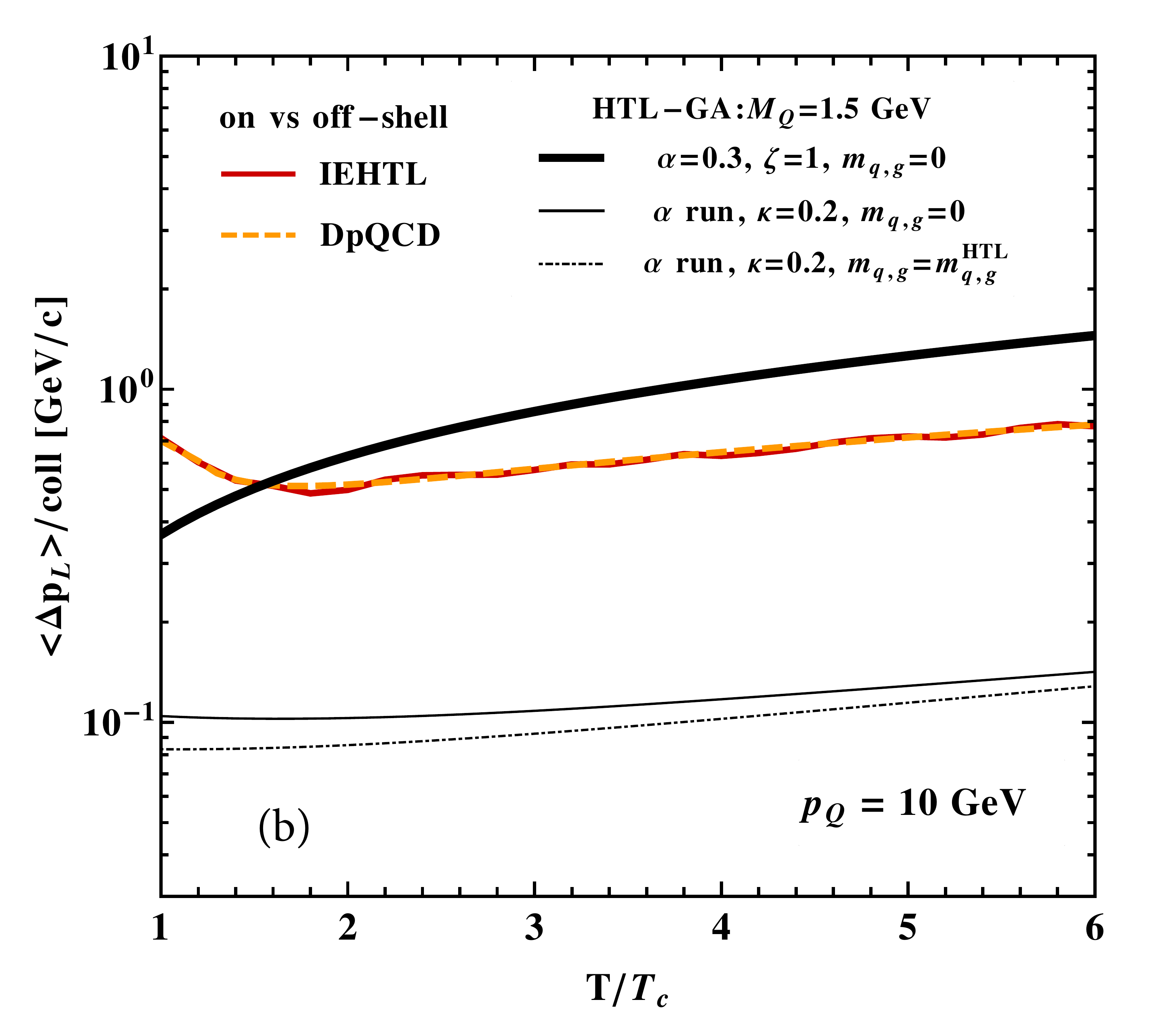}
\end{minipage}
\vspace{-0.2cm}
\caption{\emph{(Color online) The average loss of longitudinal momentum $<\! \Delta p_L\!>$ per collision for the three different approaches
as a function of the heavy quark momentum and for $T = 2 T_c$ (a) and  as a function of the medium temperature $T/T_c (T_c=0.158$ GeV$)$ for intermediate heavy
momentum $p_Q = 10$ GeV (b).}}
\label{fig:cDragOvRateDifferentModel}
\end{center}
\end{figure}

It is interesting to compare the drag coefficient with the collision rates (figs.\ref{fig:cRatevsDifferentModel}-(a) and (b).). The ratio of both
is proportional to the average longitudinal momentum change in a single collision $<\!\Delta p_L\!> \propto A/Rate$, where $A$ and the rate are
defined in eqs. (\ref{equ:Sec1.1}) and (\ref{equ:Sec1.4}), respectively. $<\!\Delta p_L\!>$ is illustrated in Fig. \ref{fig:cDragOvRateDifferentModel} as a
function of the heavy quark  momentum (a) and as a function of the temperature (b). Whereas the heavy quark drag coefficient and the heavy quark rate
are larger for the HTL-GA model with running $\alpha$,  the longitudinal momentum transfer per collision $<\!\Delta p_L\!>$ is smaller as compared
to the DpQCD/IEHTL models because a running $\alpha(Q^2)$ in HTL-GA favors collisions with a small momentum transfer. For large momenta and for a running $\alpha$
$<\!\Delta p_L\!>$  becomes almost independent of the incoming momentum.  In the  DpQCD/IEHTL models (with a temperature dependent coupling) as well as
in HTL-GA with a fixed coupling   $<\!\Delta p_L\!>$ increases with $p_Q$ . Off-shell and on-shell approaches show an almost identical
$<\!\Delta p_L\!>$ per collision such that the differences in the drag coefficient are only due to different rates. If one regards the temperature
dependence one sees that the increasing coupling  in DpQCD/IEHTL is almost completely counterbalanced by the increasing mass of the partons.
Therefore the DpQCD/IEHTL models show only a very moderate change of $<\!\Delta p_L\!>$ per collision which falls in between the HTL-GA approach
for running and fixed coupling. Also HTL-GA with a running coupling shows only a very moderate change of $<\!\Delta p_L\!>$ per collision as a
function of the temperature. In pQCD it is known that $A \propto T^2$ and the rate $R \propto T$ \cite{Moore:2004tg}. Therefore $<\!\Delta p_L\!> \propto T$
in pQCD as shown in figure \ref{fig:cDragOvRateDifferentModel}-(b) within HTL-GA for constant $\alpha$ (thick black line).

Comparing the massless (thin black line) and massive (dotted dashed black line) parton cases in the HTL-GA model in
Figs.\ref{fig:cDragOvRateDifferentModel}-(a) and (b), one concludes that the scattering of a heavy quark on massive partons transfers less
longitudinal momentum that the scattering on massless partons, except for low heavy quark momenta $p_Q \ll T$. The finite parton mass contributes
to the increase of the c.m. momentum and therefore reduces the threshold and opens the phase space for the total cross section. We will discuss
this now for the case of the scattering of a heavy quark on a light quark. From eq.(\ref{equ:Sec3.4}) we find that for $p \gg M_Q$ the rate in the
rest system of the heavy quark is given by

{\setlength\arraycolsep{0pt}
\begin{eqnarray}
\label{equ:Sec9.24}
{\cal R}\propto \frac{g_p}{8 \pi^2} \ \frac{T}{u^0u} \int dq \frac{q^2}{q^0} \sigma\left(\frac{q M_Q}{\sqrt{s}}\right) e^{-\frac{u^0 q^0-u q}{T}}
\end{eqnarray}}
which yields for  $m_q \gg T$
{\setlength\arraycolsep{0pt}
\begin{eqnarray}
\label{equ:Sec9.24b}
{\cal R} \propto (m_q T)^{\frac{3}{2}} e^{-\frac{m_q}{T}} \sigma_{as},
\end{eqnarray}}
 where $\sigma_{as}$ is the asymptotic value of $\sigma$. For the drag coefficient, one gets from eq.(\ref{equ:Sec8.14})
{\setlength\arraycolsep{0pt}
\begin{eqnarray}
\label{equ:Sec9.24c}
{\cal A} \propto \frac{g_p}{8 \pi^2} \ \frac{T}{u} \int_0^{+\infty} \frac{q dq}{q^0} e^{-\frac{u^0 q^0-u q}{T}} \times \langle \sigma \, |t|\rangle
\end{eqnarray}}
and finds in the same limit, $m_q \gg T$,
{\setlength\arraycolsep{0pt}
\begin{eqnarray}
\label{equ:Sec9.24d}
{\cal A} \propto \sqrt{m_q T^3}\times e^{-m_q/T} \times \langle \sigma \, |t|\rangle_{q=\frac{m_q P}{M_Q}}.
\end{eqnarray}}
Consequently, if $m_q\gg T$, both the collisions rate as well as the drag decrease with increasing mass $m_q$. Besides, from eqs.(\ref{equ:Sec9.24b}) and (\ref{equ:Sec9.24d}) one sees directly that in this regime
{\setlength\arraycolsep{0pt}
\begin{eqnarray}
\label{equ:Sec9.24e}
\Delta p_L \propto {\cal A}/{\cal R} \propto \frac{\langle |t| \rangle}{m_q}, \ \ \textrm{with} \ \ \langle|t|\rangle = \frac{\langle\sigma \, |t|\rangle_{q=\frac{m_q P}{M_Q}}}{\sigma_{as}}.
\end{eqnarray}}
 From eq.(\ref{equ:Sec9.24e}) one deduces that for $m_q \gg T$, $\Delta p_L$ decreases with increasing  light quark mass, which explains the trends observed in
 Fig. \ref{fig:cDragOvRateDifferentModel}.
 
\begin{figure}[h!] 
\begin{center}
   \begin{minipage}[c]{.495\linewidth}
     \includegraphics[height=8.45cm, width=8.9cm]{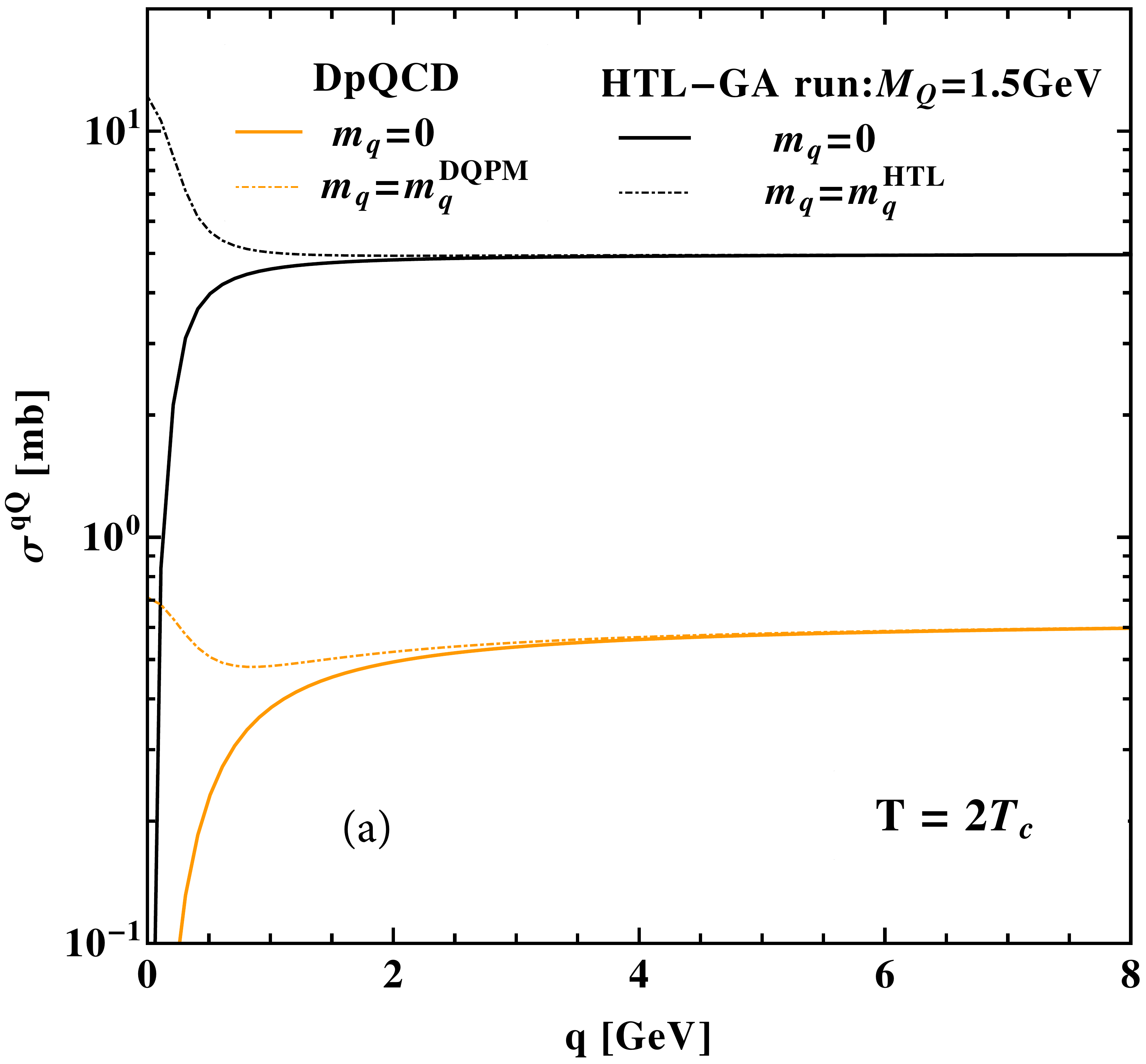}
   \end{minipage} 
   \begin{minipage}[c]{.495\linewidth}
      \includegraphics[height=8.45cm, width=8.9cm]{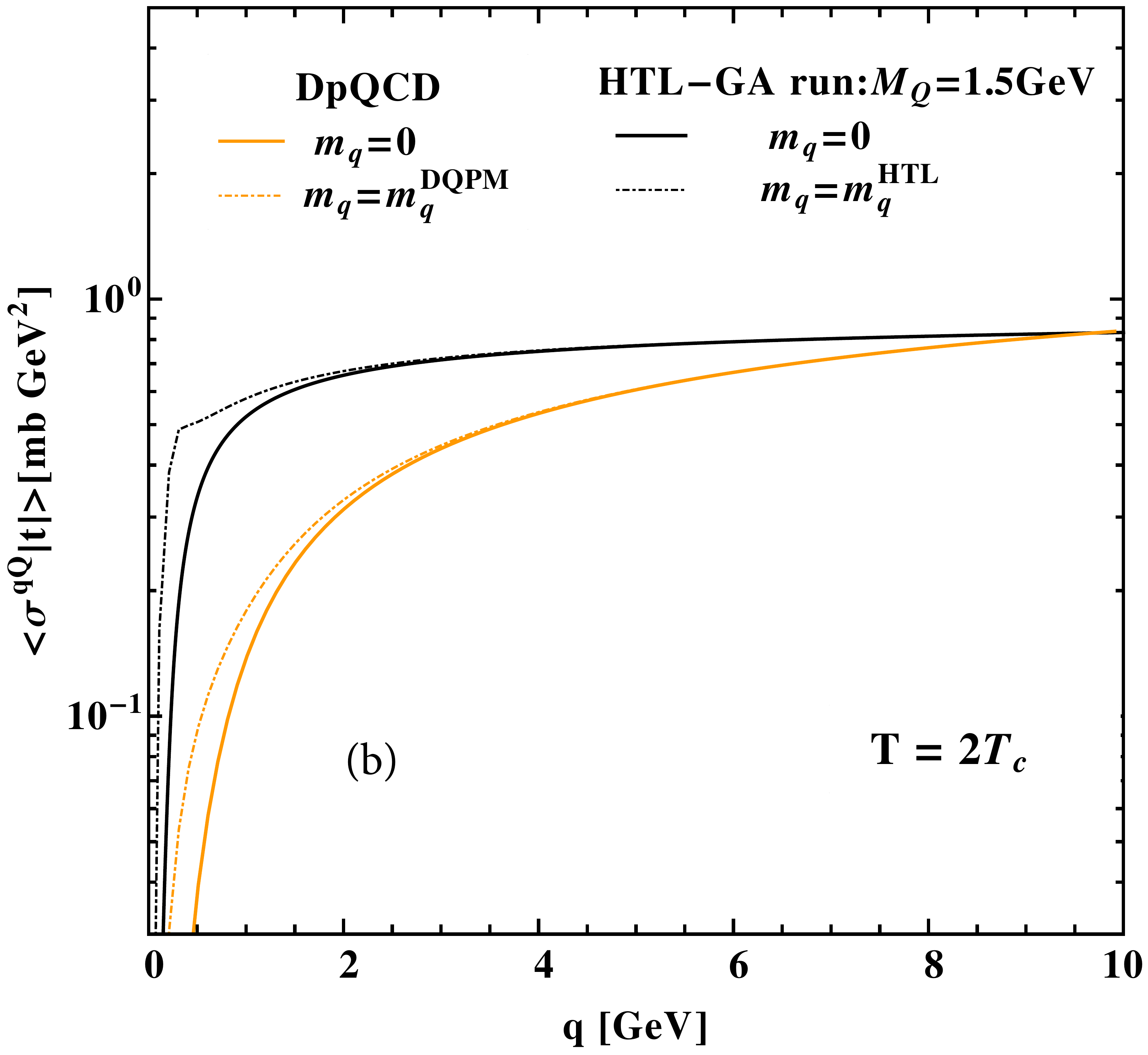}
\end{minipage}
\vspace{-0.2cm}
\caption{\emph{(Color online) The total cross section $\sigma^{qQ}$ for the elastic $qQ$ scattering (left) and $<\sigma \, |t|> = \int \frac{d \sigma}{d t} |t| dt$ for the models HTL-GA with running $\alpha$ and DpQCD as a function of $q$ which is related to $s$ by $s=M_Q^2 + m_q^2 + 2 M_Q E_q$.}}
\label{fig:SigtotSigT}
\end{center}
\end{figure}

  From Eqs. (\ref{equ:Sec9.24b}) and (\ref{equ:Sec9.24d}), we conclude that the differences in the asymptotic values between the HTL-GA and DpQCD are related to the total cross section for the rate and are related to the quantity $<\sigma \ |t|> = \int \frac{d \sigma}{d t} |t| dt$ for the drag coefficient. $<\sigma \ |t|>$ for the elastic $qQ$ scattering is plotted in figure \ref{fig:SigtotSigT}-(b) and compared to the total cross section $\sigma^{qQ}$ presented in fig.\ref{fig:SigtotSigT}-(a) following the models HTL-GA with $\alpha$ running and DpQCD. The gap in $\sigma^{qQ}$ between HTL-GA and DpQCD is reduced in $<\sigma \, |t|>$, which explains that the difference between HTL-GA and DpQCD is more important on the level of the rate compared to the drag coefficient.

\begin{figure}[h!] 
\begin{center}
   \begin{minipage}[c]{.57\linewidth}
     \includegraphics[height=9cm, width=9.5cm]{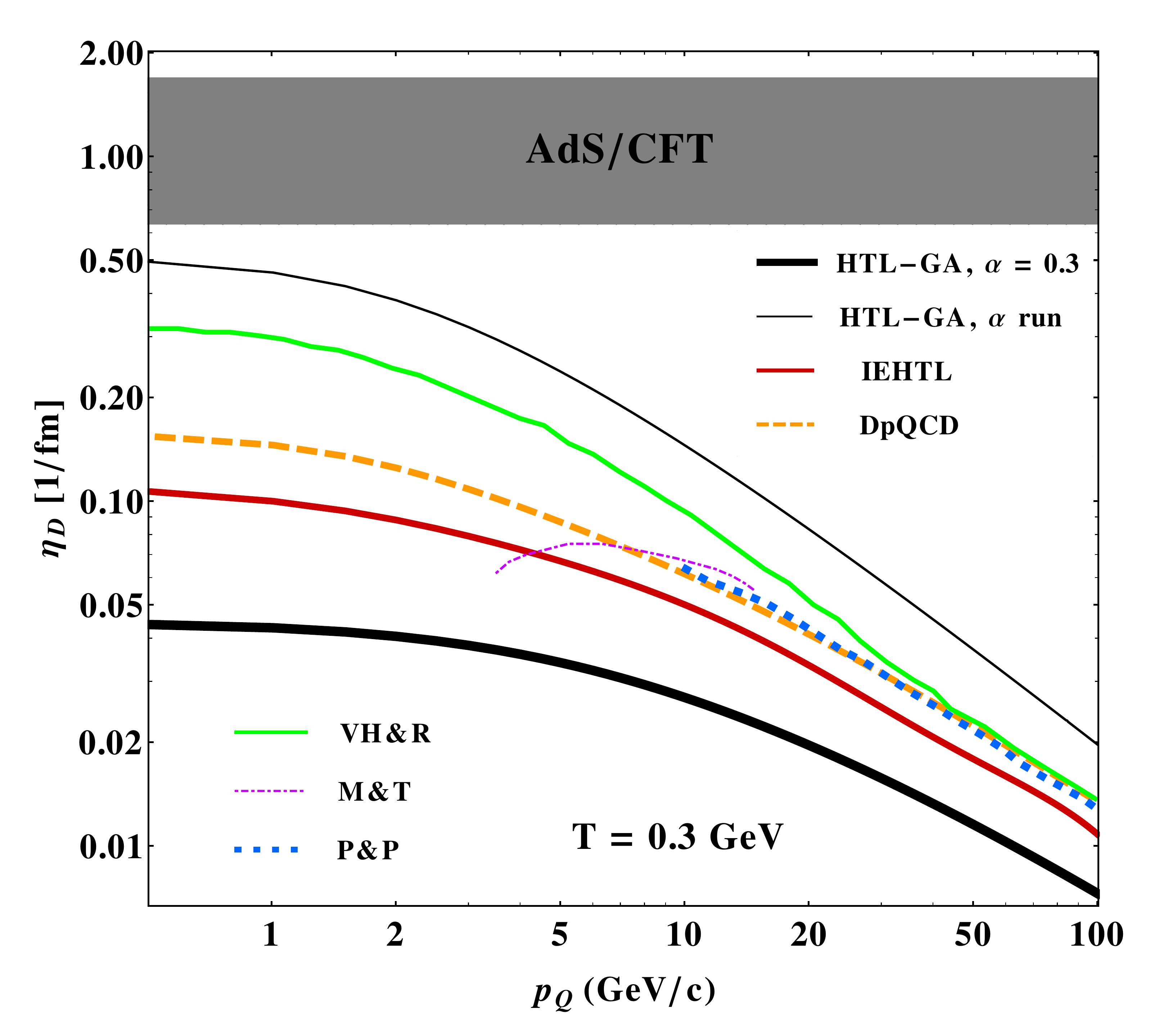}
   \end{minipage} \hfill
   \begin{minipage}[c]{.4\linewidth}
\caption{\emph{(Color online) $\eta_D$, the drag coefficient $A$ over $p_Q$ for c-quarks for different models as a function of the heavy quark momentum. The temperature of the scattering partners is 300 MeV. M\&T refers to Moore and Teaney (equation (B.31) with $\alpha = 0.3$ of \cite{Moore:2004tg}), VH\&R to van Hees and Rapp \cite{vanHees:2004gq,vanHees:2005wb,Greco:2007sz} (with a resonance width of = 400 MeV), P\&P to Peshier and Peigne \cite{Peigne:2008nd} and AdS/CFT to the drag coefficient calculated in the framework of the anti de Sitter/conformal field theory by Gubser \cite{Horowitz:2008ig,Gubser:2006qh}. HTL-GA (with $\alpha = 0.3$) and HTL-GA (with $\alpha$ run) refer to the two parameter sets of the HTL-GA model for constant and running coupling constant defined in \cite{PhysRevC.78.014904}.} \label{fig:cDragTheoreticalApproaches}}
\end{minipage}
\end{center}
\end{figure}

Figure \ref{fig:cDragTheoreticalApproaches} compares the results for charm quarks in our approach with other models for the coefficient
$\eta_D$ related to the drag coefficient by $\eta_D=A/p_Q$ at small momentum. We want to stress that due to the non-trivial $p$-dependence,
$\eta_D$ only represents the transport properties at small momentum. However, since several results for this coefficient have been presented in the literature
it may serve for a comparison between different models. For all calculations we have assumed that the heavy quark interacts with a particle
(light quarks and gluons) of the plasma at temperature $T=$300 MeV
and the calculations have been performed with the coupling constants of the corresponding publications. M\&T refers to Moore and
Teaney (equation (B.31) with $\alpha = 0.3$ of \cite{Moore:2004tg}), VH\&R to van Hees and Rapp \cite{vanHees:2004gq,vanHees:2005wb,Greco:2007sz}
(with a resonance width of 400 MeV), P\&P to Peshier and Peigne \cite{Peigne:2008nd} and AdS/CFT to the drag coefficient calculated
in the framework of the anti de Sitter/conformal field theory by Gubser \cite{Horowitz:2008ig,Gubser:2006qh}. HTL-GA (with $\alpha = 0.3$) and HTL-GA
(with $\alpha$ run) refer to the two parameter sets of the HTL-GA model for constant and running coupling as defined in \cite{PhysRevC.78.014904}.

The largest coefficient $\eta_D$ is observed for the AdS/CFT approach in which $\eta_D$ is momentum independent. All perturbative and non-perturbative
QCD-based coefficients $\eta_D$ decrease with increasing $p_Q$ which can be interpreted by the fact that with increasing momentum the relative momentum
change becomes lower. The values of $\eta_D$ vary substantially due to different assumptions on the infrared regulators and couplings and due to
different ingredients, like the presence of $qQ$ resonances in the plasma for the VH\&R model and off-shell dressed partons for IEHTL. One may ask the
question why the transverse momentum dependence of $R_{AA}$ and $v_2$, observed in heavy-ion reactions, can reproduced by different theories despite
of quite different drag coefficients.  As Gossiaux et al. \cite{Gossiaux:2009hr} have pointed out the drag coefficient is only one of two key ingredients
in heavy-ion transport approaches. The other is the expansion of the plasma, described either by hydrodynamical equations, a (tuned) fireball model or
by the parton-hadron-string dynamics (PHSD) transport approach. Therefore the measured $R_{AA}$ and $v_2$ are not a direct image of  $\eta_D$ or the drag coefficient $A$.

\subsection{Spatial diffusion coefficient}

In this section we will discuss an approximate calculation of the diffusion constant in coordinate space, $D_s$, which is related to the coefficient
$\eta_D$ by $D_s = T/(M_Q \eta_D)$ \cite{Moore:2004tg}. This relation, defined by Moore and Teaney \cite{Moore:2004tg}, is strictly valid only in the
non-relativistic limit, i.e. for velocities $\gamma \ v < \alpha/\sqrt{s}$ to leading logarithm in $T/m_D$, where $m_D$ is the Debye mass. Therefore it is
a good approximation to model the interaction of thermal heavy quarks, $M_Q \gg T$, with a typical thermal momentum $p \sim \sqrt{M T}$ and
a velocity $v \sim \sqrt{T/M} \ll 1$. Since $p \gg T$, it takes many collisions for the heavy quarks to change their momentum substantially.
Even for hard collisions with a momentum transfer $q \sim T$, it takes $\sim M/T$ collisions to change the momentum by a factor of the order one.
Therefore, we may model the interaction of the heavy quarks with the medium as uncorrelated momentum kicks.

\begin{figure}[h!] 
\begin{center}
   \begin{minipage}[c]{.47\linewidth}
     \begin{center}
     \includegraphics[height=9cm, width=9.5cm]{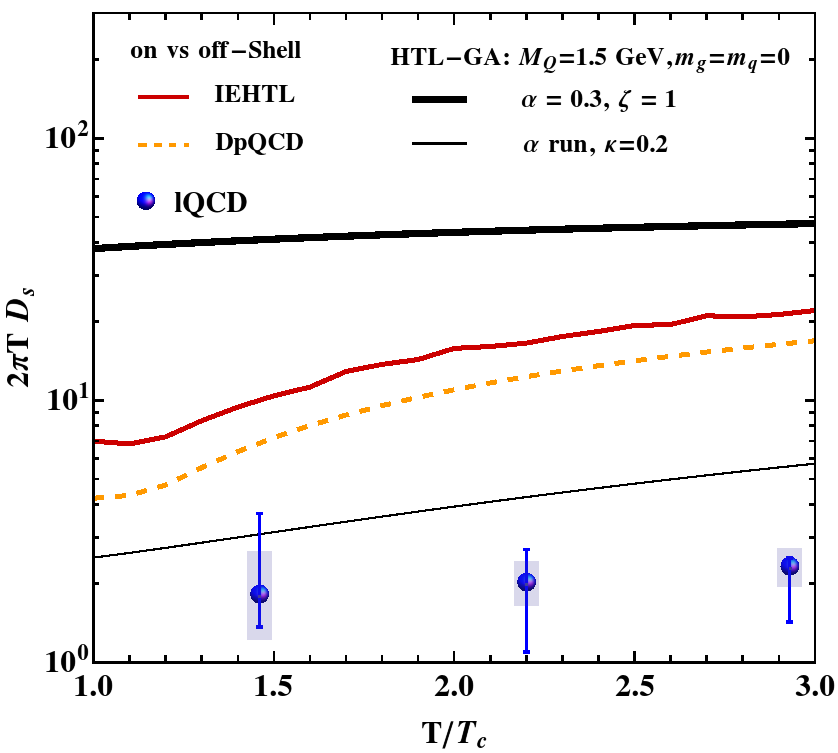}
     \end{center}
   \end{minipage} \hfill
\begin{minipage}[c]{.42\linewidth}
\caption{\emph{(Color online) The spatial diffusion coefficient $D_s$ for $c$-quarks, for HTL-GA with constant and running coupling
\cite{PhysRevC.78.014904}, for DpQCD/IEHTL in comparison with the results from lQCD \cite{Ding:2012sp} as a function of the temperature.
For the lQCD points the boxes show the statistical error estimated by the  'Jackknife method' while the bars stand for systematic uncertainties
from the MEM analyses. \label{fig:DiffusionCoef}}}
\end{minipage}
\end{center}
\end{figure}

Fig. \ref{fig:DiffusionCoef} displays $2\pi T D_s$ for charm quarks as a function of the plasma temperature. In addition to the approaches discussed in
this article, we plot also the results from quenched lattice QCD \cite{Ding:2012sp} for the spatial charm diffusion coefficient at finite temperature.
The spatial charm quark diffusion constant is related to the charmonium spectral function via the Kubo formula, Ref. \cite{Ding:2012sp}. The estimate
given by lQCD for the charm diffusion coefficient is $\approx 1/\pi T$ in the range $1.5 T_c \lesssim T \lesssim 3 T_c$. Note that the spatial charm
diffusion coefficient obtained at $1.46 T_c$ is more reliable than that obtained at higher temperatures, shown in Fig. \ref{fig:DiffusionCoef}.
At $2.20 T_c$ and $2.93 T_c$, due to the lack of precise information on the spectral function and also due to a smaller number of data points that can be used
in the MEM analyses \cite{Ding:2012sp}, the uncertainties of the spatial charm diffusion coefficient might be underestimated.

At first sight, Fig. \ref{fig:DiffusionCoef} seems to  show that all the approaches presented in our study fail to agree with the lQCD results, even if
the HTL-GA with running $\alpha$ spatial diffusion coefficient is in the vicinity of the lQCD results. Indeed,  for the models HTL-GA with constant $\alpha$ and for DpQCD/IEHTL
the spacial diffusion coefficients $D_s$ are very large compared to the one given by lQCD. However, one should consider this comparison with caution
since, as said, the relation used to determine $D_s$ \cite{Moore:2004tg} is an  approximation.
 In the DpQCD/IEHTL model the momentum transfer in a single collision is larger than $\sqrt{M T}$ as shown in Fig. \ref{fig:cDragOvRateDifferentModel}.
 Therefore a more detailed calculation of $D_s$ is necessary in order to arrive at a solid comparison between the lQCD result and our different approaches.
 Such a calculation, however, is beyond the scope of this study.

\section{Dynamical collisional energy loss}
\label{EnergyLoss}


The collisional energy loss $dE/dt$ has been formulated by Bjorken (\ref{equ:Sec1.1}) and is calculated in Sec.\ref{DragCoefficient} eq.  (\ref{equ:Sec8.17}). Similar to the drag coefficient the collisional energy loss for on-shell partons eq.(\ref{equ:Sec8.17}) can be easily extended for the off-shell case
using (\ref{equ:Sec1.3}). Going through similar calculations we arrive at


{\setlength\arraycolsep{-6pt}
\begin{eqnarray}
\label{equ:Sec10.6}
& & \frac{d E^{\textrm{off}}}{d t} \ (T) = \Biggl(A_{rest}^{0,\textrm{off}} - \tilde{A}_{rest}^{\nu,\textrm{off}} \Biggr) \hat{p},
\end{eqnarray}}
 where $A_{rest}^{0,\textrm{off}}$ and $\tilde{A}_{rest}^{0,\textrm{off}}$ are defined by

\begin{figure}[h!] 
\begin{center}
   \begin{minipage}[c]{.495\linewidth}
     \includegraphics[height=8.9cm, width=9cm]{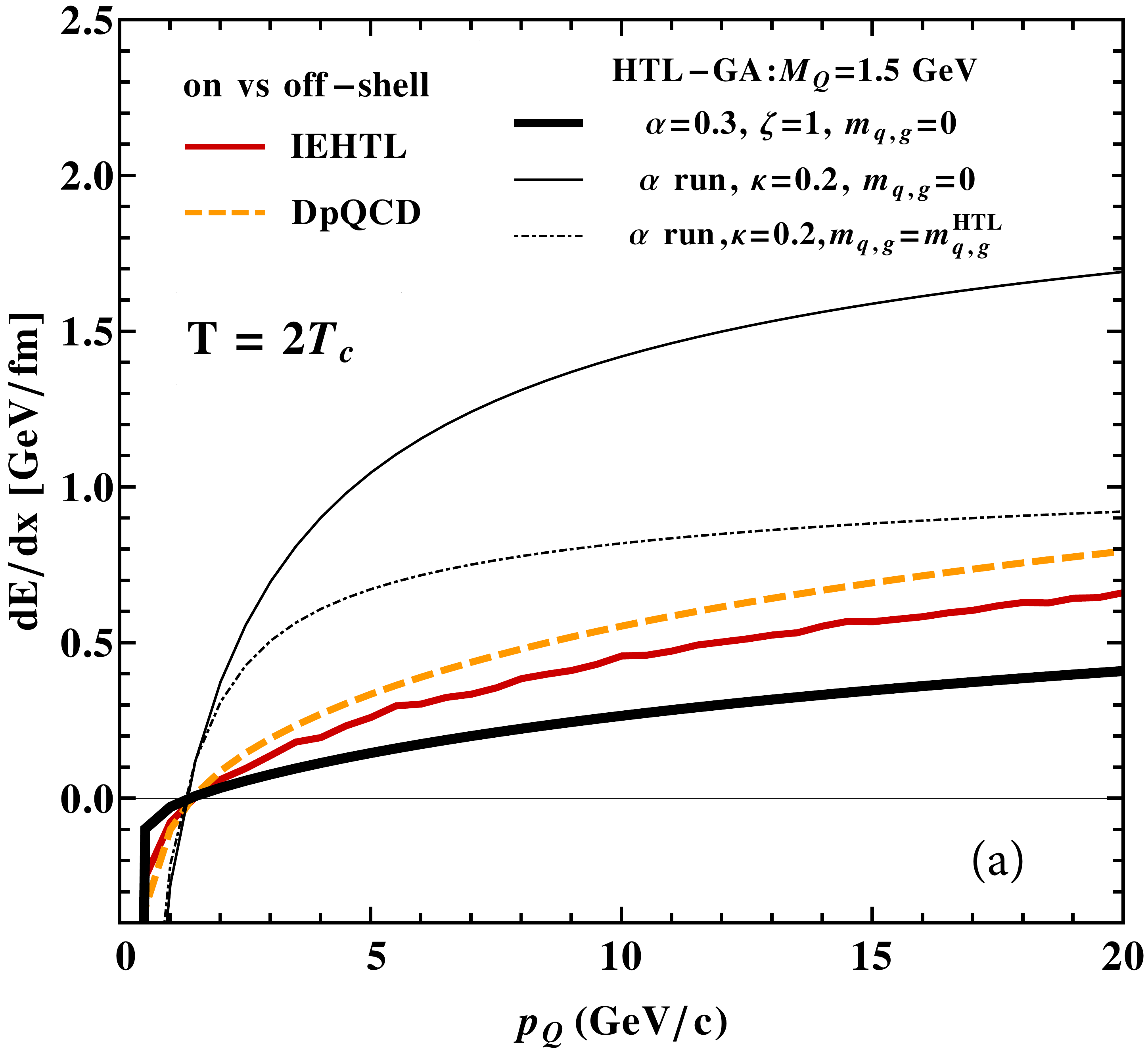}
   \end{minipage} 
   \begin{minipage}[c]{.495\linewidth}
      \includegraphics[height=8.9cm, width=9cm]{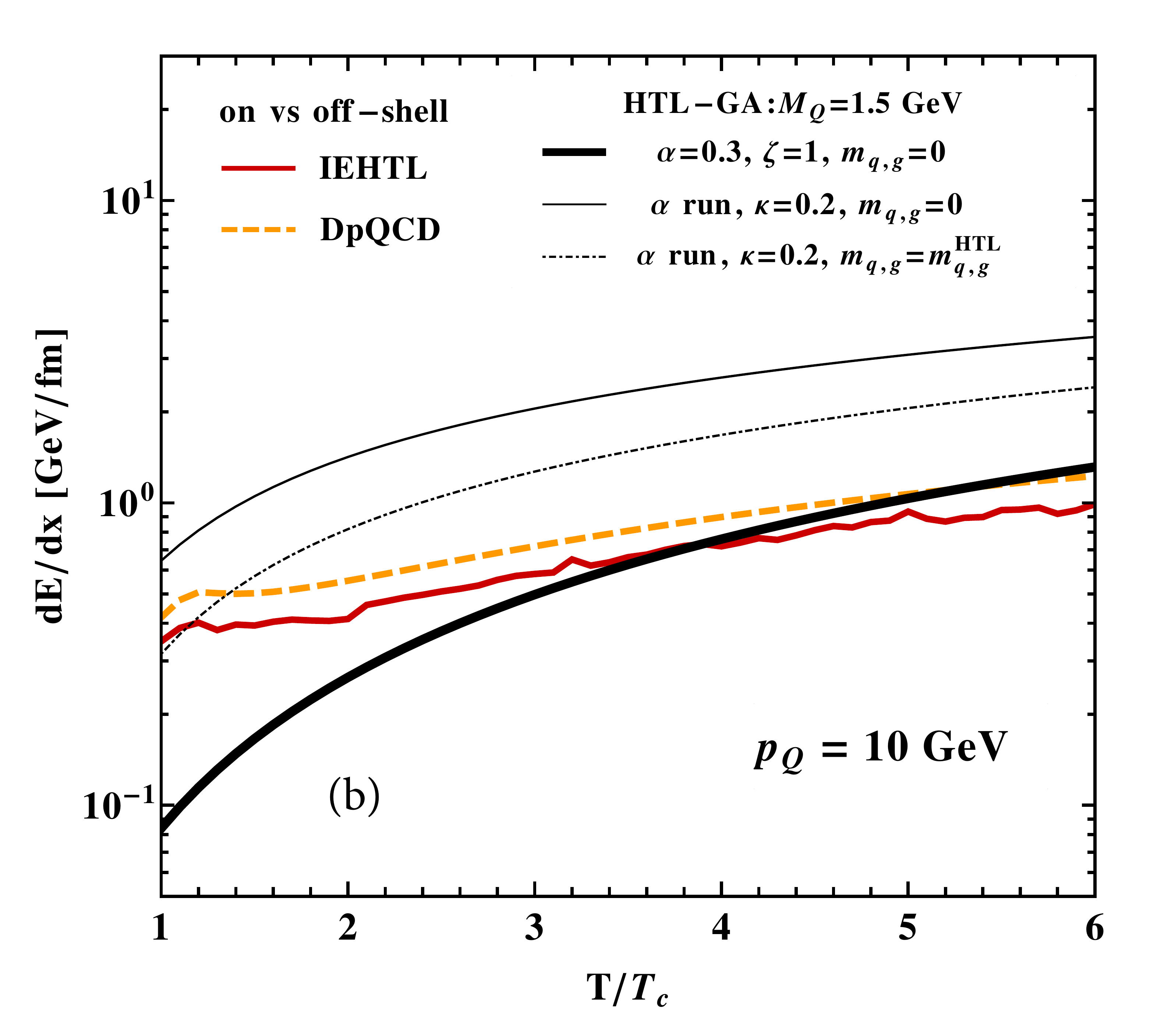}
\end{minipage}
\caption{\emph{(Color online) $c$-quark energy loss, $dE/dx$, in the plasma rest frame from the three different approaches as a function
of the heavy quark momentum $p_Q$ for $T = 2 T_c$ ($T_c = 0.158$ GeV)  (a) and as a function of the medium temperature $T$ for a heavy quark
momentum of $p_Q = 10$ GeV  (b). The heavy quark mass and momentum are given in $GeV$.}}
\label{fig:cdEdxDifferentModel}
\end{center}
\end{figure}

{\setlength\arraycolsep{0pt}
\begin{eqnarray}
\label{equ:Sec10.7}
& & A_{rest}^{0,\textrm{off}} = \frac{4}{(2 \pi)^7} \ \Pi_{i \in{p,q,p',q'}} \int m_p  m_i d m_i \ \rho_i^{WB} (m_i) \ \int_0^{\infty} \frac{q^5 m_1 (s) f_0 (q)}{s^2  E_q} d q,
\\
& & {} \tilde{A}_{rest}^{\nu,\textrm{off}} = \frac{4}{(2 \pi)^7} \ \Pi_{i \in{p,q,p',q'}} \int \frac{p m_p}{E_p}  m_i d m_i \ \rho_i^{WB} (m_i) \ \int_0^{\infty} \frac{q^4 (M_Q + E_q) m_1 (s) f_1 (q)}{s^2 E_q} d q.
\nonumber
\end{eqnarray}}
  More details are given in the appendix \ref{AppendixC}. The heavy quark energy loss (eq.\ref{equ:Sec8.17})
(as a function of the heavy quark momentum) is illustrated in Fig. \ref{fig:cdEdxDifferentModel}-(a) for the HTL-GA model for a constant and for a
running $\alpha$ and for massless and massive light quarks and gluons. The conclusions we have drawn for the drag coefficient remain also  valid
for the collisional energy loss, especially concerning  the reduction of $dE\!/\!dx$ for finite light quark masses in the HTL-GA model for a given temperature.
The running coupling leads to more energy loss in the HTL-GA approach. Fig \ref{fig:cdEdxDifferentModel}-(b) shows that with increasing temperature
-which corresponds to the increase of the parton masses in the DpQCD/IEHTL and the massive HTL-GA models-  the energy loss becomes larger.
This is more pronounced for large heavy quark momenta. $\displaystyle dE/dx$ can be linked to the drag coefficient
by 
{\setlength\arraycolsep{0pt}
\begin{eqnarray}
\label{equ:Sec10.7a}
& & \displaystyle \frac{d <\!\!E\!\!>}{d t} = \bsbt{\beta} . \frac{d <\!\! \bsp{p} \!\!>}{dt} + (1 - \beta^2) \ A_{rest}^0,
\end{eqnarray}}
 where the second term $(1 - \beta^2) \ A_{rest}^0$ corresponds to a gain from the medium to arrive at the average energy in the heat bath. This gain is seen in
Fig. \ref{fig:cdEdxDifferentModel}-(a) for small heavy quark momenta where $dE\!/\!dx$ is negative.

  The temperature dependence of the collisional energy loss for an intermediate heavy quark  momentum ($p_Q = 10$ GeV/c) is displayed in
Fig. \ref{fig:cdEdxDifferentModel}-(b) for the HTL-GA model with massless/massive light quarks and gluons and for the DpQCD/IEHTL models. The ratio
$\frac{dE}{dx}$(massive $q$,$g$)/$\frac{dE}{dx}$(massless $q$,$g$) shows almost no temperature dependence in the HTL-GA model.
Fig. \ref{fig:cdEdxDifferentModel}-(b) shows that the energy loss is increasing as a function of the temperature for both, the HTL-GA and DpQCD/IEHTL models.
For low heavy quark momenta and higher medium temperatures, the heavy quark looses less energy by elastic collisions. At  low momentum heavy quarks start
to gain energy to approach thermal equilibrium.

  Fig. \ref{fig:cdEdxDifferentModel}-(a) and (b) quantify also the influence of the finite parton mass and width on the heavy quark collisional energy loss.
Massive heavy quark scattering on massless partons, described by the HTL-GA approach (with $\alpha$ running), loose more energy as compared to the scattering
on massive partons. Comparing the DpQCD and IEHTL models, one deduces that the energy loss for partons with finite width is smaller as for the corresponding
on shell particles. This difference becomes
more important with increasing $p_Q$.  The large $dE/dx$ at low temperature observed in DpQCD/IEHTL is again due to the strong increase of the coupling.

\section{Diffusion Coefficient and Transverse Momentum Fluctuations \ $\hat{q}$}
\label{DiffusionCoefficient}

The covariant diffusion coefficient $\mathcal{B}^{\mu \nu} = \frac{E_p}{M_Q} B^{\mu \nu} (\bsp{p})$ has a tensorial structure and is given by
the explicit expression

{\setlength\arraycolsep{-6pt}
\begin{eqnarray}
\label{equ:Sec11.1}
& & \mathcal{B}^{\mu \nu} (\bsp{p}) =  \frac{E_p}{M_Q} \frac{1}{2} <(p - p')^\mu (p - p')^\nu> \ = \frac{1}{2 (2\pi)^5 M_Q} \int \frac{d^3 q}{2E_q} f(\bsp{q}) \ b^{\mu \nu} (\bsp{p}),
\nonumber\\
& & {} \textrm{with:} \ b^{\mu \nu} (\bsp{p}) := \int \frac{d^3 q'}{2E_{q'}} \int \frac{d^3 p'}{2E_{p'}} \ \frac{(p - p')^{\mu} (p - p')^{\nu}}{2} \delta^{(4)} (P_{in} - P_{fin}) \ \frac{1}{g_p g_Q} \sum_{k,l} |\mathcal{M}_{2,2}^{\textrm{off}} (p, q; i, j |p', q'; k, l)|^2 (s,t).
\end{eqnarray}}
 Using the projection in longitudinal and transverse direction (with respect to the incoming momentum $\bsp{p}$)

{\setlength\arraycolsep{-6pt}
\begin{eqnarray}
\label{equ:Sec11.4}
& & (\hat{\Pi}_{L} (\bsp{p}))_{i j} := \frac{p_i p_j}{p^2}, \hspace{1cm} (\hat{\Pi}_{T} (\bsp{p}))_{i j} := I_{i j} - \frac{p_i p_j}{p^2}.
\end{eqnarray}}

   where $\hat{\Pi}_{L}$ and $\hat{\Pi}_{T}$ are the projectors along $\bsp{p}$ and perpendicular to $\bs{p}$, respectively,
the spatial part of the diffusion coefficient $B^{i j} (\bsp{p})$ can be decomposed into a  longitudinal and a transverse component,

{\setlength\arraycolsep{-6pt}
\begin{eqnarray}
\label{equ:Sec11.3}
& & B^{i j} (\bsp{p}) =  B_{L} (p) \ \hat{p}^{i} \hat{p}^j + B_{T} (p) \ (\delta^{i j} - \hat{p}^{i} \hat{p}^j)
\nonumber\\
& & {} \textrm{with:} \hspace{0.3cm}  B_{L} (p) = \frac{1}{2} \frac{\Delta <\! p_{L}^2>}{\Delta t} := tr \left(B . \hat{\Pi}_{L} (\bsp{p}) \right); \hspace{0.7cm} B_{T} = \frac{1}{4} \ \frac{\Delta <\! p_{T}^2>}{\Delta t} := \frac{1}{2} tr \left(B . \hat{\Pi}_{T} (\bsp{p})\right),
\end{eqnarray}}
  where $<\!\Delta p_{L}^2\!>$ ($<\!\Delta p_{T}^2\!>$) is the longitudinal (transverse) averaged squared momentum acquired
by the heavy quark in a single collision with the medium constituents during its propagation through the plasma. The coefficients
$B_{L}$ and $B_{T}$ measure the second moment of the longitudinal and transverse momentum distribution of the heavy quark.
$<\!\Delta \! p_T \!>=0$ and $<\!\Delta p_L\!>$  have been studied in Sec.\ref{DragCoefficient}. The transverse second moment
$B_{T}$ is related to $\hat{q}$ which is defined as the average squared transverse momentum gain per unit
length by \cite{PhysRevC.78.014904,Beraudo:2009pe}:

{\setlength\arraycolsep{-6pt}
\begin{eqnarray}
\label{equ:Sec11.5}
& & \hat{q} = \frac{\Delta <\!p_{T}^2\!>}{\Delta x} =\frac{<\!p_{T}^2\!>_{single\ coll}}{\ell} = \frac{4 E_Q}{p_Q} \ B_{T}.
\end{eqnarray}}
 $\hat{q}$ quantifies the increase of the variance $<\!p_{T}^2\!>$ per unit length during the heavy quark propagation. In the following we will present $\hat{q}$ for massless/massive on-shell and off-shell partons.


\subsection{$\hat{q}$ \ for on-shell partons}

  For massive on-shell partons  $\hat{q}$ is given by:

{\setlength\arraycolsep{-6pt}
\begin{eqnarray}
\label{equ:Sec11.6}
& & \hat{q}^{\textrm{on}} (p) = \frac{M_Q^3}{2(2\pi)^3 p_Q} \int_{0}^{+ \infty} \frac{q^5}{s^2 E_q} \Biggl[\frac{m_1^{\textrm{on}} (s) - m_2^{\textrm{on}} (s)}{2} \left(f_{0} + f_{2} \right) + \frac{(E_q + M_Q)^2}{s} m_2^{\textrm{on}} (s) \left(f_{0} - f_{2} \right) \Biggr] \ d q
\end{eqnarray}}
  with:

{\setlength\arraycolsep{-6pt}
\begin{eqnarray}
\label{equ:Sec11.7}
& & m_1^{\textrm{on}} (s) = \int_{-1}^1 d \cos \theta \frac{1 - \cos \theta}{2} \sum |\mathcal{M}^{\textrm{on}}|^2 = \frac{1}{8 p_{cm}^4} \int_{- 4 p_{cm}^2}^{0} \ \sum |\mathcal{M}^{\textrm{on}}|^2 (s, t) \times (- t) d t
\nonumber\\
& & {} m_2^{\textrm{on}} (s) = \int_{-1}^1 d \cos \theta  \left(\frac{1 - \cos \theta}{2} \right)^2 \sum |\mathcal{M}^{\textrm{on}}|^2 = \frac{1}{32 p_{cm}^6} \int_{- 4 p_{cm}^2}^{0} \ \sum |\mathcal{M}^{\textrm{on}}|^2 (s, t) \ t^2 d t,
\end{eqnarray}}
and:
{\setlength\arraycolsep{-6pt}
\begin{eqnarray}
\label{equ:Sec11.8}
& & f_{i} = \frac{1}{2} \int d \cos \theta f \left(\frac{u^0 E_q - u q \cos \theta}{T} \right) \ \cos^i \theta, \hspace{0.3cm} i \in [0,1,2].
\end{eqnarray}}
  The details are given in the appendix \ref{AppendixD}.

\subsection{$\hat{q}$ \ for off-shell partons}

  The off-shell $\hat{q}^{\textrm{off}}$ (using the Breit-Wigner spectral functions for the off-shell partons) is deduced by extending the
  on-shell $\hat{q}^{\textrm{on}}$ as

\begin{figure}[h!] 
\begin{center}
   \begin{minipage}[c]{.495\linewidth}
     \includegraphics[height=8.9cm, width=9cm]{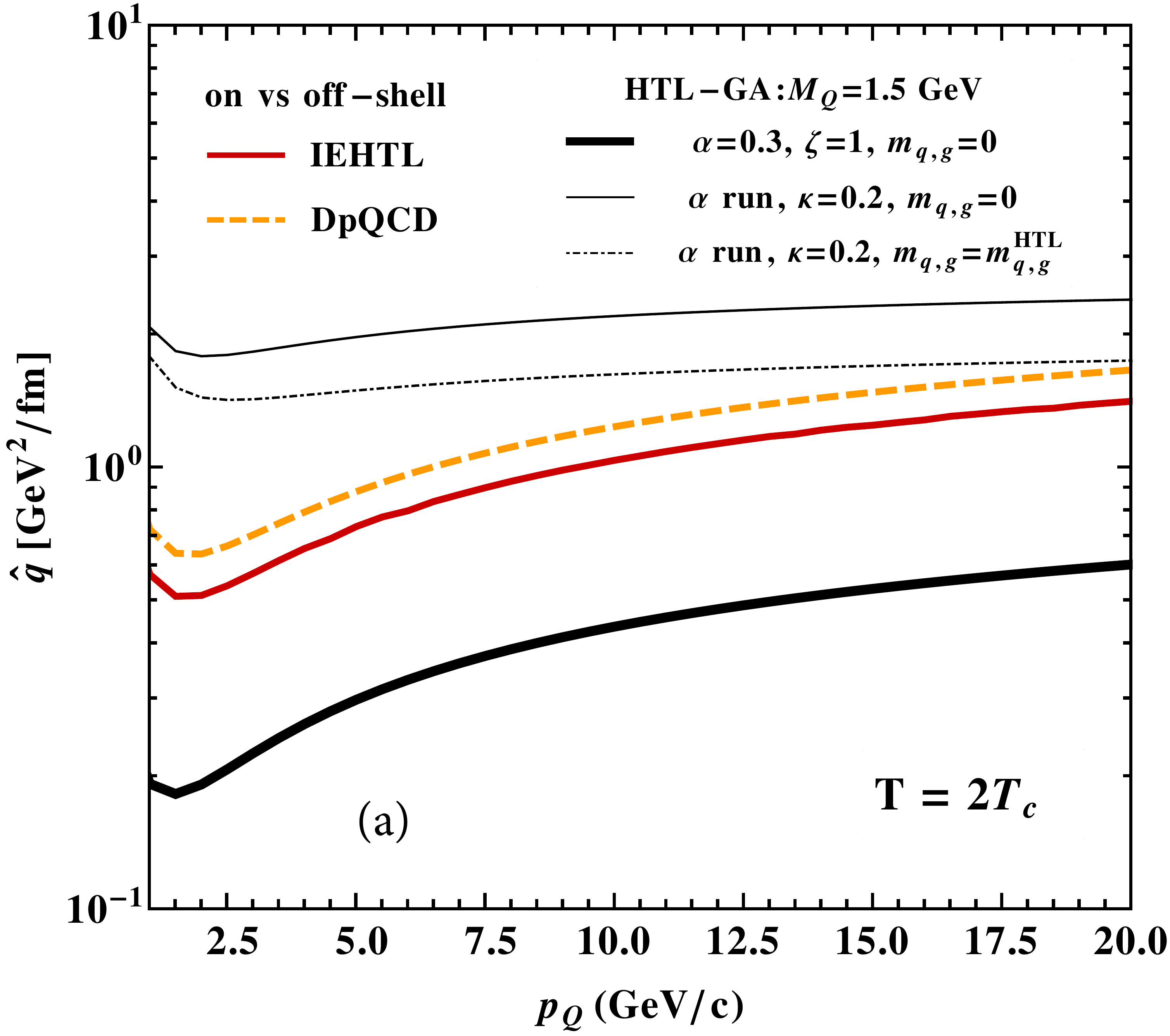}
   \end{minipage} 
   \begin{minipage}[c]{.495\linewidth}
      \includegraphics[height=8.9cm, width=9cm]{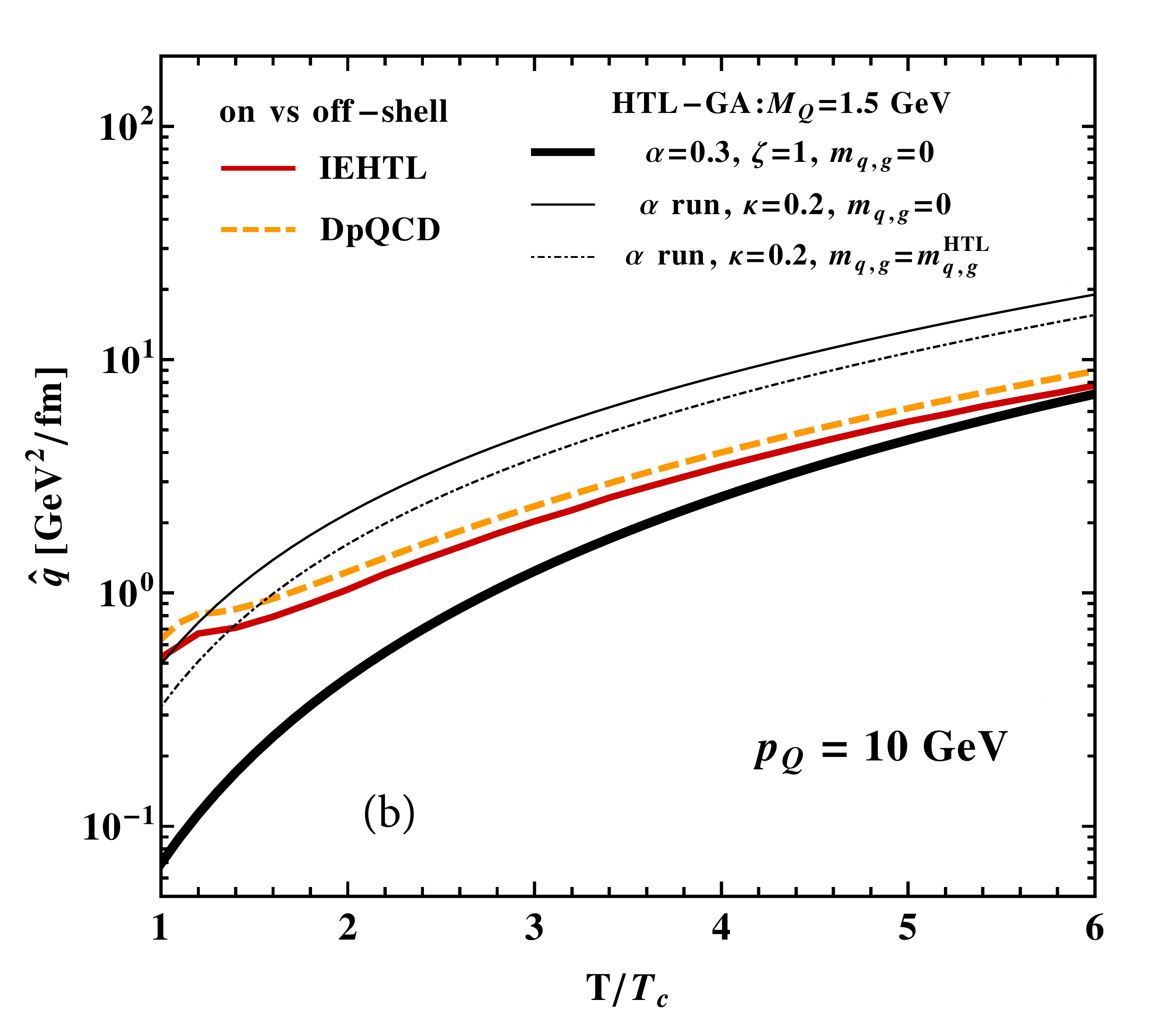}
\end{minipage}
\caption{\emph{(Color online) $\hat{q}$ for $c$-quarks in the plasma rest frame for the three different approaches. (a) $\hat{q}$ as a function of the
heavy quark momentum $p_Q$ for $T = 2 T_c$ ($T_c = 0.158$ GeV). (b) $\hat{q}$ as a function of the temperature $T/T_c$ for an  intermediate
heavy quark momentum of $p_Q = 10$ GeV.}}
\label{fig:cqhatDifferentModel}
\end{center}
\end{figure}

{\setlength\arraycolsep{-10pt}
\begin{eqnarray}
\label{equ:Sec11.9}
& & \hspace*{-1.2cm} \hat{q}^{\textrm{off}} (p) \!=\! \frac{1}{2 (2 \pi)^3} \Pi_{i \in{p,q,p',q'}} \!\!\!\! \int \frac{m_p^2}{p_Q} m_i d m_i \rho_p^{WB} (m_i) \!\!\!\! \int \frac{q^5 m_p}{s^2 E_q} \!\! \Biggl[\!\frac{m_1^{\textrm{off}} (s) - m_2^{\textrm{off}} (s)}{2} \left(f_{0} + f_{2} \right) + \frac{(E_q + m_p)^2}{s} m_2^{\textrm{off}} (s) \left(f_{0} - f_{2} \right) \!\!\Biggr] \! d q
\end{eqnarray}}
with:

{\setlength\arraycolsep{-6pt}
\begin{eqnarray}
\label{equ:Sec11.10}
& & m_1^{\textrm{off}} (s) = \int_{-1}^1 d \cos \theta \frac{1 - \cos \theta}{2} \sum |\mathcal{M}^{\textrm{off}}|^2 = \frac{1}{4 p_Q^i p_Q^f} \int_{t_{min}}^{t_{max}} \ \sum |\mathcal{M}^{\textrm{off}}|^2 (s, t) \Biggl[1 - \frac{t - (M_Q^i)^2 + (M_Q^f)^2 - 2 E_Q^i E_Q^f}{2 p_Q^i p_Q^f} \Biggr] d t
\\
& & {} m_2^{\textrm{off}} (s) = \int_{-1}^1 d \cos \theta  \left(\frac{1 - \cos \theta}{2} \right)^2 \sum |\mathcal{M}^{\textrm{off}}|^2 = \frac{1}{8 p_Q^i p_Q^f} \int_{t_{min}}^{t_{max}} \ \sum |\mathcal{M}^{\textrm{off}}|^2 (s, t) \Biggl[1 - \frac{t - (M_Q^i)^2 + (M_Q^f)^2 - 2 E_Q^i E_Q^f}{2 p_Q^i p_Q^f} \Biggr]^2 d t,
\nonumber
\end{eqnarray}}
 where $p_Q^i$ ($M_Q^i$) is the initial heavy momentum (mass) and $p_Q^f$ ($M_Q^f$) is the final heavy momentum (mass). Fig. \ref{fig:cqhatDifferentModel}-(a) and (b)
 compare  $\hat{q}$ for the on- and off-shell calculations as a function of the heavy quark momentum and of the medium temperature, respectively. The results
for the HTL-GA model are given for constant or running $\alpha$ and for massles/massive light quarks and gluons. These figures give also the corresponding
$\hat{q}$ for the DpQCD/IEHTL approaches. For all the models, after a minimum at $p_Q \in [1,2]$ GeV, $\hat{q}$ increases as a function of the heavy quark momentum.
Heavy quarks which collide with the running $\alpha$ in the HTL-GA approach show a much larger transverse momentum transfer per unit length as compared to
calculations with a constant coupling  which does not depend on the momentum transfer. The calculations in the HTL-GA  approach (with a fixed coupling)
and in the DpQCD approach (with a temperature dependent coupling)  show a similar $\hat q$ above $T=6T_c$ but differ largely for temperatures close to $T_c$.
Employing a spectral function lowers the transverse momentum transfer by about 20\% independent of momentum or temperature.

We find that $\hat q$ decreases with increasing the mass of the light partons and gluons. As for the drag coefficient, a system with massless quarks has a shorter mean-free-path than one with massive partons. This is not compensated by a larger momentum transfer in an individual collision if the partons carry a mass. For the case of the scattering of a heavy quark on a light quark we find from eq.(\ref{equ:Sec11.6}) that for $p \gg M_Q$ $\hat{q}$ in the rest system of the heavy quark is given by

{\setlength\arraycolsep{0pt}
\begin{eqnarray}
\label{equ:Sec11.11}
{\hat{q}} \propto \frac{g_p}{4 \pi^2} \ \frac{T}{u^2} \int_0^{+\infty} \frac{q^2 dq}{q^0} e^{-\frac{u^0 q^0-u q}{T}} \times \langle \sigma \, |t|\rangle
\end{eqnarray}}
and find in the limit, $m_q \gg T$,
{\setlength\arraycolsep{0pt}
\begin{eqnarray}
\label{equ:Sec11.12}
{\hat{q}} \propto \left(m_q T \right)^{3/2} \times e^{-m_q/T} \times \langle \sigma \, |t|\rangle_{q=\frac{m_q P}{M_Q}}.
\end{eqnarray}}
Consequently, if $m_q\gg T$, $\hat{q}$ decreases with increasing mass $m_q$.         

  The temperature dependence of $\hat{q}$ for an intermediate (10 GeV/c) heavy quark momentum, displayed in Fig. \ref{fig:cqhatDifferentModel}-(b), shows the
same form for all the models except near $T_c$ where DpQCD/IEHTL has a slight bump related to the temperature dependence of DQPM partons masses.
Fig. \ref{fig:cqhatDifferentModel}-(b) shows that at high temperatures $\hat{q}$ increases strongly with the temperature for both, the  HTL-GA and the
DpQCD/IEHTL models.

\begin{figure}[h!] 
\begin{center}
   \begin{minipage}[c]{.495\linewidth}
     \includegraphics[height=8.9cm, width=9cm]{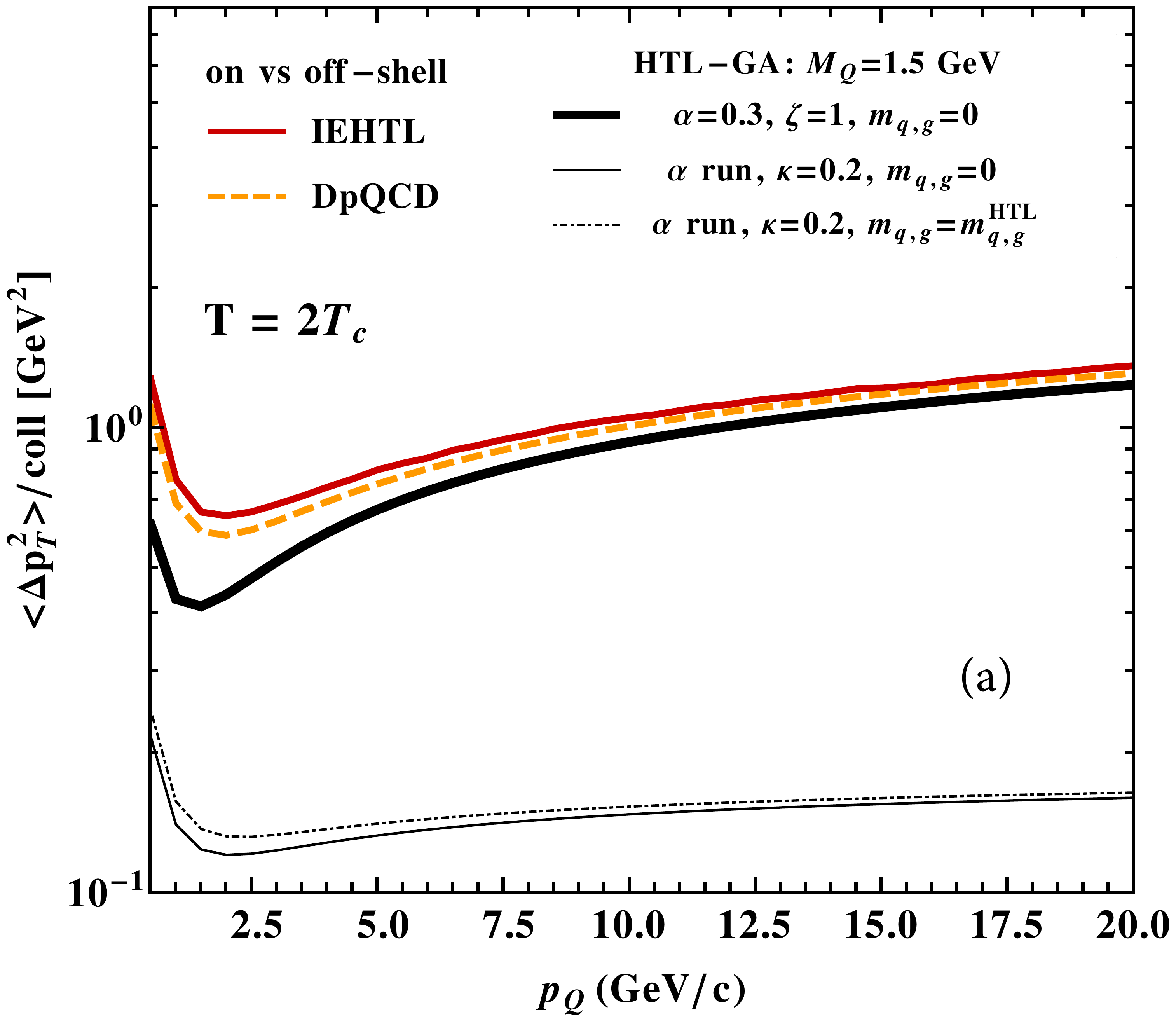}
   \end{minipage} 
   \begin{minipage}[c]{.495\linewidth}
      \includegraphics[height=8.9cm, width=9cm]{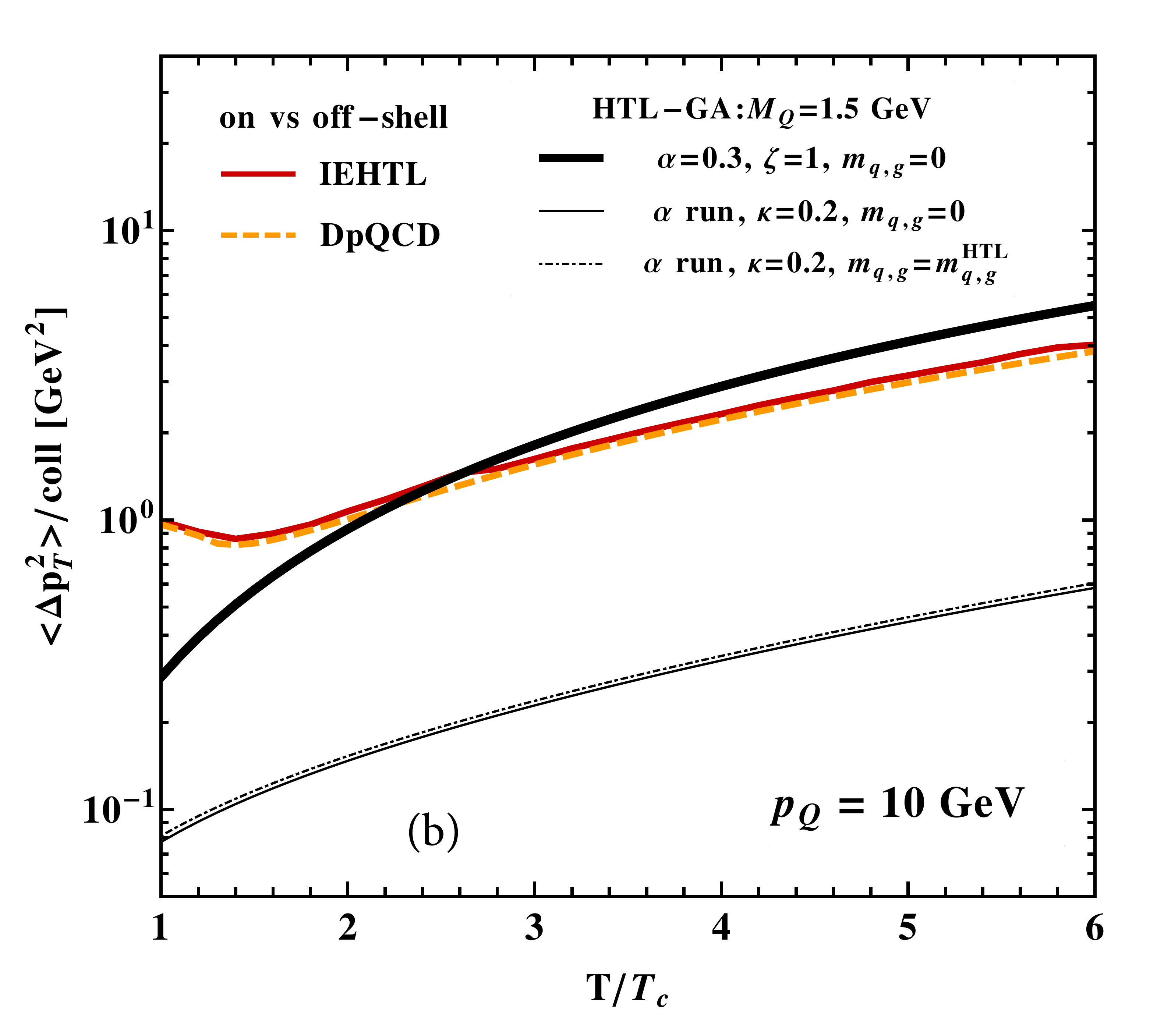}
\end{minipage}
\caption{\emph{(Color online) The transverse momentum transfer squared $<\!\Delta p_T^2\!>$ of $c$-quarks per collision in the plasma rest
frame for the three different approaches: (a) as a function of the heavy quark momentum $p_Q$ for $T = 2 T_c$ ($T_c = 0.158$ GeV), (b)
as a function of the temperature $T/T_c$ for an  intermediate heavy quark momentum of $p_Q = 10$ GeV.}}
\label{fig:cqhatOverRateDifferentModel}
\end{center}
\end{figure}

  As for the longitudinal momentum transfer, we study the transverse momentum transfer in a single collision, i.e. $<\!\Delta p_T^2\!>$ which is proportional to the ratio between $\hat{q}$ and the interaction rate. $<\!\Delta p_T^2\!>$ is illustrated in Fig. \ref{fig:cqhatOverRateDifferentModel} as a function of the heavy quark momentum (a) and as a function of the temperature (b). Whereas the heavy quark $\hat{q}$ and the heavy quark rate are larger for the HTL-GA model with running $\alpha$, the transverse momentum transfer per collision $<\!\Delta p_T^2\!>$ is smaller as compared to the DpQCD/IEHTL models. $<\!\Delta p_T^2\!>$ increases as a function of the temperature which shows that the increase of the temperature leads to more transverse momentum change and almost constant longitudinal momentum transfer, as seen in the temperature dependence of $<\!\Delta p_L\!>$. Comparing the massless (thin black line) and massive (dotted dashed black line) parton cases in the HTL-GA model in Figs. \ref{fig:cqhatOverRateDifferentModel}-(a) and (b), one deduces that in contrast to the longitudinal momentum transfers the scattering of a heavy quark on massive partons transfers more transverse momentum than the scattering on massless partons for all heavy quark momenta. This can be seen from eqs.(\ref{equ:Sec9.24b}) and (\ref{equ:Sec11.12}) where in the regime $m_q \gg T$
{\setlength\arraycolsep{0pt}
\begin{eqnarray}
\label{equ:equ:SecSec11.13}
\Delta p_T^2 \propto {\hat{q}}/{\cal R} \propto \langle |t| \rangle, \ \ \textrm{with} \ \ \langle|t|\rangle = \frac{\langle\sigma \, |t|\rangle_{q=\frac{m_q P}{M_Q}}}{\sigma_{as}},
\end{eqnarray}}
 and the trends observed in figs. \ref{fig:cqhatOverRateDifferentModel} are explained.

\begin{figure}[h!] 
\begin{center}
   \begin{minipage}[c]{.57\linewidth}
     \includegraphics[height=9cm, width=9.5cm]{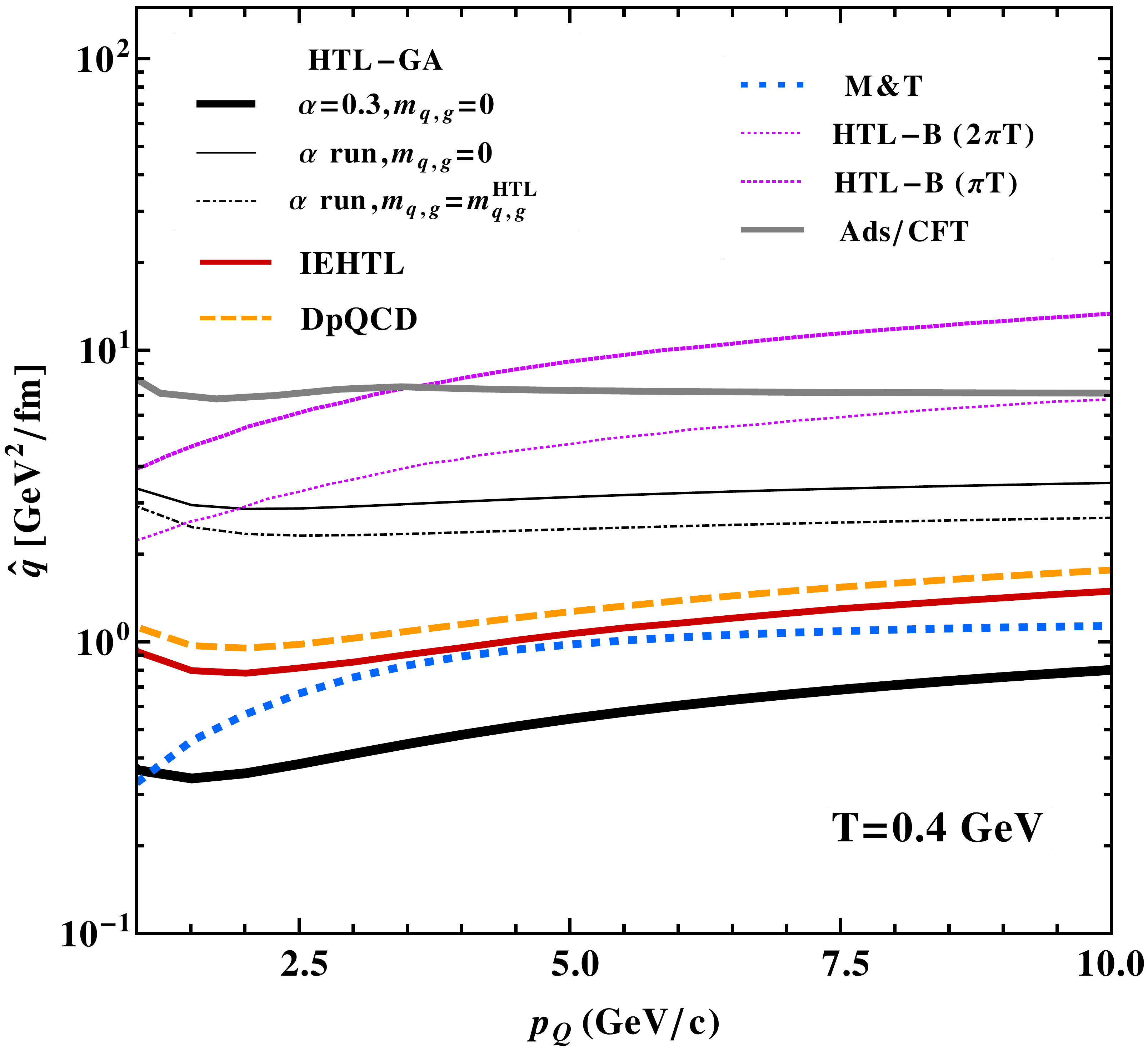}
   \end{minipage} \hfill
   \begin{minipage}[c]{.4\linewidth}
\caption{\emph{(Color online) $\hat{q}$ for $c$-quarks for different approaches as a function of the heavy quark momentum. The temperature
of the scattering partners is $T=400$ MeV. M\&T refers to Moore and Teaney (equation (B.32) with $\alpha = 0.3$ of \cite{Moore:2004tg}),
HTL-B referring to Beraudo and al. \cite{Beraudo:2009pe}, with the two choices $\mu= 2 \pi T$ and $\mu = \pi T$ ($\mu$ is a parameter in
the coupling constant) and AdS/CFT to $\hat{q}$ calculated in the framework of the anti de Sitter/conformal field $\mathcal{N} = 4$ SYM theory
by Refs.\cite{CasalderreySolana:2007qw,Gubser:2006nz,Liu:2008tz}. HTL-GA (with $\alpha = 0.3$) and HTL-GA (with running $\alpha$) refer to the
two parameter sets of the HTL-GA model for constant and running coupling  defined in \cite{PhysRevC.78.014904}.} \label{fig:cqhatTheoreticalApproaches}}
\end{minipage}
\end{center}
\end{figure}

  We extend our model comparison for $\hat{q}$, as for $\eta_D$ in Fig.\ref{fig:cDragTheoreticalApproaches}, by including in Fig.
  \ref{fig:cqhatTheoreticalApproaches} the standard pQCD result given by M\&T referring to Moore and Teaney (equation (B.32) with $\alpha = 0.3$
  of \cite{Moore:2004tg}), HTL-B referring to Beraudo and al \cite{Beraudo:2009pe}, with the two choices $\mu= 2 \pi T$ and $\mu = \pi T$ ($\mu$
  is a parameter in the coupling constant) and AdS/CFT to $\hat{q}$ calculated in the framework of the anti de Sitter/conformal field
  $\mathcal{N} = 4$ SYM theory by Refs.\cite{CasalderreySolana:2007qw,Gubser:2006nz,Liu:2008tz}. HTL-GA (with $\alpha = 0.3$) and HTL-GA (with running $\alpha$)
  refer to the two parameter sets of the HTL-GA model for constant and running coupling defined in \cite{PhysRevC.78.014904}. For all calculations
  we have assumed the plasma temperature to be  fixed to $T=400$ MeV.

The largest value for $\hat{q}$ is given by Ads/CFT and HTL-B. The calculations in HTL-B are given in the spirit of the Hard Thermal Loop for the week coupling limit of heavy quarks with the plasma partons. AdS/CFT predicts $\hat{q} = \frac{2}{v} \gamma^{1/2} \sqrt{\lambda \pi} T_{SYM}^3$, with $\gamma=1/\sqrt{1-v^2}$ and $\lambda = g_{SYM}^2 N_c$;  $\hat{q}$ given by M\&T in the framework of pQCD is slightly above the results from the DpQCD/IEHTL model.  
\section{Transport Cross Section and critical analysis of the model applicability}
\label{TransportCrossSection}

  In potential-scattering theory, the so called transport cross section is defined as the cross section weighted with $1 - \cos \theta$: 
{\setlength\arraycolsep{-6pt} 
\begin{eqnarray}
\label{equ:Sec7.1}
& & \sigma^{tr} (\bsp{p}, T) = \int d \sigma (1 - \cos \theta) = \int \frac{d \sigma}{dt} (1 - \cos \theta (t)) \ d t, 
\end{eqnarray}}  
where $\theta$ is the angle between the momentum of the outgoing and incoming particle measured in the lab frame. It is closely related to the transport coefficients \cite{PBGossiauxHD,KhrapakTrCrossSection}. It has been introduced to have a measure for the momentum transfer in a single collision and gives a better possibility to compare different theories as the cross section itself because it weights the cross section by the average momentum transfer and suppresses small angle scattering which may have a large cross section but does not change much the momentum of the incoming particle. Therefore, the transport cross section is a more meaningful parameter to describe quark relaxation since QCD $2 \rightarrow 2$ interactions are usually forward-peaked, and so a large number of scatterings may map into a small changes of the momentum, only. For this purpose, we will first generalize the definition of the transport cross section to the case of scattering on finite mass particles and then present the results for the various models under study.
  
\subsection{Relations between the transport cross section and the transport coefficients} 

In the case of a test particle propagating in a cell which contains finite mass particles distributed like $f$, different extensions of the definition (\ref{equ:Sec7.1}) are possible. Starting from the expression of the interaction rate $\Gamma$ 
{\setlength\arraycolsep{-6pt} 
\begin{eqnarray}
\label{equ:Sec7.2}
& & \Gamma = \int f(s) v_{\textrm{rel}} (s) \ \sigma^{\textrm{tot}} (s)ds = \frac{v}{\ell}, 
\end{eqnarray}}  

where $\ell$ is the mean free path, $v$ is the relative velocity between the incident particle and the center of mass of the cell and $v_{\textrm{rel}}$ is the relative velocity of this incident particle and one particles of the medium, one can define an effective total cross section as

{\setlength\arraycolsep{-6pt} 
\begin{eqnarray}
\label{equ:Sec7.3}
& & \sigma_{\textrm{eff}}^{\textrm{tot}} (v;f) = \frac{\Gamma}{n \ v} = \frac{\int f(s) v_{\textrm{rel}} (s) \ \sigma^{tot} (s)}{v \int f(s) ds}, 
\end{eqnarray}}  

where $n=  \int d^3q/(2\pi)^3 f(q)$. From eq.(\ref{equ:Sec7.3}) it follows that the mean free path $\ell := \frac{1}{n \sigma_{\textrm{eff}}^{\textrm{tot}}}$ corresponds exactly to the interaction rate $\Gamma$. In the spirit of  eq.(\ref{equ:Sec7.3})  one may define the transport cross section as,

{\setlength\arraycolsep{-6pt} 
\begin{eqnarray}
\label{equ:Sec7.4}
& & \sigma_{\textrm{I}}^{\textrm{tr}} (v;f) =  \frac{\int d^3q /(2\pi)^3 f(q) v_{\textrm{rel}} (q, p) \ \int \sigma (1 -  \hat{p} \cdot \hat{p}')}
 {v \int d^3 q/(2\pi)^3 f(q)}, 
\end{eqnarray}}  

 where $\hat{p}$ ($\hat{p}'$) is the direction of the initial (final) heavy quark momentum in the cell rest system. For a relativistic particle, $\parallel\! \bs{p} \!\parallel \approx \parallel\! \bs{p}' \!\parallel$ and at small scattering angles, $\cos\theta \approx 1 - \theta^2/2$,  we can approximate  $1 -  \hat{p} \cdot \hat{p}' = 1 - cos\theta \approx \frac{sin^2\theta}{2} \approx \frac{(\Delta p_T)^2}{2 p^2}$. Then eq.(\ref{equ:Sec7.4}) becomes 

{\setlength\arraycolsep{-6pt} 
\begin{eqnarray}
\label{equ:Sec7.5}
& & \sigma_{\textrm{I}}^{\textrm{tr}} (v;f) = \frac{\int d^3q/(2\pi)^3 f(q) v_{\textrm{rel}} (q, p) \ \int \sigma (\Delta p_T)^2}{2 n p^2}. 
\end{eqnarray}}  
The numerator of this expression is nothing but $\hat{q}$ (\ref{equ:Sec11.5}), which measures the transverse momentum transfer squared per unit length. Hence, one obtains a first generalization of the transport cross section, which is directly proportional to $\hat{q}$:

{\setlength\arraycolsep{-6pt} 
\begin{eqnarray}
\label{equ:Sec7.6}
& & \sigma_{\textrm{I}}^{\textrm{tr}} (v;f) = \frac{\hat{q}}{2 n p^2}.
\end{eqnarray}}  
to which one can associate a mean-free-path of diffusion given by $\ell^{\rm tr} = \frac{1}{n \sigma_{\textrm{I}}^{\textrm{tr}}}= \frac{\hat{q}}{2 p^2}$. As $\hat{q}$ is given by 
$\langle\Delta p_T^2\rangle/\ell $ with $\langle \Delta p_T^2\rangle$ being the transverse momentum transfer squared in a single collision and $\ell =\frac{1}{n\sigma_{tot}}$, we find the relation
{\setlength\arraycolsep{-6pt} 
\begin{eqnarray}
\label{equ:Sec7.7a}
& &\ell^{\rm tr} = \frac{2p^2}{\Delta p_T^2}  \ell \ \gg \ell.
\end{eqnarray}}  
This mean free path of diffusion Eq.~(\ref{equ:Sec7.7a}) is the typical length after which the ultrarelativistic particle is deflected by a finite angle. It can thus be interpreted as a relaxation time of this angular evolution. For the purpose of describing heavy-quark transport, one can consider that an average heavy quark is undergoing a rather soft transverse momentum exchange of the order of  $\langle \Delta p_T^2\rangle$ every mean free path $\ell$ or, alternatively, a rather hard momentum exchange $\langle \Delta p_T^2\rangle \times \frac{\ell^{\rm tr}}{\ell}$ every $\ell^{\rm tr}$, both leading to the same $\hat{q}$ value.
 
 
  For the case of a potential scattering one has $\parallel\! \bsp{p} \! \parallel = \parallel\! \bsp{p}' \! \parallel$ and therefore an alternative generalization of the transport cross section is possible \cite{PBGossiauxHD},

{\setlength\arraycolsep{-6pt} 
\begin{eqnarray}
\label{equ:Sec7.7}
& & \sigma_{\textrm{II}}^{\textrm{tr}} (v;f) = \frac{\int d^3 q /(2\pi)^3 f(q) v_{\textrm{rel}} (q, p) \ \int \sigma (1 -  \frac{p'_L}{p})}{v \int d^3 q/(2\pi)^3 f(q)}. 
\end{eqnarray}}  

 The numerator in (\ref{equ:Sec7.7}) is directly related to the drag coefficient (\ref{equ:Sec8.5}) and $\sigma_{\textrm{II}}^{\textrm{tr}} (v;f)$ becomes  

{\setlength\arraycolsep{-6pt} 
\begin{eqnarray}
\label{equ:Sec7.8}
& &  \sigma_{\textrm{II}}^{\textrm{tr}} (v;f) = \frac{\hat{p}}{p} . \frac{\bsA{A}}{n \ v} = \frac{A}{p \ n \ v} = \frac{\eta_D}{n v}.
\end{eqnarray}}    
  The relaxation time related to this transport coefficient is given by

{\setlength\arraycolsep{-6pt} 
\begin{eqnarray}
\label{equ:Sec7.9}
& &  \tau = \eta_D^{-1} = \frac{1}{n \sigma_{\textrm{II}}^{\textrm{tr}} v},
\end{eqnarray}}    
  Therefore, the transport cross section $\sigma_{\textrm{II}}^{\textrm{tr}}$ allows for the natural interpretation of a cross section that permits to evaluate the typical time after which a significant longitudinal momentum change has occurred.
  From eqs.(\ref{equ:Sec7.6}) and (\ref{equ:Sec7.8}), one concludes that the transport cross section might be related to the longitudinal momentum transfer, via the drag coefficient and to the transverse momentum transfer, via $\hat{q}$. The first interpretation is related to the slowing down of the heavy quark and the second to its transverse deflection. In pQCD, one has $A \propto \alpha^2 T^2 v$ and $\hat{q} \propto \alpha^2 T^3$ \cite{Moore:2004tg}, so that

{\setlength\arraycolsep{-6pt} 
\begin{eqnarray}
\label{equ:Sec7.10}
& &  \sigma_{\textrm{I}}^{\textrm{tr}} \propto \frac{\alpha^2}{p^2} \ll \sigma_{\textrm{II}}^{\textrm{tr}} \propto \frac{\alpha^2}{p T} \ll \sigma \propto \frac{\alpha^2}{\mu^2}, 
\end{eqnarray}}    

 and one concludes that in pQCD most of the energy loss processes do not imply a large deflection. Let us insist that  $\sigma_{\textrm{I}}^{\textrm{tr}} = \sigma_{\textrm{II}}^{\textrm{tr}}$ in potential scattering theory.

\subsection{Transport cross sections in our models} 

\begin{figure}[h!] 
\begin{center}
   \begin{minipage}[c]{.495\linewidth}
     \includegraphics[height=8.5cm, width=8.9cm]{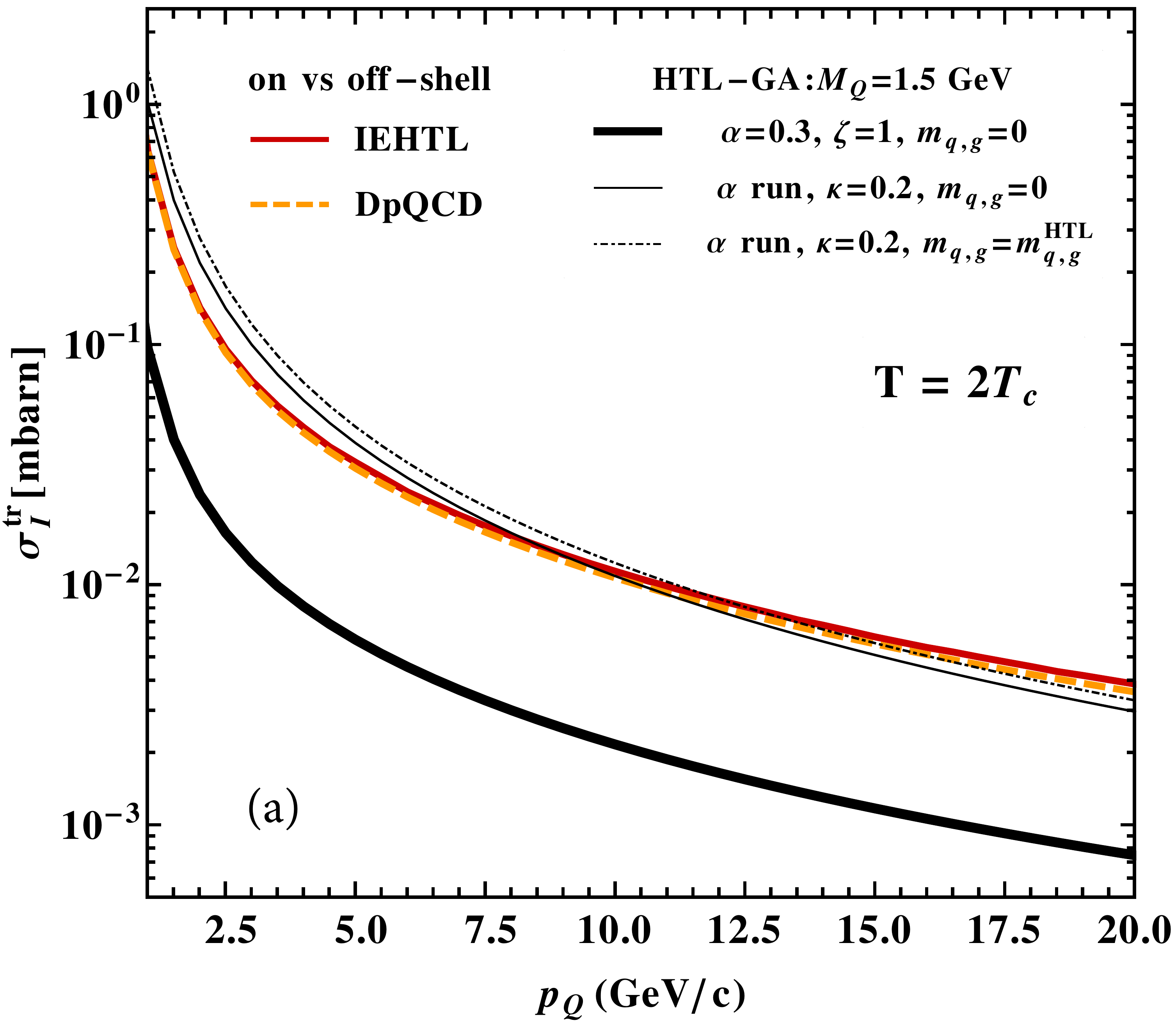}
   \end{minipage} 
   \begin{minipage}[c]{.495\linewidth}
      \includegraphics[height=8.5cm, width=8.9cm]{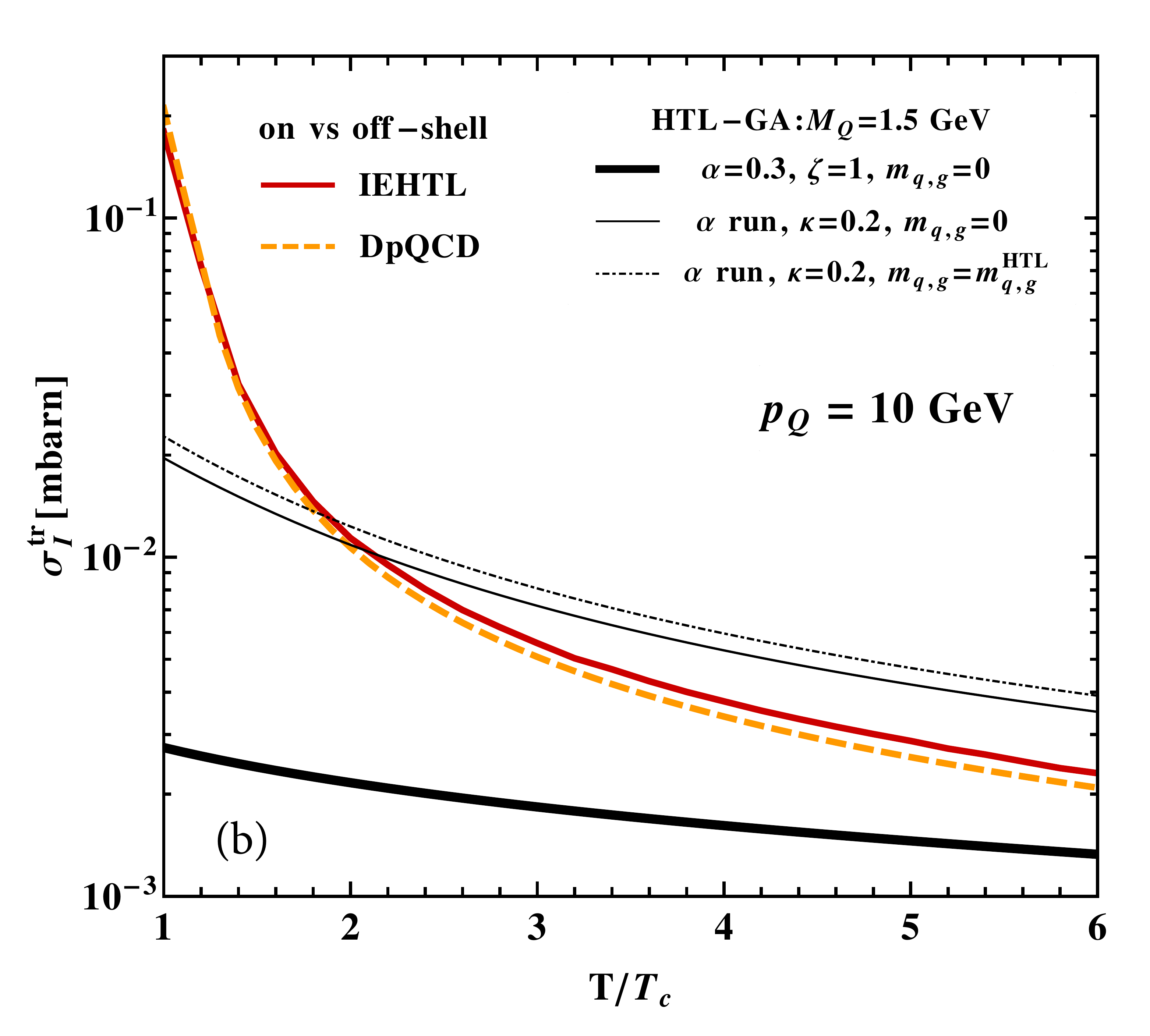}
\end{minipage}
\vspace{-0.2cm}
\caption{\emph{(Color online) The total transport cross section $\sigma_{\textrm{I}}^{\textrm{tr}}$ (\ref{equ:Sec7.6}) for the $q(g) c \rightarrow q(g)$ processes for the three different approaches as a function of the heavy quark momentum $p_Q$ for $T = 2 T_c$  (left) and as a function of the temperature $T/T_c$ ($T_c = 0.158$ GeV) for an intermediate heavy quark momentum, $p_Q = 10$ GeV (right).}}
\label{fig:TransportCrossSectionI}
\end{center}
\end{figure}

\begin{figure}[h!] 
\begin{center}
   \begin{minipage}[c]{.495\linewidth}
     \includegraphics[height=8.5cm, width=8.9cm]{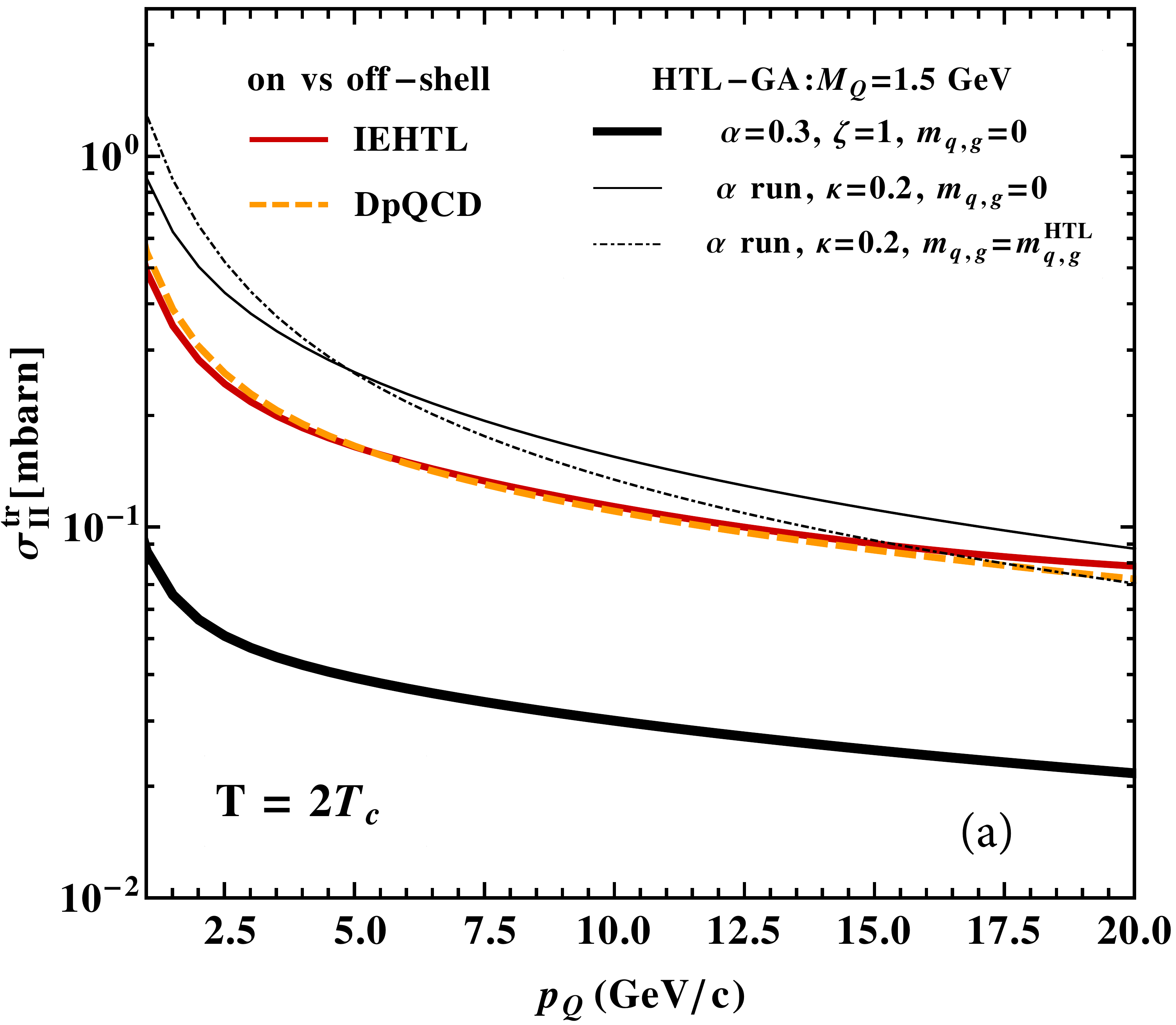}
   \end{minipage} 
   \begin{minipage}[c]{.495\linewidth}
      \includegraphics[height=8.5cm, width=8.9cm]{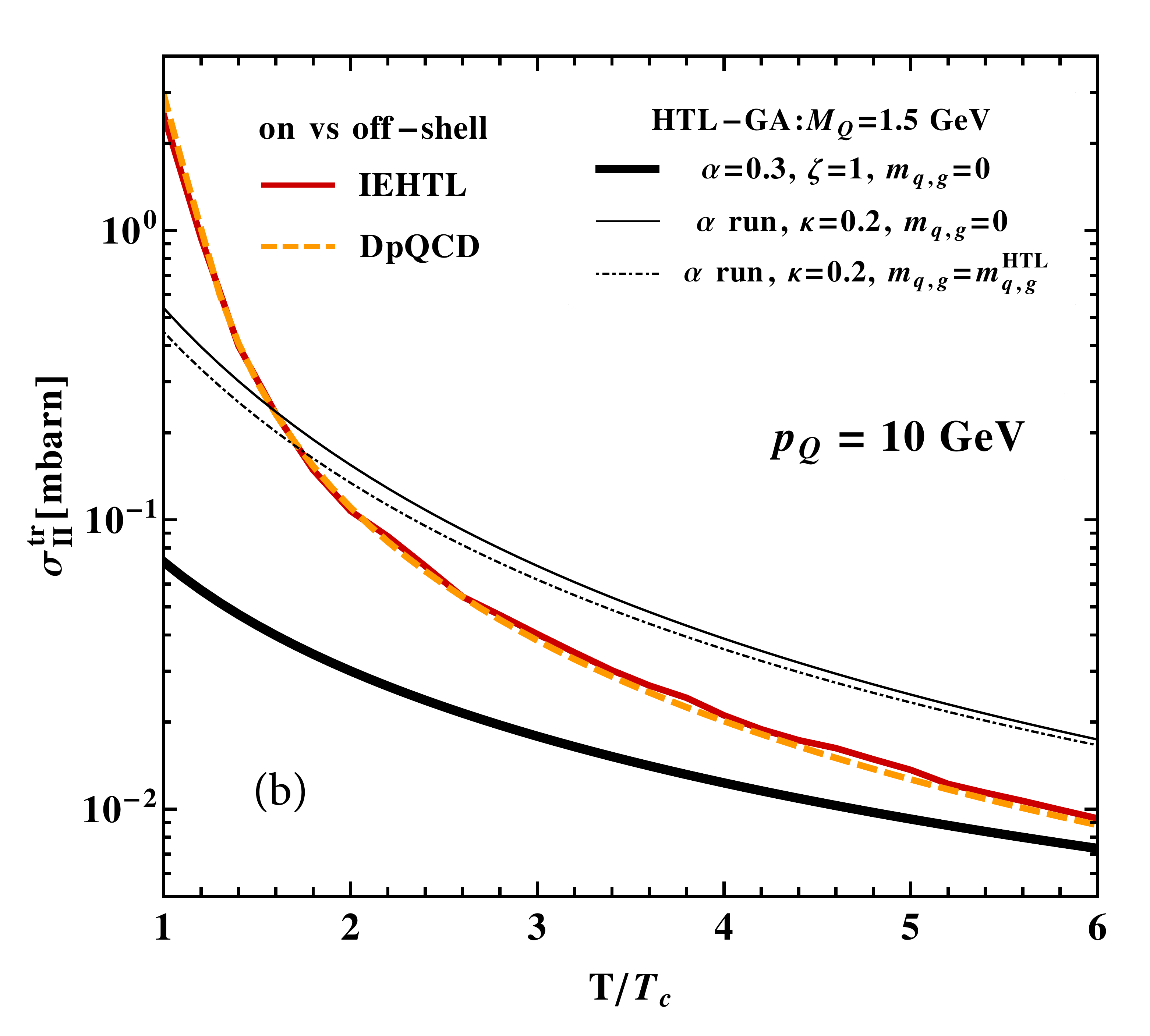}
\end{minipage}
\vspace{-0.2cm}
\caption{\emph{(Color online) The total transport cross section $\sigma_{\textrm{II}}^{\textrm{tr}}$ (\ref{equ:Sec7.8}) for the $q(g) c \rightarrow q(g)$ processes for the three different approaches as a function of the heavy quark momentum $p_Q$ for $T = 2 T_c$  (left) and as a function of the temperature $T/T_c$ ($T_c = 0.158$ GeV) for an intermediate heavy quark momentum, $p_Q = 10$ GeV (right).}}
\label{fig:TransportCrossSectionII}
\end{center}
\end{figure}

  We evaluated $\sigma_{\textrm{I}}^{\textrm{tr}}$ and $\sigma_{\textrm{II}}^{\textrm{tr}}$, as given by (\ref{equ:Sec7.6}) and (\ref{equ:Sec7.8}), for both the processes $q(g) c \rightarrow q(g) c$ for the models described in our study and report the values in figs. \ref{fig:TransportCrossSectionI}-(a) and \ref{fig:TransportCrossSectionII}-(a) as a function of the heavy quark momentum $p_Q$ for $T=2T_c$ and in figs. \ref{fig:TransportCrossSectionI}-(b) and \ref{fig:TransportCrossSectionII}-(b) as a function of the temperature for $p_Q = 10$ GeV. These figures show that the large difference seen in the total cross sections between the HTL-GA and DpQCD/IEHTL models is reduced significantly for the transport cross section. Indeed, whereas the total cross section is dominated by an infrared divergence, the transport cross section weights the square of the matrix element with $|t|$. As for the previous transport coefficients (drag and $\hat{q}$), the increase of the parton mass leads to a decrease of $\sigma_{\textrm{II}}^{\textrm{tr}}$ and to an increase of $\sigma_{\textrm{I}}^{\textrm{tr}}$. Figs. \ref{fig:TransportCrossSectionI}-(b) and \ref{fig:TransportCrossSectionII}-(b) show that for temperatures near $T_c$, the transport cross section is larger in DpQCD/IEHTL compared to the HTL-GA model. Due to the enhancement of the coupling constant, the DpQCD/IEHTL model gives the largest value of the transport cross sections for temperatures smaller than $2 T_c$. For $T>2 T_c$, the HTL-GA transport cross section is the largest one for all heavy quark momenta but the gap is smaller as compared to the total cross section (cf. Ref.\cite{Berrehrah:2013mua}). Figs. \ref{fig:TransportCrossSectionI}-(a), (b) and \ref{fig:TransportCrossSectionII}-(a), (b) show also that the off-shell effects are almost washed out for the transport cross section as in the case of the total cross section. 
  
\subsection{Microscopic interpretation of $\sigma^{\textrm{tr}}$ and applicability of Boltzmann transport} 

  In a microscopic approach based on the diffusion of a heavy quark with a particle of effective mass $m_{q/g}^{\rm eff}=\sqrt{m_{q/g}^2+T^2}$ at rest, one can show that the cross section eq.(\ref{equ:Sec7.6}) for $p \gg m = \{m_q, m_g,M_Q,T\}$ is given by

{\setlength\arraycolsep{-6pt} 
\begin{eqnarray}
\label{equ:Sec7.11}
& &  \sigma_{\textrm{I}}^{\textrm{tr}} \sim \frac{1}{p^2} \int_{-|t|_{max}}^0 |t| \frac{d \sigma}{d t} \ dt,
\end{eqnarray}}    
whereas  $\sigma_{\textrm{II}}^{\textrm{tr}}$ (\ref{equ:Sec7.8}) can be expressed  by  
{\setlength\arraycolsep{-6pt} 
\begin{eqnarray}
\label{equ:Sec7.12}
& &  \sigma_{\textrm{II}}^{\textrm{tr}} \sim \frac{1}{2 m_{q/g}^{\rm eff} \ p} \int_{-|t|_{max}}^0 |t| \frac{d \sigma}{d t} \ dt.
\end{eqnarray}}      
Both transport cross sections are related by 
{\setlength\arraycolsep{-6pt} 
\begin{eqnarray}
\label{equ:Sec7.13}
& &  \sigma_{\textrm{I}}^{\textrm{tr}} \sim \frac{2 m_{q/g}^{\rm eff}}{p} \sigma_{\textrm{II}}^{\textrm{tr}}.
\end{eqnarray}}       
  Therefore it follows: 
{\setlength\arraycolsep{-6pt} 
\begin{eqnarray}
\label{equ:Sec7.14}
& &  \sigma_{\textrm{I}}^{\textrm{tr}} \ll \sigma_{\textrm{II}}^{\textrm{tr}} \ll \sigma^{\textrm{tot}},
\end{eqnarray}}     
  and the following relation holds between $\hat{q}$ and $A$  
{\setlength\arraycolsep{-6pt} 
\begin{eqnarray}
\label{equ:Sec7.15}
& &  \hat{q} \sim 2 m_{q/g}^{\rm eff} A.
\end{eqnarray}}     
  
  From eqs.(\ref{equ:Sec7.11}) and (\ref{equ:Sec7.12}), we conclude that the differences in the transport cross sections between the HTL-GA and DpQCD/IEHTL are related to the quantity $<\sigma \ |t|> = \int \frac{d \sigma}{d t} |t| dt$. $<\sigma \ |t|>$ for the elastic $qQ$ scattering is plotted in figure \ref{fig:SigtotSigT}-(b) and compared to the total cross section $\sigma^{qQ}$ presented in fig.\ref{fig:SigtotSigT}-(a) following the models HTL-GA with $\alpha$ running and DpQCD. Whereas $\sigma^{qQ}$ is much larger in HTL-GA, the difference between HTL-GA and DpQCD for $<\sigma \, |t|>$ is reduced, which explains the profiles seen in figs. \ref{fig:TransportCrossSectionI}-(a), (b) and \ref{fig:TransportCrossSectionII}-(a), (b).
  
\begin{figure}[h!] 
\begin{center}
   \begin{minipage}[c]{.495\linewidth}
     \includegraphics[height=8.5cm, width=8.9cm]{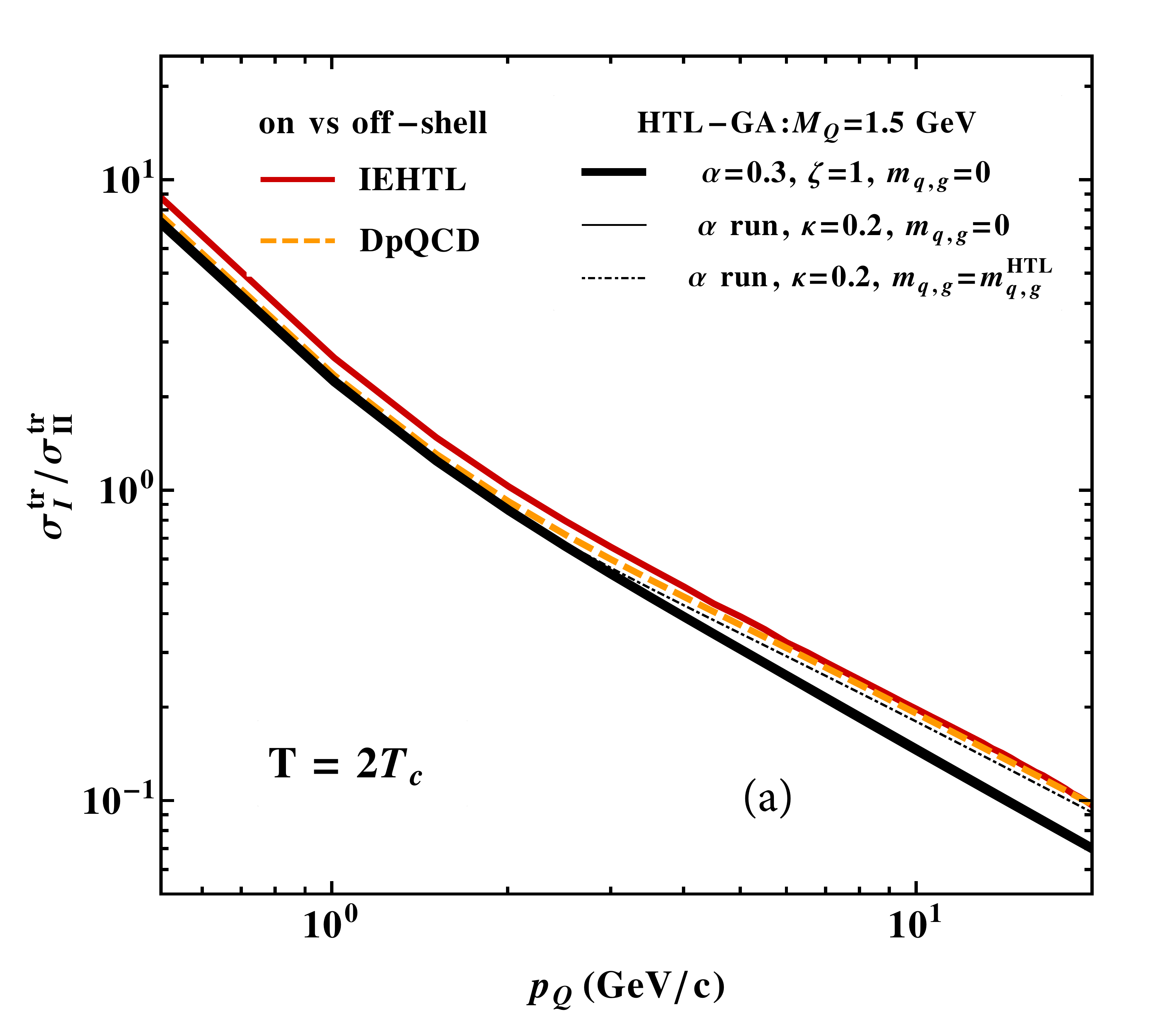}
   \end{minipage} 
   \begin{minipage}[c]{.495\linewidth}
      \includegraphics[height=8.5cm, width=8.9cm]{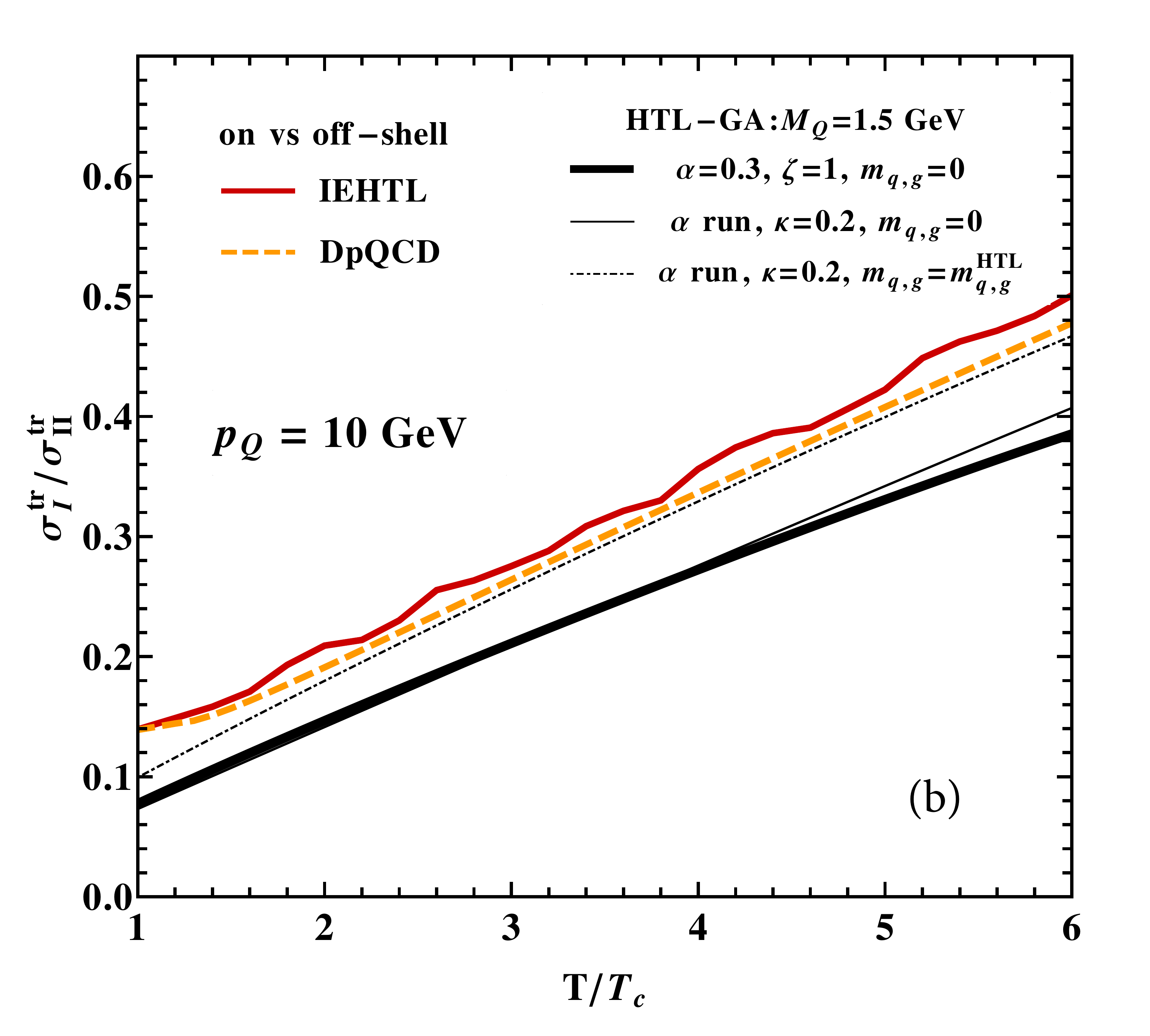}
\end{minipage}
\vspace{-0.2cm}
\caption{\emph{(Color online) The ratio $\sigma_{\textrm{I}}^{\textrm{tr}}/\sigma_{\textrm{II}}^{\textrm{tr}}$ as a function of the heavy-quark momentum $p_Q$ for $T=2 T_c$ (left) and as a function of the temperature for $p_Q = 10$ GeV (right). We note that $\sigma_{\textrm{I}}^{\textrm{tr}}/\sigma_{\textrm{II}}^{\textrm{tr}} = \sigma_{\textrm{I,qQ}}^{\textrm{tr}}/\sigma_{\textrm{II,qQ}}^{\textrm{tr}} + \sigma_{\textrm{I,gQ}}^{\textrm{tr}}/\sigma_{\textrm{II,gQ}}^{\textrm{tr}}$.}}
\label{fig:SigTrIOvSigTrII}
\end{center}
\end{figure}

  To check the relation between $\sigma_{\textrm{I}}^{\textrm{tr}}$ and $\sigma_{\textrm{II}}^{\textrm{tr}}$ (\ref{equ:Sec7.13}), we plot the ratio $\sigma_{\textrm{I}}^{\textrm{tr}}/\sigma_{\textrm{II}}^{\textrm{tr}}$ as a function of the heavy quark momentum $p_Q$ in figure \ref{fig:SigTrIOvSigTrII}-(a) and as a function of the temperature in fig.\ref{fig:SigTrIOvSigTrII}-(b). From these figures, we see clearly that $\sigma_{\textrm{I}}^{\textrm{tr}}/\sigma_{\textrm{II}}^{\textrm{tr}} \propto p_Q^{-1} = p^{-1}$ for a given temperature and $\sigma_{\textrm{I}}^{\textrm{tr}}/\sigma_{\textrm{II}}^{\textrm{tr}} \propto T$ for a given heavy quark momentum and therefore the relation (\ref{equ:Sec7.13}) is verified as $m_{q/g}^{\rm eff} \propto T$. Furthermore, we can approximate from eqs.(\ref{equ:Sec9.24d}) and (\ref{equ:Sec11.12}) the ratio $\sigma_{\textrm{I}}^{\textrm{tr}}/\sigma_{\textrm{II}}^{\textrm{tr}}$ for the interaction of a heavy quark with a massive light quark to 
{\setlength\arraycolsep{-6pt} 
\begin{eqnarray}
\label{equ:Sec7.16}
& &  \sigma_{\textrm{I,qQ}}^{\textrm{tr}}/\sigma_{\textrm{II,qQ}}^{\textrm{tr}} = \frac{\hat{q}}{2 E A} \propto \frac{m_q}{2 E}.
\end{eqnarray}}    
Since $m_q$ and $m_g$ are proportional to the temperature and in addition $m_q^{DQPM}>m_q^{HTL}$, the relation in eq.(\ref{equ:Sec7.16}) is well reproduced in fig. \ref{fig:SigTrIOvSigTrII}-(b) including the mass hierarchy between the models HTL-GA with finite $m_{q,g}$ and DpQCD/IEHTL.
  
  Present transport approaches describing the propagation of heavy flavor particles inside the QGP rely either on the phenomenological Langevin dynamics or on the Boltzmann equation. This equation is legitimate under the assumptions of diluteness and of molecular chaos \cite{Reif1965}, achieved provided the mean free path $\ell$ is much larger than the interaction range $d\simeq \mu^{-1}$. An equivalent criterion is to require that the relaxation time $\tau=\langle R\rangle^{-1}$ (\ref{equ:Sec3.6}) should be larger than the collision time $\tau_{\rm int}$ for a $Qq\to Qq$ or $Qg\to Qg$ process, with $\tau_{\rm int}\simeq (\mu \langle v_{\rm rel}\rangle)^{-1} \approx \mu^{-1}$.
   
  
  Comparing fig. \ref{fig:RelxationTime} (left) with fig. \ref{fig:MassDQPM-Mu2} (right), one can check that this criteria is indeed satisfied for all values of $T$ in the case of the DpQCD/IEHTL models, while it is never satisfied for the HTL-GA model with $\alpha$ running. 
  Although this could cast doubt upon the applicability of the Boltzmann transport in this case, one has to remind that most of the collisions in the GA model do not lead to large momentum transfer (as seen from fig. \ref{fig:cDragOvRateDifferentModel}) and are thus inefficient for the deceleration of  heavy quarks. One therefore obtains a more suitable criteria by considering, instead of the relaxation time (\ref{equ:Sec3.6}), the one associated with the transport cross section $\sigma_{II}^{\rm tr}$, as defined in Eq. (\ref{equ:Sec7.9}). For $p\approx 0$, one obtains $\eta_D/\mu\approx 0.3$ (see Fig. \ref{fig:cDragTheoreticalApproaches}) in the HTL-GA model with $\alpha$ running, and this ratio even decreases $\propto p^{-1}$ at finite $p$. This implies that the time interval between 2 subsequent collisions relevant for the momentum transfers is indeed larger than the interaction time, which legitimates the use of the Boltzmann equation for this model as well.

\section{Summary}
\label{Summary}

We have studied the transport coefficients for three different effective theories (or models) which are based on cross sections calculated in the Born approximation.
The latter cross sections have perturbative or non-perturbative elements of QCD:
\begin{itemize}\setlength{\itemsep}{-4mm}
\item[$a)$] the HTL-GA model which employs a running coupling and an effective infrared regulator whose value is adjusted to hard thermal loop
calculations \cite{PhysRevC.78.014904}. In this model quarks and gluons have either no mass or a mass which depends linearly on the temperature.\\
\item[$b)$] the DpQCD approach in which on-shell quarks and gluons have a thermal mass which is adjusted to reproduce lattice gauge calculation in
equilibrium at finite temperature $T$. This model employs a running coupling (in $T/T_c$) and is in accord with lattice QCD correlators
for the shear and bulk viscosities for finite $T$ \cite{Marty:2013ita}.\\
\item[$c)$] the IEHTL which takes fully into account the off-shell character of the quarks and gluons and contains the DpQCD masses as pole masses.
\end{itemize}
For each of these models we have calculated the interaction rates, the $c$-quark drag and diffusion coefficient, the longitudinal and transverse
 momentum broadening and the transport cross section.

 Our study has demonstrated that even if the influence of the finite width of the quasi-particles on heavy quark scattering is negligible
 for the off-shell heavy quark $Q$ scattering cross sections \cite{Berrehrah:2013mua}, a noticeable effect is seen in the off-shell transport
 coefficients (IEHTL model) compared to their on-shell counterparts (DpQCD model). This is essentially due to the reduced kinematical threshold
 in the off-shell cross sections and mass effects. The finite width of the IEHTL model lowers the interaction rate and therefore as well the drag
 and diffusion coefficients. Also $\hat q$ becomes smaller but all these differences are on the level of 10-20\%.  For temperatures
 ($T> 3 T_c$) the drift coefficient in the HTL-GA model with a fixed coupling $\alpha$ and zero mass partons and that of DpQCD are not very different.
 This is also true for the energy loss $dE/dx$. For higher temperatures the HTL-GA model shows a larger stopping because the increasing
 mass in the DpQCD approach (with temperature) lowers the interaction rate substantially without leading to a much higher stopping in an individual
 collision of a heavy quark with the plasma particles.


  Our survey has provided a comprehensive comparison between perturbative and non-perturbative QCD based models on the determination of the heavy quark
  transport coefficients. Furthermore, the phenomenological consequences of using HTL-GA and DpQCD/IEHTL cross sections (as determined
  in Ref.\cite{Berrehrah:2013mua}) on the heavy quark collective transport have been presented in detail. A significant differences between HTL-GA and
  DpQCD/IEHTL models has been pointed out for the different heavy-quark transport characteristics.
  However, the comparisons have to be carried out with some notes of caution. Indeed, even if the HTL-GA for a running coupling gives the largest heavy quark
  drag coefficient and interaction rates compared to the other models, the heavy quark longitudinal and transverse momentum fluctuations are smaller.

  The final goal of our investigation is to figure out the main differences between two models based on perturbative or non-perturtbative effective QCD
  models on one side and two concepts for the heavy quark diffusion in a medium with many massless light quarks and gluons or a few massive/off-shell $q$ and $g$,
  respectively. Our finding is that explicit transport calculations in comparison to experimental data are needed to figure out the appropriate
  scenario as a function of collision centrality and bombarding
  energy, since the microscopic ingredients (fixed or running couplings in momentum or temperature) and the QGP background description
  (massless or massive degrees of freedom)  are strongly coupled and specific to each model.

  The heavy quark scattering cross sections obtained in Ref.\cite{Berrehrah:2013mua} and the heavy quark transport properties illustrated in this study
  will form the basis of a consistent calculation of the heavy quark propagation and dynamics in heavy-ion collisions at SPS, RHIC and LHC energies by
  implementing the partonic processes into the transport approach of the PHSD collaboration to study heavy-quark dynamics at SPS, RHIC and LHC energies
  as a function of centrality. These future studies will also   address the momentum correlations of $c \bar{c}$ pairs (or $D
  \bar{D}$ pairs) as a function of rapidity and transverse   momentum.

  \vspace{-3mm}
\section*{Acknowledgment}
\vskip -3mm

 H. Berrehrah thanks R. Marty for fruitful discussions and his continuous interest. He also acknowledges the ``HIC for FAIR'' framework
 of the ``LOEWE'' program and the 'Together project of the region Pays de la Loire' for support of this work.
 The computational resources were provided by the LOEWE-CSC.


  \section*{Appendix}\appendix
\label{appendix}
\vskip -3mm

\section{Elastic interaction rate for off-shell partons}
\label{AppendixA}

The covariant off-shell rate $\mathcal{R}^{\textrm{off}} = \frac{d N_{\textrm{coll}}^{\textrm{off}}}{d \tau}$ is given by

{\setlength\arraycolsep{-6pt}
\begin{eqnarray}
\label{equ:SecA.1}
& & \mathcal{R}^{\textrm{off}} \ (T) = \frac{1}{(2\pi)^{9}} \int d \omega \frac{E_p}{M_Q} \ \rho_p (p) \Theta (\omega_p) \ \int d^4 q \ f(\bsq{q}) \rho_q (q) \Theta (\omega_q) \ r^{\textrm{off}} (s, T),
\nonumber\\
& & {} r^{\textrm{off}} (s, T) := \int d^4 q' \int d^4 p'   \rho_{p'} (p') \rho_{q'} (q') \Theta (\omega_{p'}) \Theta (\omega_{q'}) \delta^{(4)} (P_{in} - P_{fin}) \ \frac{1}{g_p g_Q} \sum_{k,l} |\mathcal{M}_{2,2}^{\textrm{off}} (p, q; i, j |p', q'; k, l)|^2.
\end{eqnarray}}
 We proceed following the same procedure as for the on-shell case. Nevertheless, the derivation of the explicit expression for
 $\mathcal{R}^{\textrm{off}} \ (T)$ and $r^{\textrm{off}} (s, T)$ depends on the choice of the parton spectral function which we will consider.
 We will give the expressions for both, the Lorentzian-$\omega$ form and the Breit-Wigner-$m$ for of the parton spectral functions.

\begin{itemize}
\item For the Lorentzian-$\omega$ form of the spectral function:

  In a complete off-shell picture $\mathcal{R}^{\textrm{off}} (T)$ depends only on the medium temperature since we convolute with the mass and momentum
  spectral functions. $r^{\textrm{off}}$ is explicitly given by

{\setlength\arraycolsep{0pt}
\begin{eqnarray}
\label{equ:SecA.2}
\hspace{-1cm} r^{\textrm{off}} (s, T) & & \ := \int E_{q'} d \omega_{q'}  \rho_{q'}^{L} (\omega_{q'}) \int E_{p'} d \omega_{p'}  \rho_{p'}^{L} (\omega_{p'}) \int \frac{d^3 q'}{E_{q'}} \int \frac{d^3 p'}{E_{p'}} \delta^{(4)} (P_{in} - P_{fin}) \ \frac{1}{g_p g_Q} \sum_{k,l} |\mathcal{M}_{2,2}^{\textrm{off}} (p, q; i, j |p', q'; k, l)|^2 (s,t).
\nonumber\\
& & {} \ = \int E_{q'} d \omega_{q'}  \rho_{q'}^{L} (\omega_{q'}) \int  E_{p'} d \omega_{p'}  \rho_{p'}^{L} (\omega_{p'}) \frac{\pi}{\sqrt{s} p_{cm} (s)}  \int_{t_{min}}^{t_{max}} \frac{1}{g_p g_Q} \sum_{k,l} |\mathcal{M}_{2,2}^{\textrm{off}} (p, q; i, j |p', q'; k, l)|^2 (s,t) \ d t,
\end{eqnarray}}
 where $t_{min}$ and $t_{max}$ are given by:

{\setlength\arraycolsep{0pt}
\begin{eqnarray}
\label{equ:SecA.3}
& & t_{min}^{max} = - \frac{s}{2} \Biggl(1 - (\beta_1 + \beta_2 + \beta_3 + \beta_4) + (\beta_1 -  \beta_2)(\beta_3 - \beta_4) \pm \sqrt{(1 - \beta_1 - \beta_2)^2 - 4 \beta_1 \beta_2} \sqrt{(1 - \beta_3 - \beta_4)^2 - 4 \beta_3 \beta_4} \Biggr)
\nonumber\\
& & {} \textrm{with:} \beta_1 = (m_q^i)^2/s, \beta_2 = (M_Q^i)^2/s, \beta_3 = (m_q^f)^2/s, \beta_4 = (M_Q^f)^2/s.
\end{eqnarray}}
  The remaining integral in (\ref{equ:SecA.1}) is evaluated in the rest frame of the heavy quark and can be calculated in a similar way as for the
  on-shell case with a special care on the convolution with the parton spectral functions.

\item For the Breit-Wigner-$m$ form of the spectral function:

 Considering the Breit-Wigner-$m$ for the spectral functions, valid in the non-relativistic limit, the expression for
 $\displaystyle Tr_{(i)}^{\textrm{on}} = \sum_i \int \frac{d^3 p_i}{(2 \pi)^3} \ \frac{1}{2 E_{p_i}}$ becomes in the off-shell
 case $\displaystyle Tr_{(i)}^{\textrm{off}} = \sum_i \int \frac{d m_i}{(2 \pi)} \frac{m_i}{\sqrt{m_i^2 + \bs{p}_i^2}} \ \rho_i^{BW} (m_i)
 \ \int \frac{d^3 p}{(2 \pi)^3}$. Then $\mathcal{R}^{\textrm{off}} \ (T)$ and $r^{\textrm{off}} (s, T)$ are simply obtained from the
 on-shell case where we should convolute with the different parton spectral functions in mass. Therefore, for a Breit-Wigner-$m$ parton spectral
 function the expressions in (\ref{equ:SecA.1}) become

{\setlength\arraycolsep{-6pt}
\begin{eqnarray}
\label{equ:SecA.4}
& & \mathcal{R}^{\textrm{off}} \ (T) = \frac{1}{(2\pi)^{9}} \int d m_p \  \rho_p^{BW} (m_p) \ \int m_q d m_q  \rho_q^{BW} (m_q) \ \int \frac{d^3 q}{E_q}f(\bsq{q}) \ r^{\textrm{off}} (s,T),
\\
& & {} r^{\textrm{off}} (s, T) := \int m_{q'} d m_{q'}  \rho_{q'}^{BW} (m_{q'}) \int m_{p'} d m_{p'}  \rho_{p'}^{BW} (m_{p'}) \int \frac{d^3 q'}{E_{q'}} \int \frac{d^3 p'}{E_{p'}} \delta^{(4)} (P_{in} - P_{fin}) \ \frac{1}{g_p g_Q} \sum_{k,l} |\mathcal{M}_{2,2}^{\textrm{off}} (p, q; i, j |p', q'; k, l)|^2 (s,t).
\nonumber\\
& & {} \hspace*{1.3cm} = \int m_{q'} d m_{q'}  \rho_{q'}^{BW} (m_{q'}) \int m_{p'} d m_{p'}  \rho_{p'}^{BW} (m_{p'}) \frac{\pi}{\sqrt{s} p_{cm} (s)}  \int_{t_{min}}^{t_{max}}  \frac{1}{g_p g_Q} \sum_{k,l} |\mathcal{M}_{2,2}^{\textrm{off}} (p, q; i, j |p', q'; k, l)|^2 (s,t) \ d t.
\nonumber
\end{eqnarray}}
By direct analogy to the on-shell case, $R^{\textrm{off}} (\bs{p})$ becomes

{\setlength\arraycolsep{0pt}
\begin{eqnarray}
\label{equ:SecA.5}
R^{\textrm{off}} (\bs{p}) = \Pi\displaystyle_{i \in{p,q,p',q'}} \ \int m_i d m_i \rho_i^{BW} (m_i) \frac{m_p}{8 (2 \pi)^3 E_p} \int \frac{q^3 m_0^{\textrm{off}} (s) f_0 (\frac{p}{M_Q}; q)}{s \ E_q} \ d q,
\end{eqnarray}}
 where

{\setlength\arraycolsep{0pt}
\begin{eqnarray}
\label{equ:SecA.6}
& & m_0^{\textrm{off}} (s) = \int_{-1}^{+1} \!\! d \cos \theta \sum |\mathcal{M}_{2,2}^{\textrm{off}} (p, q; i, j |p', q'; k, l)|^2 (s,t) = \frac{1}{2 p_{cm}^2 (s)} \int_{t_{min}}^{t_{max}} \sum |\mathcal{M}_{2,2}^{\textrm{off}} (p, q; i, j |p', q'; k, l)|^2 (s,t) \ d t,
\nonumber\\
& & {} f_0 \left(\frac{p}{M_Q}; q \right) = \frac{1}{2} \int \frac{d \cos \theta_r}{\exp (u^0 E_q - q \ u \cos \theta_r) \pm 1},
\end{eqnarray}}
  and  $\sum |\mathcal{M}_{2,2}^{\textrm{off}} (p, q; i, j |p', q'; k, l)|^2 (s,t)$ the off-shell transition amplitude defined in IEHTL approach,
  Ref.\cite{Berrehrah:2013mua}.

\end{itemize}

\section{Drag coefficient for off-shell partons} 
\label{AppendixB}

Similarly to the transition from the on-shell to the off-shell limits for the study of the interaction rates, one deduces the corresponding
expression for $A_i^{\textrm{off}}$, the drag coefficient for the off-shell case,

{\setlength\arraycolsep{-6pt}
\begin{eqnarray}
\label{equ:SecB.1}
A_i^{\textrm{off}} (T) = \frac{1}{(2\pi)^{9}} \int d \omega \int d^4 q f(\bsq{q}) \int d^4 q' & & \int d^4 p' \  \rho_p (p) \rho_q (q) \rho_{p'} (p') \rho_{q'} (q') \Theta (\omega_p) \Theta (\omega_q) \Theta (\omega_{p'}) \Theta (\omega_{q'})
\nonumber\\
& & {} \times (p - p')_i \delta^{(4)} (P_{in} - P_{fin}) \ \frac{1}{g_p g_Q} \sum_{k,l} |\mathcal{M}_{2,2}^{\textrm{off}} (p, q; i, j |p', q'; k, l)|^2 (s,t),
\end{eqnarray}}

where, by writing $\displaystyle \frac{1}{2 E_p} = \int_0^{\infty} \frac{d \omega}{2 \pi} 2 \pi \delta (\omega^2 - E_p^2)$, the transition from
the on-shell case to the off-shell case is characterized by $\displaystyle \frac{1}{2 E_p} \rightarrow \int_0^{\infty}
\frac{d \omega}{2 \pi} \rho_p (\omega, \bsp{p}) = \int \frac{d \omega}{2 \pi} \rho_p (\omega, \bsp{p}) \theta (\omega)$.
Therefore, $\mathcal{A}^{\textrm{off}} = \frac{d p_i^{\textrm{off}}}{d \tau}$ becomes

{\setlength\arraycolsep{-6pt}
\begin{eqnarray}
\label{equ:SecB.2}
& & \mathcal{A}_{\textrm{off}}^{\mu} \ (T) = - \frac{1}{(2\pi)^{9}} \int d \omega \frac{E_p}{M_Q} \ \rho_p (p) \Theta (\omega_p) \ \int d^4 q \ f(\bsq{q}) \rho_q (q) \Theta (\omega_q) \ a_{\textrm{off}}^{\mu} (\bsp{p}),
\\
& & {} a_{\textrm{off}}^{\mu} (\bsp{p}) := \int d^4 q' \int d^4 p' \ \rho_{p'} (p') \ \rho_{q'} (q') \Theta (\omega_{p'}) \Theta (\omega_{q'}) (q - q')^{\mu} \delta^{(4)} (P_{in} - P_{fin}) \ \frac{1}{g_p g_Q} \sum_{k,l} |\mathcal{M}_{2,2}^{\textrm{off}} (p, q; i, j |p', q'; k, l)|^2 (s,t).
\nonumber
\end{eqnarray}}
 We then proceed following the same procedure as for the on-shell case. Nevertheless, the derivation of the explicit expression for
 $\mathcal{A}_{\textrm{off}}^{\mu} \ (T)$ and $a_{\textrm{off}}^{\mu} (\bsp{p})$ depends on the choice of the parton spectral function.
 If the Lorentzian-$\omega$ form of the parton spectral function is considered then we should derive a new expression
 for $\mathcal{A}_{\textrm{off}}^{\mu} \ (T)$ and $a_{\textrm{off}}^{\mu} (\bsp{p})$ since the running energy involves a running of the different parton momenta.
 On the other hand, if we consider the Breit-Wigner-$m$ form of the spectral functions valid in the non-relativistic limit,
 then $\mathcal{A}_{\textrm{off}}^{\mu} \ (T)$ and $a_{\textrm{off}}^{\mu} (\bsp{p})$ are simply obtained from the on-shell case by convoluting
 this expression with the different parton spectral functions and the transition from the on- to off-shell case can be factorized.
 Therefore, for a Breit-Wigner-$m$ parton spectral function the expressions (\ref{equ:SecB.2}) become

{\setlength\arraycolsep{-6pt}
\begin{eqnarray}
\label{equ:SecB.3}
& & \mathcal{A}_{\textrm{off}}^{\mu} \ (T) = - \frac{1}{(2\pi)^{9}} \int d m_p \ \rho_p^{BW} (m_p) \ \int m_q d m_q  \rho_q^{BW} (m_q) \ \int \frac{d^3 q}{E_q}f(\bsq{q}) \ a_{\textrm{off}}^{\mu} (\bsp{p}),
\\
& & {} a_{\textrm{off}}^{\mu} (\bsp{p}) := \int m_{q'} d m_{q'} \rho_{q'}^{BW} (m_{q'}) \int m_{p'} d m_{p'} \rho_{p'}^{BW} (m_{p'}) \int \frac{d^3 q'}{E_{q'}} \int \frac{d^3 p'}{E_{p'}} (q - q')^{\mu} \delta^{(4)} (P_{in} - P_{fin}) \ \frac{1}{g_p g_Q} \sum_{k,l} |\mathcal{M}_{2,2}^{\textrm{off}} (p, q; i, j |p', q'; k, l)|^2 (s,t).
\nonumber
\end{eqnarray}}
In the cm frame, we find for $a_{cm}^{\textrm{off}}$

{\setlength\arraycolsep{-6pt}
\begin{eqnarray}
\label{equ:SecB.4}
& & \bsa{a}_{cm}^{\textrm{off}} = \int m_{q'} d m_{q'} \rho_{q'}^{BW} (m_{q'}) \int m_{p'} d m_{p'} \rho_{p'}^{BW} (m_{p'}) \frac{4 \pi p_{cm}^2 (s)}{\sqrt{s}} \  m_1^{\textrm{off}} (s) \times \hat{\bsq{q}}_{cm},
\nonumber\\
& & {} \textrm{with:} \hspace{0.3cm}  m_1^{\textrm{off}} (s) := \frac{1}{4 p_Q^i p_Q^f} \int_{t_{min}}^{t_{max}} \ \sum |\mathcal{M}_{2,2}^{\textrm{off}} (p, q; i, j |p', q'; k, l)|^2 (s,t) \ \Biggl[1 - \frac{t - (M_Q^i)^2 + (M_Q^f)^2 - 2 E_Q^i E_Q^f}{2 p_Q^i p_Q^f} \Biggr] d t,
\end{eqnarray}}
  where $p_Q^i$ ($M_Q^i$) is the initial heavy quark momentum (mass) and $p_Q^i$ ($M_Q^i$) is the final heavy quark momentum (mass). In the heavy quark
  rest frame $a_{rest}^{\textrm{off}}$ is given by

{\setlength\arraycolsep{-6pt}
\begin{eqnarray}
\label{equ:SecB.5}
a_{rest}^{\textrm{off}} = \int m_{q'} d m_{q'} \rho_{q'}^{BW} (m_{q'}) \int m_{p'} d m_{p'} \rho_{p'}^{BW} (m_{p'}) \frac{4 \pi p_{cm}^2 (s)}{s} \  m_1^{\textrm{off}} (s) \, \begin{pmatrix}
q_{rest} \\
(E_{q_{rest}} + m_Q) \ \hat{\bsq{q}}_{cm}
\end{pmatrix},
\end{eqnarray}}
which leads to the following expressions of $A_{rest}^{0,\textrm{off}}$ and $A_{rest}^{\nu,\textrm{off}}$

{\setlength\arraycolsep{-6pt}
\begin{eqnarray}
\label{equ:SecB.6}
& & A_{rest}^{0,\textrm{off}} = \frac{4}{(2 \pi)^7} \ \Pi_{i \in{p,q,p',q'}} \int m_p m_i d m_i \ \rho_i^{WB} (m_i) \ \int_0^{\infty} \frac{q^5 m_1 (s) f_0 (q)}{s^2  E_q} d q,
\\
& & {} A_{rest}^{\nu,\textrm{off}} = \frac{4}{(2 \pi)^7} \ \Pi_{i \in{p,q,p',q'}} \int m_p m_i d m_i \ \rho_i^{WB} (m_i) \ \int_0^{\infty} \frac{q^4 (M_Q + E_q) m_1 (s) f_1 (q)}{s^2 E_q} d q.
\nonumber
\end{eqnarray}}

   Finally the drag coefficient for the off-shell partons in the heat bath rest frame is given by

{\setlength\arraycolsep{-6pt}
\begin{eqnarray}
\label{equ:SecB.7}
& & \bsA{A}_{\textrm{heat bath}}^{\textrm{off}} = \Biggl(A_{rest}^{\nu,\textrm{off}} - \tilde{A}_{rest}^{0,\textrm{off}} \Biggr) \hat{\bsp{p}},
\end{eqnarray}}
 where $\tilde{A}_{rest}^{0,\textrm{off}}$ and $\tilde{A}_{rest}^{\nu,\textrm{off}}$ are defined by

{\setlength\arraycolsep{0pt}
\begin{eqnarray}
\label{equ:SecB.8}
& & \tilde{A}_{rest}^{0,\textrm{off}} = \frac{4}{(2 \pi)^7} \ \Pi_{i \in{p,q,p',q'}} \int \frac{p \ m_p}{E_p} m_i d m_i \ \rho_p^{WB} (m_i) \ \int_0^{\infty} \frac{q^5 m_1 (s) f_0 (q)}{s^2  E_q} d q,
\\
& & {} \tilde{A}_{rest}^{\nu,\textrm{off}} = \frac{4}{(2 \pi)^7} \ \Pi_{i \in{p,q,p',q'}} \int \frac{p \ m_p}{E_p} m_i d m_i \ \rho_p^{WB} (m_i) \ \int_0^{\infty} \frac{q^4 (M_Q + E_q) m_1 (s) f_1 (q)}{s^2 E_q} d q.
\nonumber
\end{eqnarray}}

\section{Dynamical collisional energy loss for off-shell partons}
\label{AppendixC}

The collisional energy loss for on-shell partons (\ref{equ:Sec1.1}) can be easily extended to the off-shell case. The covariant
$\frac{d E^{\textrm{off}}}{d \tau}$ is given by

{\setlength\arraycolsep{-6pt}
\begin{eqnarray}
\label{equ:SecC.1}
\frac{d E^{\textrm{off}}}{d \tau} (T) = \frac{1}{(2\pi)^{9}} \int d \omega \frac{E_p}{M_Q} \int d^4 q f(\bsq{q}) \int d^4 q' & & \int d^4 p' \  \rho_p (p) \rho_q (q) \rho_{p'} (p') \rho_{q'} (q') \Theta (\omega_p) \Theta (\omega_q) \Theta (\omega_{p'}) \Theta (\omega_{q'})
\nonumber\\
& & {} \times (E - E') \delta^{(4)} (P_{in} - P_{fin}) \ \frac{1}{g_p g_Q} \sum_{k,l} |\mathcal{M}_{2,2}^{\textrm{off}} (p, q; i, j |p', q'; k, l)|^2 (s,t),
\end{eqnarray}}
 therefore, $\frac{d E^{\textrm{off}}}{d \tau}$ becomes

{\setlength\arraycolsep{-6pt}
\begin{eqnarray}
\label{equ:SecC.2}
& & \frac{d E^{\textrm{off}}}{d \tau} \ (T) = - \frac{1}{(2\pi)^{9}} \int d \omega \frac{E_p}{M_Q} \ \rho_p (p) \Theta (\omega_p) \ \int d^4 q \ f(\bsq{q}) \rho_q (q) \Theta (\omega_q) \ e^{\textrm{off}} (\bsp{p}),
\\
& & {} e^{\textrm{off}} (\bsp{p}) := \int d^4 q' \int d^4 p' \ \rho_{p'} (p') \ \rho_{q'} (q') \Theta (\omega_{p'}) \Theta (\omega_{q'}) (E - E') \delta^{(4)} (P_{in} - P_{fin}) \ \frac{1}{g_p g_Q} \sum_{k,l} |\mathcal{M}_{2,2}^{\textrm{off}} (p, q; i, j |p', q'; k, l)|^2 (s,t).
\nonumber
\end{eqnarray}}
As for the drag coefficient the explicit expressions of $\frac{d E^{\textrm{off}}}{d \tau}$ and $e^{\textrm{off}} (\bsp{p})$ depend on the choice of the
parton spectral function. For this study we consider the Breit-Wigner-$m$ form of the spectral functions, then
$\frac{d E^{\textrm{off}}}{d \tau}$ and $e^{\textrm{off}} (\bsp{p})$ are simply obtained from the on-shell case by convoluting these expressions
with the different parton spectral functions and the transition from the on- to off-shell case can be factorized again. Using Breit-Wigner-$m$ spectral
functions (\ref{equ:SecC.2}) becomes

{\setlength\arraycolsep{-6pt}
\begin{eqnarray}
\label{equ:SecC.3}
& & \frac{d E^{\textrm{off}}}{d \tau} \ (T) = - \frac{1}{(2\pi)^{9}} \int d m_p \ \rho_p^{BW} (m_p) \ \int m_q d m_q  \rho_q^{BW} (m_q) \ \int \frac{d^3 q}{E_q}f(\bsq{q}) \ e^{\textrm{off}} (\bsp{p}),
\\
& & {} e^{\textrm{off}} (\bsp{p}) := \int m_{q'} d m_{q'} \rho_{q'}^{BW} (m_{q'}) \int m_{p'} d m_{p'} \rho_{p'}^{BW} (m_{p'}) \int \frac{d^3 q'}{E_{q'}} \int \frac{d^3 p'}{E_{p'}} (E - E') \delta^{(4)} (P_{in} - P_{fin}) \ \frac{1}{g_p g_Q} \sum_{k,l}  |\mathcal{M}_{2,2}^{\textrm{off}} (p, q; i, j |p', q'; k, l)|^2 (s,t).
\nonumber
\end{eqnarray}}
 Proceeding as for the drag coefficient for the off-shell partons, $\frac{d E^{\textrm{off}}}{d t} \ (T)$ is given by

{\setlength\arraycolsep{-6pt}
\begin{eqnarray}
\label{equ:SecC.4}
& & \frac{d E^{\textrm{off}}}{d t} \ (T) = A_{rest}^{0,\textrm{off}} - \tilde{A}_{rest}^{\nu,\textrm{off}},
\end{eqnarray}}
 where $A_{rest}^{0,\textrm{off}}$ and $\tilde{A}_{rest}^{\nu,\textrm{off}}$ are defined in (\ref{equ:SecB.6}) and (\ref{equ:SecB.8}).

\section{$\hat{q}$ \ for on-shell partons}
\label{AppendixD}

 For a massive on-shell parton, $B_{T}$, and then $\hat{q}$ is obtained following the same procedure as for the evaluation of the drag coefficient.
 Using (\ref{equ:Sec11.1}) and (\ref{equ:Sec11.3}), one finds that $b^{i j}$, written in the c.m frame by

{\setlength\arraycolsep{-6pt}
\begin{eqnarray}
\label{equ:SecD.1}
b_{cm}^{i j} (\bsp{p}) & & \ = \frac{p_{cm}}{\sqrt{s}} \int \frac{d^3 q'}{2E_{q'}} \int \frac{d^3 p'}{2E_{p'}} \ \frac{(q - q')^{i}(q - q')^{j}}{2} \ \delta^{(4)} (P_{in} - P_{fin}) \ \frac{1}{g_p g_Q} \sum_{k,l}  |\mathcal{M}_{2,2}^{\textrm{on}} (p, q; i, j |p', q'; k, l)|^2 (s,t)
\end{eqnarray}}
can be decomposed in its transverse $b_{cm,T}$ and longitudinal $b_{cm,L}$ components

{\setlength\arraycolsep{-6pt}
\begin{eqnarray}
\label{equ:SecD.2}
b_{cm}^{i j} (\bsp{p}) & & \ = b_{cm,T} (I - |\hat{q}_{cm}><\hat{q}_{cm}|) + b_{cm,L} |\hat{q}_{cm}><\hat{q}_{cm}|,
\end{eqnarray}}
 with

{\setlength\arraycolsep{-6pt}
\begin{eqnarray}
\label{equ:SecD.3}
b_{T,cm} (\bsp{p}) & & \ = \frac{p_{cm}}{2\sqrt{s}} \int \frac{d^3 q'}{2E_{q'}} \int \frac{d^3 p'}{2E_{p'}} \ \left(\frac{\parallel \bsq{q} - \bsq{q}' \parallel^2}{2} - \frac{((\bsq{q} - \bsq{q}').\hat{q})^2}{2} \right) \delta^{(4)} (P_{in} - P_{fin}) \ \frac{1}{g_p g_Q} \sum_{k,l}  |\mathcal{M}_{2,2}^{\textrm{on}} (p, q; i, j |p', q'; k, l)|^2 (s,t)
\nonumber\\
& & {} = \frac{p_{cm}}{2\sqrt{s}} \int d \Omega_{rel} \ \left(\frac{\parallel \bsq{q} - \bsq{q}' \parallel^2}{2} - \frac{p_{cm}^2}{2} (1 - \cos \theta)^2\right) \ \frac{1}{g_p g_Q} \sum_{k,l}  |\mathcal{M}_{2,2}^{\textrm{on}} (p, q; i, j |p', q'; k, l)|^2 (s,t),
\nonumber\\
& & {} = \frac{2 \pi p_{cm}^3}{\sqrt{s}} \biggl( m_1 (s) - m_2 (s) \biggr),
\end{eqnarray}}
  and

{\setlength\arraycolsep{-6pt}
\begin{eqnarray}
\label{equ:SecD.4}
b_{L,cm} (\bsp{p}) & & \ = \frac{p_{cm}}{\sqrt{s}} \int \frac{d^3 q'}{2E_{q'}} \int \frac{d^3 p'}{2E_{p'}} \ \left(\frac{((\bsq{q} - \bsq{q}').\hat{q})^2}{2} \right) \delta^{(4)} (P_{in} - P_{fin}) \ \frac{1}{g_p g_Q} \sum_{k,l}  |\mathcal{M}_{2,2}^{\textrm{on}} (p, q; i, j |p', q'; k, l)|^2 (s,t)
\nonumber\\
& & {} = \frac{p_{cm}}{\sqrt{s}} \int d \Omega_{rel} \ \left(\frac{p_{cm}^2}{2} (1 - \cos \theta)^2\right) \ \frac{1}{g_p g_Q} \sum_{k,l}  |\mathcal{M}_{2,2}^{\textrm{on}} (p, q; i, j |p', q'; k, l)|^2 (s,t),
\nonumber\\
& & {} = \frac{4 \pi p_{cm}^3}{\sqrt{s}} m_2 (s),
\end{eqnarray}}
  where we have defined:

{\setlength\arraycolsep{-6pt}
\begin{eqnarray}
\label{equ:SecD.5}
& & m_1^{\textrm{on}} (s) = \int_{-1}^1 d \cos \theta \frac{1 - \cos \theta}{2} \sum |\mathcal{M}^{\textrm{on}}|^2 = \frac{1}{8 p_{cm}^4} \int_{- 4 p_{cm}^2}^{0} \ \sum |\mathcal{M}^{\textrm{on}}|^2 (s, t) \times (- t) d t
\nonumber\\
& & {} m_2^{\textrm{on}} (s) = \int_{-1}^1 d \cos \theta  \left(\frac{1 - \cos \theta}{2} \right)^2 \sum |\mathcal{M}^{\textrm{on}}|^2 = \frac{1}{32 p_{cm}^6} \int_{- 4 p_{cm}^2}^{0} \ \sum |\mathcal{M}^{\textrm{on}}|^2 (s, t) \ t^2 d t.
\end{eqnarray}}
  Using a Lorentz boost we transform $b_{cm}^{i j}$ in eq.(\ref{equ:SecD.2}), with $b_{T,cm} (\bsp{p})$ and $b_{L,cm} (\bsp{p})$
  given by (\ref{equ:SecD.3}) and (\ref{equ:SecD.4}), respectively, to the heavy-quark rest frame. Together with the remaining integral in
  $\mathcal{B}^{\mu \nu}$ (\ref{equ:Sec11.1}), one can prove that $\mathcal{B}^{\mu \nu}$ is expressed in the heavy-quark rest frame by

{\setlength\arraycolsep{-6pt}
\begin{eqnarray}
\label{equ:SecD.6}
\mathcal{B}_{rest}^{\mu \nu} & & \ = \frac{M_Q^2}{16(2\pi)^4} \int \frac{q^5 d q}{s^2 E_q} \ \int d \Omega_{rel} \ f_r (\bsq{q}) \Biggl[\biggl(m_1(s) - m_2(s)\biggr) (I - |\hat{q}_{cm}><\hat{q}_{cm}|)^{\mu \nu} + 2 m_2 (s) \ \tilde{a}^{\mu} \tilde{a}^{\nu} \Biggr]
\end{eqnarray}}
with $\tilde{a}^{\mu}$ is the a 4-vector defined by

{\setlength\arraycolsep{-6pt}
\begin{eqnarray}
\label{equ:SecD.7}
\tilde{a}^{\mu} = \begin{pmatrix}
\frac{q}{\sqrt{s}} \vspace*{0.3cm} \\
\frac{M_Q + E_q}{\sqrt{s}} \ \hat{\bsq{q}}_{cm}
\end{pmatrix}.
\end{eqnarray}}
Averaging over $\hat{\bsq{q}}_{cm}$, $\mathcal{B}_{rest}^{\mu \nu}$ can be written in the basis $|\hat{u}><\hat{u}|$, where $u$ is the fluid
velocity in the heavy quark rest frame as

{\setlength\arraycolsep{-6pt}
\begin{eqnarray}
\label{equ:SecD.8}
\mathcal{B}_{rest}^{\mu \nu} = \frac{M_Q^2}{8(2\pi)^3} \int \frac{q^5 d q}{s^2 E_q} & & \ \Biggl\{ 2 \ m_2(s) \begin{pmatrix}
\frac{q^2}{s} f_{0} &  \hspace*{0.5cm} f_{1} \frac{q(M_Q + E_q)}{s} \hat{u} \vspace*{0.3cm} \\
f_{1} \frac{q(M_Q + E_q)}{s} \ \hat{u}  & \hspace*{0.3cm}  0
\end{pmatrix}
\nonumber\\
& & {} + \Biggl[\biggl(\frac{m_1(s) - m_2(s)}{2}\biggr) (f_{0} + f_{2}) + \frac{(M_Q + E_q)^2}{s} \ m_2 (s) (f_{0} - f_{2}) \Biggr] (I - |\hat{u}><\hat{u}|)
\nonumber\\
& & {} + \Biggl[\biggl(m_1(s) - m_2(s)\biggr) (f_{0} - f_{2}) + \frac{2 (M_Q + E_q)^2}{s} \ m_2 (s) f_{2} \Biggr] |\hat{u}><\hat{u}| \Biggr\},
\end{eqnarray}}
  with

{\setlength\arraycolsep{-6pt}
\begin{eqnarray}
\label{equ:SecD.9}
& & f_{i} = \frac{1}{2} \int d \cos \theta f \left(\frac{u^0 E_q - u q \cos \theta}{T} \right) \ \cos^i \theta, \hspace*{0.3cm} i \in[0,1,2].
\end{eqnarray}}
  Finally, a Lorentz boost of $\mathcal{B}_{rest}^{\mu \nu}$ eq.(\ref{equ:SecD.8}) with $(\gamma_{u} \beta_{u} = \bsp{p}/M_Q, \gamma_u = E_p/M_Q)$ allows
  to determine $\mathcal{B}^{\mu \nu}$ in the heat bath rest frame. For the partial parts of $B_{\textrm{heat bath}}^{\mu \nu}$, one finds

{\setlength\arraycolsep{-6pt}
\begin{eqnarray}
\label{equ:SecD.10}
& & B_{T,\textrm{heat bath}} (\bsp{p}) = \frac{M_Q^3}{8(2\pi)^3 E_Q} \int \frac{q^5 d q}{s^2 E_q}  \Biggl[\frac{m_1(s) - m_2(s)}{2} (f_{0} + f_{2}) + \frac{(E_q + M_Q)^2}{s} m_2 (s) (f_{0} - f_{2}) \Biggr]
\\
& & {} B_{L,\textrm{heat bath}} (\bsp{p}) = \frac{M_Q^3}{8(2\pi)^3 E_Q} \int \frac{q^5 d q}{s^2 E_q}  \Biggl[\!\! \biggr(m_1(s) - m_2(s)\biggl) (f_{0} - f_{2}) u_0^2 + \frac{2 m_2 (s)}{s} \Biggl((E_q + M_Q)^2 u_0^2 f_{2} - 2 u_0 u \ q (E_q + M_Q) f_{1} + u^2 q^2 f_{0} \Biggr) \!\! \Biggr].
\nonumber
\end{eqnarray}}
$\hat{q}^{on}$ which is proportional to $B_{T}$ (cf. eq.(\ref{equ:Sec11.5})) is then given in the heat bath rest frame by \cite{PBGossiauxHD}:

{\setlength\arraycolsep{-6pt}
\begin{eqnarray}
\label{equ:SecD.11}
& & \hat{q}^{\textrm{on}} (p) = \frac{M_Q^3}{2(2\pi)^3 p_Q} \int_{0}^{+ \infty} \frac{q^5}{s^2 E_q} \Biggl[\frac{m_1^{\textrm{on}} (s) - m_2^{\textrm{on}} (s)}{2} \left(f_{0} + f_{2} \right) + \frac{(E_q + M_Q)^2}{s} \ m_2^{\textrm{on}} (s) \left(f_{0} - f_{2} \right) \Biggr] \ d q.
\end{eqnarray}}

  \bibliography{References}

\end{document}